\documentclass[english, aip, pop, amsmath,amssymb, reprint]{revtex4-1}

\pdfoutput=1  

\usepackage{graphicx}
\usepackage{dcolumn}
\usepackage{bm}

\makeatletter

\providecommand{\tabularnewline}{\\}

\@ifundefined{textcolor}{}
{%
 \definecolor{BLACK}{gray}{0}
 \definecolor{WHITE}{gray}{1}
 \definecolor{RED}{rgb}{1,0,0}
 \definecolor{GREEN}{rgb}{0,1,0}
 \definecolor{BLUE}{rgb}{0,0,1}
 \definecolor{CYAN}{cmyk}{1,0,0,0}
 \definecolor{MAGENTA}{cmyk}{0,1,0,0}
 \definecolor{YELLOW}{cmyk}{0,0,1,0}
}

\makeatother

\begin{document}

\title{Fast-ignition design transport studies: realistic electron source,
integrated PIC-hydrodynamics, imposed magnetic fields}

\author{D. J. Strozzi, M. Tabak, D. J. Larson, L. Divol, A. J. Kemp, C. Bellei,
M. M. Marinak, M. H. Key}

\affiliation{Lawrence Livermore National Laboratory, 7000 East Ave., Livermore,
CA 94550}

\date{5/7/12}
\begin{abstract}
Transport modeling of idealized, cone-guided fast ignition targets
indicates the severe challenge posed by fast-electron source divergence.
The hybrid particle-in-cell {[}PIC{]} code Zuma is run in tandem with
the radiation-hydrodynamics code Hydra to model fast-electron propagation, fuel heating, and thermonuclear burn. The fast electron source
is based on a 3D explicit-PIC laser-plasma simulation with the PSC
code. This shows a quasi two-temperature energy spectrum, and a divergent
angle spectrum (average velocity-space polar angle of $52^{\circ}$).
Transport simulations with the PIC-based divergence do not ignite
for $>$ 1 MJ of fast-electron energy, for a modest (70 $\mu\mathrm{m}$)
standoff distance from fast-electron injection to the dense fuel.
However, artificially collimating the source gives an ignition energy
of 132 kJ. To mitigate the divergence, we consider imposed axial magnetic
fields. Uniform fields $\sim$50 MG are sufficient to recover the
artificially collimated ignition energy. Experiments at the Omega
laser facility have generated fields of this magnitude by imploding a capsule in
seed fields of 50-100 kG. Such imploded fields are however more compressed in the transport region than in the laser absorption
region. When fast electrons encounter increasing field strength,
magnetic mirroring can reflect a substantial fraction of them and
reduce coupling to the fuel. A hollow magnetic pipe, which peaks at
a finite radius, is presented as one field configuration which circumvents
mirroring. 
\end{abstract}


\maketitle

\section{Introduction}

\global\long\def\Efast{E_{\mathrm{fast}}}
 \global\long\def\Estop{E_{\mathrm{stop}}}
\global\long\def\rspot{r_{\mathrm{spot}}}
 \global\long\def\micron{\mu\mathrm{m}}

The fast ignition approach to inertial fusion exploits a short-pulse,
ultra-intense laser to heat an isochoric hot spot to ignition conditions
\citep{tabak-fi-pop-1994}. Unlike the central hot-spot approach,
fast ignition separates dense fuel assembly from hot-spot formation
\citep{basov-fastign-1992}. This opens the prospect of high energy
gain with less laser energy, and may be an attractive avenue
for inertial fusion energy. The first integrated but sub-ignition scale experiments were
performed at Osaka in 2001-2002, and explored the cone-in-shell geometry
\citep{kodama-fastig-nature-2001,kodama-fastig-nature-2002}. Subsequent
similar experiments were done at Vulcan \citep{key-fastign-pop-2008},
Omega EP \citep{theobald-fastign-pop-2011}, and Osaka in 2009 and
2010 \citep{shiraga-gekko-ppcf-2011,fujioka-sofe-2011}. The 2002
Osaka experiments were interpreted to show high coupling of short-pulse
laser energy to the fuel, of order 20\%. All the later experiments show lower
coupling, the best being 10-20\% coupling in the 2010 Osaka experiments
\citep{shiraga-gekko-ppcf-2011,fujioka-sofe-2011}. This work suggests
the reduced coupling seen in 2009 at Osaka was due to higher pre-pulse in the short-pulse
laser. Pre-pulse energy creates an underdense pre-plasma in which the laser converts
to over-energetic electrons, and this source is farther away from the fuel. The negative
impact of pre-plasma on fast-electron generation inside a cone has
been reported, e.g., in
Refs.~\onlinecite{baton-cone-pop-2008,macphee-prepulse-prl-2010,ma-conewire-prl-2012}. Coupling
efficiences at small scale do not directly apply at ignition scale.

This paper presents integrated fast-ignition modeling studies at
ignition scale, which is well beyond parameters currently accessible
by experiment. We utilize a new, hybrid PIC code Zuma \citep{larson-zuma-dpp-2010}
to model fast electron transport through a collisional plasma, with
self-consistent return current and electric and magnetic field generation.
To alleviate the need to resolve light waves or background Langmuir
waves, Zuma does not include the displacement current in Amp\`ere's
law, and employs an Ohm's law (obtained from the inertialess limit
of the background electron momentum equation) to find the electric
field. We recently coupled Zuma to the radiation-hydrodynamics code
Hydra \citep{marinak-hydra-pop-2001}, which has been widely used
to model inertial fusion and other high-energy-density systems. 

We do not model the short-pulse laser, but instead inject electrons
with a specified distribution into Zuma. The source electron spectrum
is a key element of this approach. We obtain the spectrum
by using the particle-in-cell code PSC \citep{bonitz-psc-manybody,kemp-pop-2010} to
perform a 3D full-PIC simulation of the laser-plasma interaction (LPI).
This gives a quasi two-temperature energy spectrum, with a (cold,
hot) temperature of (19, 130)\% of the so-called ponderomotive temperature as defined below \citep{wilks-pond-prl-1992} at
the nominal laser intensity.  It is generally seen in PIC simulations that
LPI occurring at lower density produces more energetic electrons. The experimental understanding of fast
electron energy spectra is not entirely clear. Ma \emph{et al.}~recently
reported experimental evidence indicating a two-temperature energy
spectrum \citep{ma-conewire-prl-2012}, albeit at lower energies and shorter
pulses than considered here for ignition.

The PIC-based angle spectrum is very divergent, with an average
polar angle in velocity space of 52$^{\circ}$ or an integrated solid angle of
4.85 sterad. A large divergence
has been reported in other PIC simulations, such as Refs.~\onlinecite{ren-aps05-pop-2006,adam-shortpulse-prl-2006}.
Experimental evidence for a significant divergence comes from modeling
by Honrubia\emph{ et al.}~\citep{honrubia-heating-lpb-2006} of $K_{\alpha}$
data obtained by Stephens \emph{et al.}~\citep{stephens-fastign-pre-2004}.
More recent work by Westover \emph{et al.}~\citep{westover-dpp-2011}
also indicates a substantial source divergence.

Our Zuma-Hydra modeling with a realistic fast electron source (both
energy and angle spectra) and an idealized fuel mass located 70 $\mu$m from the
electron source indicates poor coupling to the fuel hot
spot, with $>1$ MJ of fast electrons inadequate to ignite. Artificially
collimating the electron source dramatically improves the picture,
with an ignition energy of 132 kJ. This is much higher than the ideal
estimate of 8.7 kJ absorbed in the ignition hot spot (detailed below), due largely to the energy spectrum
being too hot for the electrons to stop fully in the hot spot. In
Ref.~\onlinecite{strozzi-ifsa-epjwc-2012}, we report in more detail the
effects of the energy spectrum, as well as E and B fields, on the
ignition requirements for an artificially-collimated fast electron
source. We merely note here that, for our particular plasma condition
profiles, using the complete Ohm's law Eq.~(\ref{eq:ebmom}) reduces
the fast-electron coupling to the fuel compared to the case of no
E or B fields, while using the resistive Ohm's law $\vec{E}=\eta\vec{J}_{b}$
increases the coupling over the no-field case. This is likely due to $\nabla n_{e}\times\nabla T_{e}$
magnetic fields that develop at the outer radius of the dense fuel
and push the fast electrons to larger radius, as observed earlier
in Ref.~\onlinecite{nicolai-xport-pre-2011}.

The focus of this paper is on mitigating the beam divergence by imposed
magnetic fields. In particular, we do not pursue here other attractive
options, such as field generation by resistivity gradients \citep{robinson-switchyard-pop-2007}.
Cylindrical \citep{knauer-bfield-pop-2010} and spherical \citep{chang-sphere-prl-2011,hohenberger-bfield-dpp-2011}
implosions at the Omega laser have compressed seed magnetic fields
$\sim$10 kG to strengths of 20-40 MG. We show that a uniform, initial
axial field of 50 MG almost recovers the ignition energy of the artificially
collimated beam. We stress that such magnetic fields do not collimate
the electrons (that is, reduce the velocity-space divergence), but
rather confine them in space: once they emerge from the confining
field, they still have their initial divergence. A magnetic field that increases
in the axial direction leads however to substantial
reflection due to magnetic mirroring. We explore a hollow magnetic
pipe, that peaks at a finite radius, as one method to circumvent mirroring.
Pipes with a peak field of 50 MG and radial FWHM of (20, 30) $\mu$m
ignite for (211, 158) kJ of fast electrons, compared with 132 kJ for
an artificially-collimated source with no imposed field. 

The paper is organized as follows. Sec.~II describes the fast electron
source derived from a full PIC simulation of the LPI. In Sec.~III
we detail the Zuma model, and how it is coupled to Hydra. Sec.~IV
presents Zuma-Hydra results with realistic fast electron divergence
and an artificially collimated source. We study ways to mitigate the
source divergence with several imposed magnetic fields (uniform, increasing,
and a hollow pipe) in Sec.~V. In Sec.~VI we make some concluding
remarks.

\section{Electron source from full-PIC laser-plasma modeling}

\label{sec:source}The distribution of fast electrons produced by
the short-pulse LPI is a crucial element of fast ignition. 
The computational scale of integrated LPI-transport-hydro simulations is currently
prohibitive. We excite a fast electron source derived from full-PIC
LPI simulations in our transport modeling. This neglects feedback
of the transport on the LPI, e.g. the exact details of the return
current, and also loses some detail in the fast-electron source. It
does call out the major design challenges, and allows for the development
of ideas to mitigate them.

Our LPI simulations are performed with the relativistic PIC code PSC
\citep{bonitz-psc-manybody}. This code has recently been extended
to include a hybrid model, valid for collisional plasmas \citep{cohen-psc-jcp-2010}.
The results presented here do not use the hybrid model, and are all explicit,
full-PIC calculations, with the complete Maxwell equations and no
collision operator. We note that more recent PSC simulations of several
ps duration show the convergence of profiles with and without an initial
pre-plasma, and the development of a third, super-hot component (temperature
several times the ponderomotive value given below) besides the two components present in
the results discussed here. Our source will be updated with these new results in
the near future.

\begin{figure}
\includegraphics[width=3in]{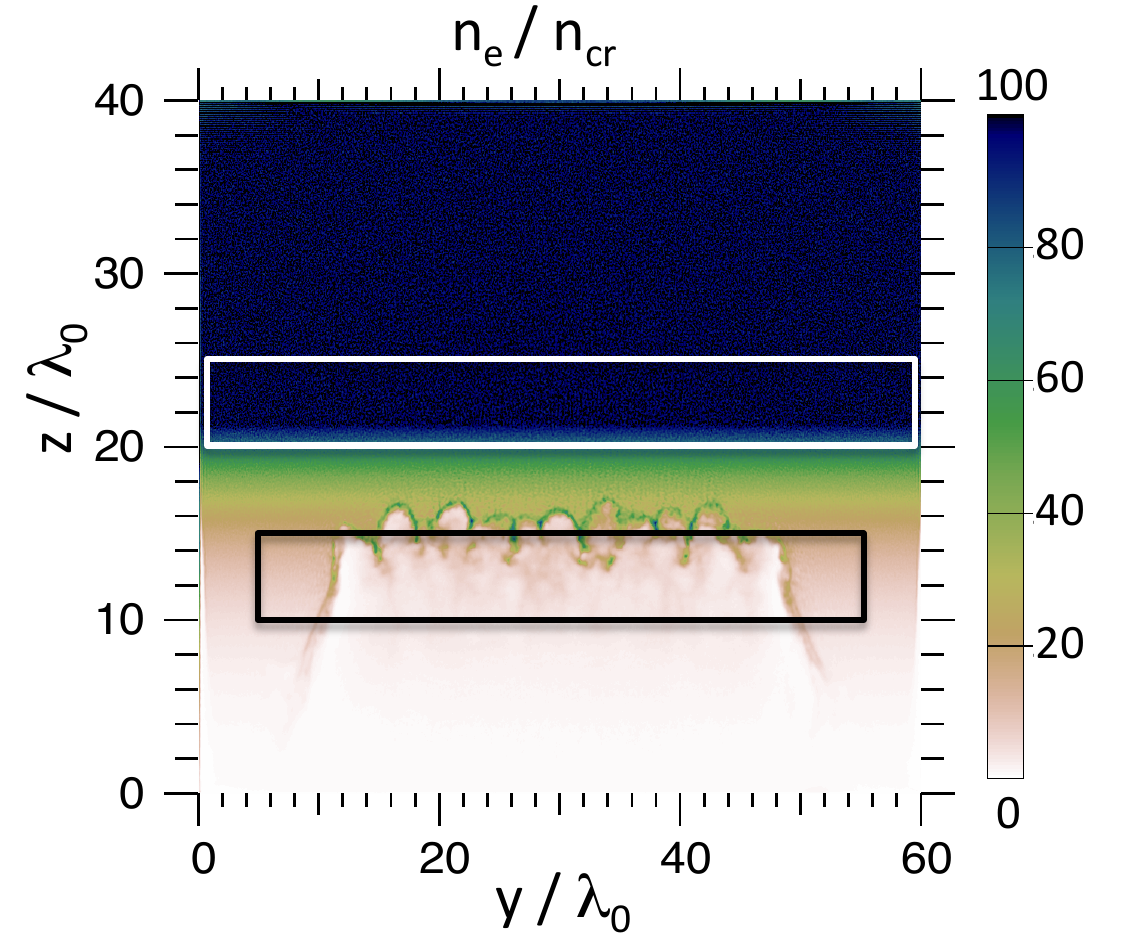}\caption{(Color online) Electron density at time 360 fs in the PSC run used
to characterize the fast electron source. The white box indicates
the extraction box, and the black box indicates the source box in
the hybrid-implicit LSP run. The laser was incident from $z=0$ with
a vacuum focus at $z=10$ $\micron$.}

\label{fig:psc}
\end{figure}

The specific PSC run used for the fast electron source was as follows.
The geometry was 3D Cartesian, and the electron density at time 360
fs over a 2D plane is shown in Fig. \ref{fig:psc}. The domain extended
from $-$30 to $+$30 $\micron$ in the two transverse ($x$ and $y$)
directions and from 0 to 40 $\micron$ in $z$ (nominal direction
of laser propagation). The particle and field boundary conditions
(BCs) were periodic in $x$ and $y$. In $z$, the particle BCs were
thermalizing re-emission, while the field BCs were radiative (outgoing-wave).
The initial plasma profile was $n_{e}/n_{cr}=100$ for $z>z_{0}$
and $\exp[(z-z_{0})/3.5\,\mu\mathrm{m}]$ for $z<z_{0}$ with $z_{0}=$20
$\micron$ and $n_{cr}=1.115\cdot10^{21}/\lambda_{0}^{2}[\mu\mathrm{m]}$
cm$^{-3}$, the non-relativistic laser critical density. This profile
was chosen to replicate the pre-plasma produced by a small pre-pulse ($\sim$1-10 mJ)
in the short-pulse laser (e.g. growing from ASE). Both electrons and
deuterium ions ($Z/A=1/2$) were mobile. The uniform cell size was
$\Delta x=\Delta y=\lambda_{0}/16$ and $\Delta z=\lambda_{0}/16.375$.
The time step was $c\,\Delta t=0.421\Delta x$. There were (twelve,
four) numerical macro-particles per cell for (electrons, ions). The
run required about 160k cpu-hours to complete.

The laser had a vacuum wavelength of $\lambda_{0}=$1 $\micron$ and
a vacuum focal spot at $z=$10 $\micron$ with radial intensity profile
$I(r)=I_{0L}\exp\left[-(r/18.3\,\mu\mathrm{m})^{8}\right]$. For a given laser power and maximum (hard-edge) spot radius, a flat
as opposed to peaked (e.g., Gaussian) profile reduces the average
intensity and gives a cooler spectrum. We chose $I_{0L}=1.37\cdot10^{20}$
W/cm$^{2}$, corresponding to a normalized vector potential $a_{0}=10$.
The ponderomotive temperature, as defined in Ref.~\onlinecite{wilks-pond-prl-1992},
for the peak laser intensity is $T_{p}/m_{e}c^{2}=\left[1+a_{0}^{2}\right]^{1/2}-1=9.05$
, or $T_{p}=4.62$ MeV. We simply use $T_{p}$ to denote an energy
scale, without any implication about what the fast electron distribution
is (which is discussed below). The laser pulse was ramped up to peak
intensity over 30 fs.

We took all electrons at the time 360 fs, in a cylindrical ``extraction
box'' from $z=20-25$ $\micron$ and radius 30 $\micron$. This box
is deep enough into the overdense region that the laser did not propagate
there. We also selected only electrons with $0.55<E\,[\mathrm{MeV]}<29.5$
(or $0.12<E/T_{p}<6.37$) and $v_{z}>0$, where $E=m_{e}c^{2}(\gamma-1)$.
This is done to eliminate the return current and background heating.
Some of this heating is unphysical grid heating, and some is a legitimate
kinetic energy transfer between the fast electrons and the background plasma at
the 100$n_{cr}$ density used in the PIC simulation for numerical reasons. This
heating is expected to play a negligible role at the densities assumed in our
hybrid simulations \citep{kemp-collisions-prl-2006}.

For our transport studies, we do \emph{not} inject a fast electron source in an
analogous extraction box. In particular, such a source would need a radially
outward drift that varies with radius. Instead, we excite fast electrons in a
``source box'' analogous to the laser absorption region, such that after
propagating a small distance into an equivalent extraction box, the
transport-code electron distribution matches that of the PIC electrons. This
method automatically handles a host of issues regarding propagation from the
source to extraction regions (e.g., finite ``view factors'' that vary with angle
and radius), and provides the overall laser to electron conversion efficiency.
The source electron intensity %
\footnote{By source intensity, we mean the injected kinetic energy per time, per
  transverse area in the injection plane. This differs from the $z$ flux of
  kinetic energy.%
} is $\alpha_{CE}I(r)$ where $I(r)$ is the vacuum laser intensity given above,
and $\alpha_{CE}=0.52$ is an overall laser-to-electron power conversion
efficiency.  $\alpha_{CE}$ was chosen so the total fast electron kinetic energy
in the PSC and transport-code extraction boxes match. The source intensity is varied in
space and time only by varying the rate of excitation, not the velocity-space
distribution.

To arrive at the transport-code distribution excited in the source
box, we performed a hybrid implicit-PIC simulation with the LSP plasma
simulation code \citep{welch-lsp-pop-2006},
with kinetic fast electrons and fluid background species (this is
the only time LSP is used in this paper).
The LSP source box was located from $z=$ 10 to 15 $\micron$, and
the plasma was uniform 10 g/cm$^{3}$ carbon at 100 eV. We used this
much denser background than the PSC simulation because the fluid model
is only valid at high collisionality, and our transport studies are
performed in compressed matter. The difference between free-particle
propagation and the full LSP results are small, indicating that forces
are not important as electrons transit from source to extraction boxes.

\begin{figure}
\includegraphics[width=3in]{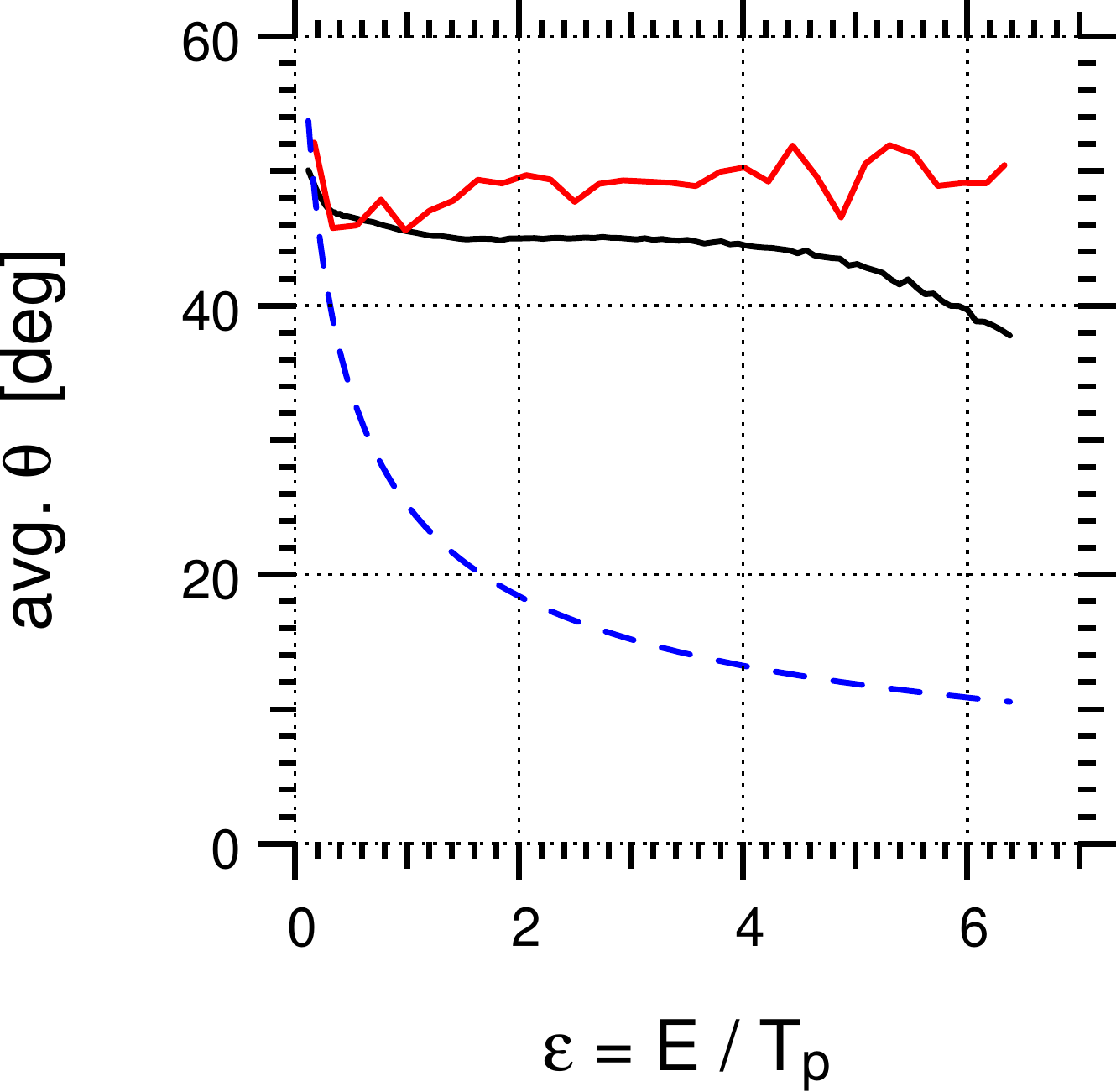}

\caption{(Color online) Average velocity-space polar angle vs.~electron kinetic
energy in the extraction box for the PSC full-PIC (solid black) and
LSP implicit-PIC (solid red) runs. The classical ejection angle $\theta_{c}$
from Eq.~(\ref{eq:cea}) is plotted in dashed blue.}

\label{fig:qvsE}
\end{figure}

The following LSP source gave electrons in the extraction box that
agreed adequately with the PSC extraction box electrons. The source
velocity-space distribution is azimuthally symmetric and given by
$d^{2}N/dEd\theta=N_{0}f_{E}(E)f_{\theta}(\theta)$, so that $(f_{E},f_{\theta})$
are proportional to the 1D distributions $(dN/dE,dN/d\theta)$. $N_{0}$ is an overall
normalization factor. $\tan\theta=\left[v_{z}/\left(v_{x}^{2}+v_{y}^{2}\right){}^{1/2}\right]$
defines the polar angle in velocity space. For PSC runs at ignition
powers and wide focal spots, we generally find the angle spectrum
does not vary much with energy (see Fig.~\ref{fig:qvsE}). This justifies
our 1D factorization; the method can be easily extended to several energy
bins each with different $f_{\theta}$. 

\begin{figure}
\includegraphics[width=2.3in]{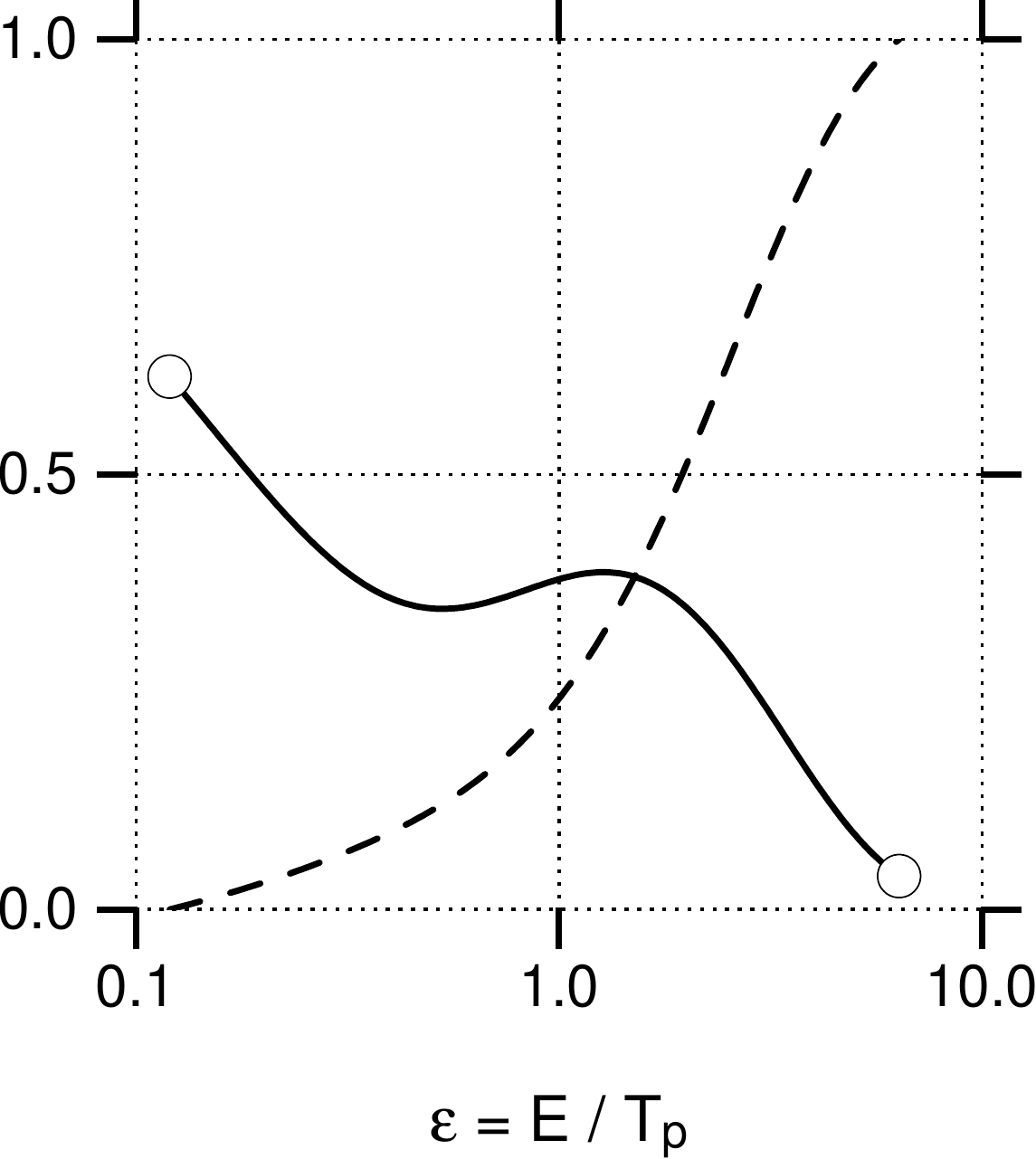}

\caption{Source energy spectrum $E\, f_{E}$ (solid) and its running integral
(dash) for the analytic form in Eq.~(\ref{eq:fE}). The circles indicate
the limits of the domain taken in the extraction box, and injected
into our Zuma-Hydra simulations.}

\label{fig:dNdE}
\end{figure}

\subsection{Fast electron energy spectrum}

For $f_{E}$ we use the 1D energy spectrum $dN/dE$ found in the PSC
extraction box. This is well-fit by a quasi two-temperature form:
\begin{equation}
f_{E}(\epsilon)=\frac{1}{\epsilon}\exp\left[-\epsilon/\tau_{1}\right]+0.82\exp\left[-\epsilon/\tau_{2}\right].\label{eq:fE}
\end{equation}
$\epsilon\equiv E/T_{p}$ is the ponderomotively scaled energy, and
we assume as we vary the laser intensity and wavelength that $dN/dE$
scales in this manner. Figure \ref{fig:dNdE} plots this analytic
form, as well as its running integral. The temperature-like parameters
have the values $\tau_{1}=0.19$ and $\tau_{2}=1.3$. These correspond
to a relatively ``cold'' component produced by LPI near or above
$n_{cr}$, and a slightly hotter than $T_{p}$ component arising from
underdense LPI. The $1/\epsilon$ factor on the cold part improves
the fit at low energy, although this may change as better ways to
eliminate return current and background heating are developed. We
only inject over the domain $0.12<\epsilon<6.37$ , which is the domain
taken from the PSC extraction box. The average injected electron energy
is $\left\langle \epsilon\right\rangle =1.02$, while only 24\% of
the injected energy is carried by electrons with $\epsilon<1$. This
is unfortunate for ignition with the laser intensities we contemplate,
as discussed in Sec. \ref{sec:noB}.

\begin{figure}[th]
\includegraphics[width=3in]{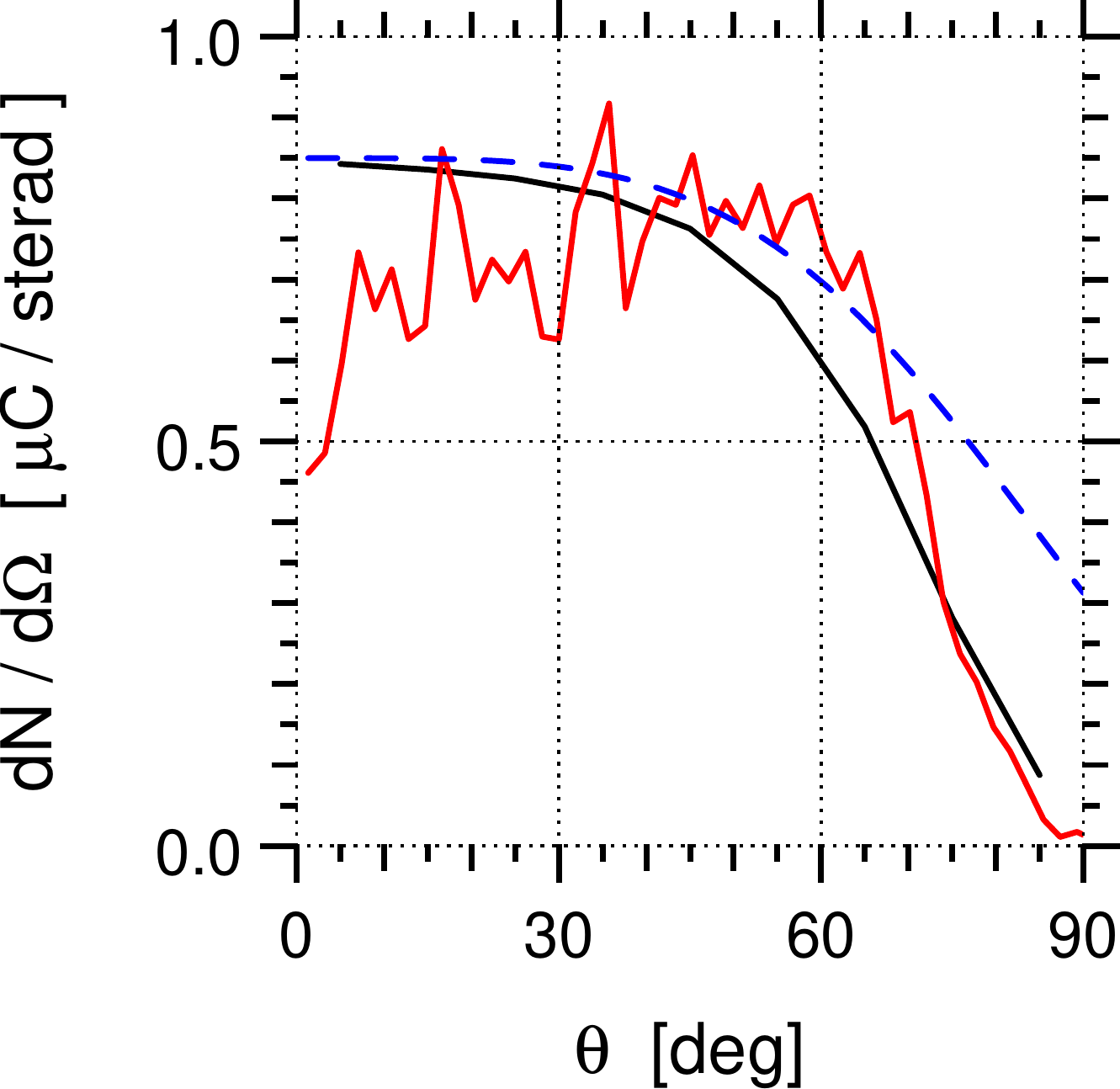}

\caption{(Color online) Solid angle spectra in extraction box for PSC run as
described in text (black), LSP run extraction box (red), and source
$f_{\Omega}$ from Eq. (\ref{eq:fq}) with an arbitrary scale factor
(blue dashed). The source spectrum is broader than those in the extraction
boxes due to limited view factor at large angles. Note that the first
two curves are given in physical units. }

\label{fig:dNdW}
\end{figure}

\subsection{Fast electron angle spectrum}

The average angle $\theta$ in the extraction box as a function of
electron energy is displayed in Fig.~\ref{fig:qvsE}. The PSC full-PIC
$\theta$ becomes slightly more collimated at higher energies, while
the LSP implicit-PIC $\theta$ is essentially independent of energy.
The agreement is excellent for $E<T_{p}$, but the LSP $\theta$ is
slightly larger at large energies. We consider the energy dependence
of $\theta$ to be weak enough to ignore and use a factorized source.
Both PIC simulations have much larger $\theta$, and much less decrease
with energy, than the classical ejection angle $\theta_{c}$
given by \citep{quesnel-elec-pre-1998}
\begin{equation}
\tan\theta_{c}=\left[\frac{2}{\gamma-1}\right]^{1/2}.\label{eq:cea}
\end{equation}
This result obtains for a single electron in a focused laser beam in vacuum, not including plasma effects.

\begin{figure}[th]
\includegraphics[width=3in]{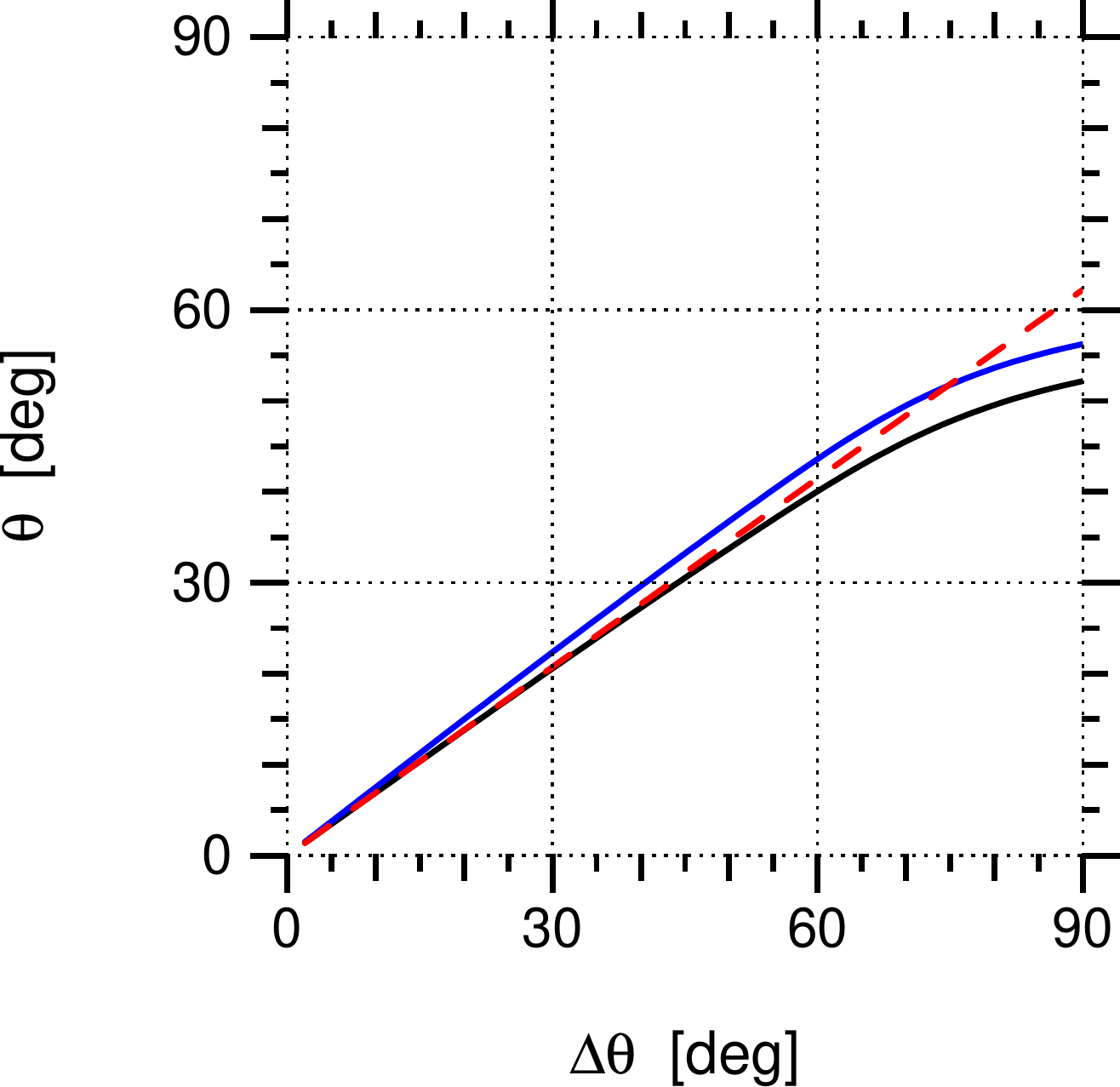}

\caption{Average (solid black) and rms (solid blue, slightly higher) polar
angle for the angle spectrum in Eq. (\ref{eq:fq}), vs. the parameter
$\Delta\theta$. The red dashed line is the approximate form $\left\langle \theta\right\rangle =0.691\Delta\theta$
from Eq. (\ref{eq:qav}).}

\label{fig:qav}
\end{figure}

The source angle spectrum we use is
\begin{equation}
f_{\theta}(\theta)=2\pi\sin(\theta)f_{\Omega}\qquad f_{\Omega}=\exp\left[-\left(\theta/\Delta\theta\right)^{4}\right].\label{eq:fq}
\end{equation}
$\Omega$ represents solid angle, related to $\theta$ by $d\Omega=2\pi\sin(\theta)d\theta$.
The value of the parameter $\Delta\theta$ that gives good agreement
with the angle spectrum in the extraction box is $\Delta\theta=90^{\circ}$.
Figure \ref{fig:dNdW} displays the source $f_{\Omega}$ as well as
the angle spectra in the PSC and LSP extraction boxes. The resulting
extraction box angle spectrum is somewhat narrower than this source,
due to limited view factor at large $\theta$. In addition, the LSP
extraction spectrum is depleted at small angle compared to the PSC
spectrum, and may slightly overstate the divergence (although there
are few particles at small $\theta$ due to the $\sin\theta$ Jacobian).
We stress that $\Delta\theta$ is a parameter in a function, and not
an observable quantity. The average $\theta$, which has physical
meaning, is 
\begin{equation}
\left\langle \theta\right\rangle \equiv\frac{\intop_{0}^{\pi/2}d\theta\, f_{\theta}\theta}{\intop_{0}^{\pi/2}d\theta\, f_{\theta}}\approx\frac{\Gamma[3/4]}{\pi^{1/2}}\Delta\theta=0.691\Delta\theta,\,\,\,\,\,\,\,\Delta\theta<1.\label{eq:qav}
\end{equation}
$\Gamma$ is the mathematical gamma function. $\left\langle \theta\right\rangle $
and the rms $\theta$ are shown vs. $\Delta\theta$ in Fig. \ref{fig:qav}.
Note that for large $\Delta\theta$, $\left\langle \theta\right\rangle $
falls below the approximate linear result given above. For $\Delta\theta=90^{\circ}$
we find $\left\langle \theta\right\rangle =52^{\circ}$ and rms $\theta=56^{\circ}$.
The integrated solid angle $\Omega_I$ is $\int_0^{2\pi}d\Omega\,f_\Omega=4.85$
sterad (for an isotropic distribution, $f_\Omega=1$ and $\Omega_I=2\pi$). This is a substantial divergence; the rest of this paper is focused
on its consequences for fast ignition requirements, and mitigation
ideas based on imposed magnetic fields. 

We briefly note that our source is symmetric in azimuthal angle. However,
it quickly develops a radially outward drift as it propagates. That
is, the average angle, $\chi$, between an electron's position and
velocity vectors decreases. For free motion, in the far-field limit
they become parallel: $\vec{r}\approx\vec{v}t$. In the extraction
box the LSP and PSC electrons have similar $\chi$ distributions vs.~radius.
Debayle \citep{debayle-diverge-pre-2010} has recently discussed the
role of an intrinsic radial drift produced by a Gaussian laser; we
find such a drift naturally develops due to propagation from a symmetric
source.

\section{Zuma-Hydra integrated modeling}

Our transport modeling is done with the hybrid-PIC code Zuma \citep{larson-zuma-dpp-2010}
coupled to the radiation-hydrodynamics code Hydra \citep{marinak-hydra-pop-2001}.
We describe here the Zuma model in some detail, and how it is coupled
to Hydra. We briefly discuss how we run Hydra, and refer the reader
to the extensive literature on this well-established code.

\subsection{Zuma hybrid-PIC code}

Zuma is a parallel, hybrid-PIC code that currently supports 3D Cartesian
and 2D cylindrical RZ geometry. It employs an explicit time-stepping
approach, treats the fast electrons by a standard, relativistic PIC
method, and models the background plasma as a collisional fluid. The
electric field is found from Ohm's law (i.e., the momentum equation
for the background electrons in the limit $m_{e}d\vec{v}_{eb}/dt\rightarrow0$),
and the background return current is found from Amp\`ere's law without
displacement current. This reduced-model approach is similar to Gremillet
\citep{gremillet-fil-pop-2002}, Honrubia \citep{honrubia-ppcf-2009},
Davies \citep{davies-mc-pre-2002}, and Cohen \textit{et al.~}\citep{cohen-psc-jcp-2010}
(although the last approach uses particles to describe the collisional,
``fluid'' background). This combination eliminates both light and
Langmuir waves, and allows stable modeling of dense plasmas without
needing to resolve these fast modes. An alternative approach to dense-plasma
modeling is implicit PIC \citep{langdon-imppic-1985,hewett-imppic-jcp-1987},
which numerically damps unresolved, high-frequency modes, and is utilized
in codes such as LSP \citep{welch-lsp-pop-2006} and ELIXIRS \citep{drouin-pic-jcp-2010}.
Of course, the reduced-model approach is inapplicable to laser-plasma
interactions, or low-density regions with, e.g., Debye sheaths. Ion
dynamics is not modeled in Zuma (although including them is consistent
with the reduced-model approach), and we assume charge neutrality:
$n_{eb}=\bar{Z}n_{I}$ where $n_{eb}$ is the number density of free
(not atomically bound) background electrons and $n_{I}=\sum_{i}n_{i}$
is the total ion density. 

Zuma advances each fast electron by
\begin{equation}
d\vec{x}/dt=\vec{v}, \qquad d\vec{p}/dt=-e\left(\vec{E}+\vec{v}\times\vec{B}\right)-\nu_{d}\vec{p}+\vec{R}
\end{equation}
where $\vec{p}=m_{e}c\gamma\vec{\beta}$ is the relativistic momentum.
The term $-\nu_d\vec p$ is frictional drag (energy loss), and the Langevin term
$\vec R$
represents a random rotation of $\vec p$ which gives
angular scattering. We uss the drag and scattering formulas of Solodov and Betti \citep{solodov-stop-pop-2008},
and Davies \textit{et al.}\ \citep{atzeni-ppcf-2009,davies-stopping-dpp-2008}.
We follow the numerical approach of Lemons \citep{lemons-collision-jcp-2009},
by applying drag directly to $|\vec{p}|$ and then
randomly rotating its direction. Manheimer \citep{manheimer-pic-jcp-1997}
presented a similar collision algorithm which acts on Cartesian velocity
components. Binary-collision algorithms like that of Takizuka and Abe \citep{takizuka-pic-jcp-1977}
have advantages like exact conservation relations and can be applied to models
like ours \citep{cohen-collisions-jcp-2012}. They generally require, however, the drag and scattering to satisfy an
Einstein relation and thus have the same ``Coulomb logarithm,'' which is \textit{not} the
case for the formulas used here.  An Einstein relation obtains when both
processes result from many small, uncorrelated kicks to the
test-particle momentum.  Our angular scattering arises from such binary collisions, but the
energy loss also contains a collective (Langmuir-wave
emission) part. The energy loss is off all background electrons $n_{eb}^{\mathrm{tot}}$
(free and atomically bound), and both types of electrons are treated
using the Solodov-Davies energy-loss formula. This strictly applies
to free electrons, or to bound electrons in the limit where the density
effect dominates \citep{icru37}. Radiative loss is neglected, although
it becomes important for high-$Z$ ions and high-energy electrons.
Specifically, 
\begin{eqnarray}
\nu_{d} &=& \frac{4\pi
  cr_{e}^{2}}{\gamma\beta^{3}}n_{eb}^{\mathrm{{tot}}}L_{d}, \\
L_{d} &=&
\ln\left[\frac{m_{e}c^{2}}{\hbar\omega_{pe}}\beta\left(\frac{\gamma-1}{2}\right)^{1/2}\right]+\frac{9}{16}
\\
&& -\left(\ln2+\frac{1}{8}\right){2\gamma-1 \over 2\gamma^2},
\end{eqnarray}
$ $with $\omega_{pe}^{2}=n_{eb}^{\mathrm{{tot}}}e^{2}/\epsilon_{0}m_{e}$
and $r_{e}=e^{2}/(4\pi\epsilon_{0}m_{e}c^{2})$ the classical electron
radius. This gives rise to a stopping power ($ds=|\vec{v}|dt$) of

\begin{equation}
\frac{dE}{ds}=-m_{e}c\gamma\beta\nu_{d}=-4\pi m_{e}c^{2}r_{e}^{2}\frac{n_{eb}^{\mathrm{{tot}}}}{\beta^{2}}L_{d}.
\end{equation}
The Langevin term $\vec{R}$ is chosen to give the following mean-square
change in pitch angle $\psi$, with respect to $\vec{p}:$
\footnote{The analogous formula, Eq.~(24) in Ref.~\onlinecite{atzeni-ppcf-2009},
contains a typo in the powers of $\beta$ and $\gamma$%
}
\begin{equation}
\frac{\left\langle \left(\Delta\psi\right)^{2}\right\rangle }{\Delta s}=\frac{8\pi cr_{e}^{2}}{\gamma^{2}\beta^{4}}n_{eb}\left(L_{e}+\sum_{i}\frac{Z_{i}^{2}}{\bar{Z}}L_{I}\right),
\end{equation}
\begin{equation}
L_{e}=\ln\Lambda-(1/2)(1+\ln[2\gamma+6]),
\end{equation}
\begin{equation}
L_{I}=\ln\Lambda-(1/2)(1+\beta^{2}).
\end{equation}
$\Lambda=2\lambda_{De}p/\hbar$ with
$\lambda_{De}^{2}=\epsilon_{0}T_{e}/n_{eb}e^{2}$. $Z_i$ is the nuclear charge,
since screening in partially-ionized ions only affects very small-angle scatters.

The fast electron current $\vec{J}_{f}$ is deposited onto the spatial
grid, and the background current $\vec{J}_{b}$ is found from Amp\`ere's
law:
\begin{equation}
\vec{J}_{b}=-\vec{J}_{f}+\mu_{0}^{-1}\nabla\times\vec{B}.\label{eq:amp}
\end{equation}
 The magnetic field is advanced by Faraday's law, $\partial_{t}\vec{B}=-\nabla\times\vec{E}$. 

The electric field is given by the Ohm's law:
\begin{eqnarray}
\vec{E} & = & \vec{E}_{C}+\vec{E}_{NC},\label{eq:ebmom}\\
\vec{E}_{C} & = & \overleftrightarrow{\eta}\cdot\vec{J}_{b}-e^{-1}\overleftrightarrow{\beta}\cdot\nabla T_{e},\nonumber \\
\vec{E}_{NC} & = & -\frac{\nabla p_{e}}{en_{eb}}-\vec{v}_{eb}\times\vec{B}.\nonumber 
\end{eqnarray}
We follow ``notation II'' of Ref.~\onlinecite{epperlein-xport-pof-1986}.
$\vec{E}_{C}$ and $\vec{E}_{NC}$ are, respectively, collisional
and collisionless effects. $\vec{J}_{b}$ is the background electron
current; if ion currents were included, $\vec{J_{b}}$ in Eqs. (\ref{eq:amp},
\ref{eq:ebmom}) should be replaced by the (total, electron) current.
Our Ohm's law neglects terms arising from advection $\sim\vec{v}_{eb}\cdot\nabla\vec{v}_{eb}$,
off-diagonal pressure terms, and collisions between background and
fast electrons. The background temperature $T_{e}$ (currently the
same for both electrons and ions) is updated due to collisional heating,
as well as fast-electron frictional energy loss $Q_{\mathrm{{fric}}}$
(all of which is deposited as heat, not directed flow): 
\begin{equation}
(3/2)(1+1/\bar{Z})n_{eb}\partial_{t}T_{e}=\vec{E}_{C}\cdot\vec{J}_{b}+Q_{\mathrm{{fric}}}.
\end{equation}
We neglect heat flow (e.g.~due to gradients) in Zuma, and rely on
the coupling to Hydra to provide that physics. For the collisional
transport coefficients $\overleftrightarrow{\eta}$ and $\overleftrightarrow{\beta}$,
we use the approach of Lee and More \citep{lee-more-pof-1984}, but
with the numerical tables of Ref.~\onlinecite{epperlein-xport-pof-1986}
to account for electron-electron collisions and background magnetization.
We utilize a modified version of Desjarlais' extension to Lee and
More \citep{desjarlais-metal-cpp-2001}, and use his extension of
Thomas-Fermi theory to find the ionization state $\bar{Z}$.

\subsection{Hydra for transport simulations}

This section describes how we run Hydra for our coupled Zuma-Hydra transport
simulations. We run in cylindrical RZ geometry on a fixed Eulerian
mesh. The radiation is modeled with implicit Monte-Carlo photonics, and
tabulated equation of state and opacity data are used. Fusion reactions occur in
all initially dense ($\rho>10$ g/cm$^{3}$) DT zones, with alpha transport and
deposition done via multi-group diffusion; no neutron deposition is done,
although this could lower the ignition energy slightly. Electron thermal
conduction uses the Lee and More model with no magnetic field
\citep{lee-more-pof-1984}. Although Hydra has an MHD package and the option for
magnetized thermal conduction, we currently do not use these features.

\subsection{Zuma-Hydra coupling}

The coupling of Zuma and Hydra is as follows. Zuma models a subset of Hydra's
spatial domain, since the fast electron source becomes unimportant far enough
from the source box. This paper reports results in cylindrical RZ geometry,
and 3D Cartesian results have been reported in \citep{marinak-zuma-dpp-2010}.
Data transfer between the codes is performed via files produced by the LLNL
Overlink code \citep{grandy-overlink-jcp-1999}, which can interpolate between
different meshes.

The two codes are run sequentially for a set of ``coupling steps''
that are long compared to a single time step of each code. A coupling
step from time $t_{0}$ to $t_{1}$ consists of:
\begin{enumerate}
\item Plasma conditions (materials, densities, temperatures) are transferred
from Hydra to Zuma. 
\item Zuma runs for several time steps from $t_{0}$ to $t_{1}$.
\item The net change in each zone's background plasma energy and momentum is transferred
from Zuma to Hydra, as external deposition rates. 
\item Hydra runs for several time steps from time $t_{0}$ to $t_{1}$. 
\end{enumerate}


Zuma calculates its own ionization state each timestep, and does not
use Hydra's value. For the results in this paper, we ran both codes
for 20 ps when the electron source was injected and then ran Hydra
for 180 ps to study the subsequent burn and ignition. Such a run,
utilizing 24 CPUs for Zuma and 48 for Hydra, takes several hours of
wall time to complete. 3D runs are much more demanding, so 2D runs
are envisioned for routine design studies.

\section{Ignition-scale modeling with pic-based fast electron source}

\label{sec:noB}The next two sections present results of Zuma-Hydra
modeling of an idealized ignition-scale, cone-guided target. This
section considers cases with no initial magnetic field, and the next
studies imposed field schemes to mitigate source divergence. Table
\ref{tab:runs} summarizes the runs, and Fig.~\ref{fig:gall} contain
RZ plots of the ion pressure and fast electron current for several
runs.

\subsection{Simulation setup}

\begin{figure}
\includegraphics[width=3.3in]{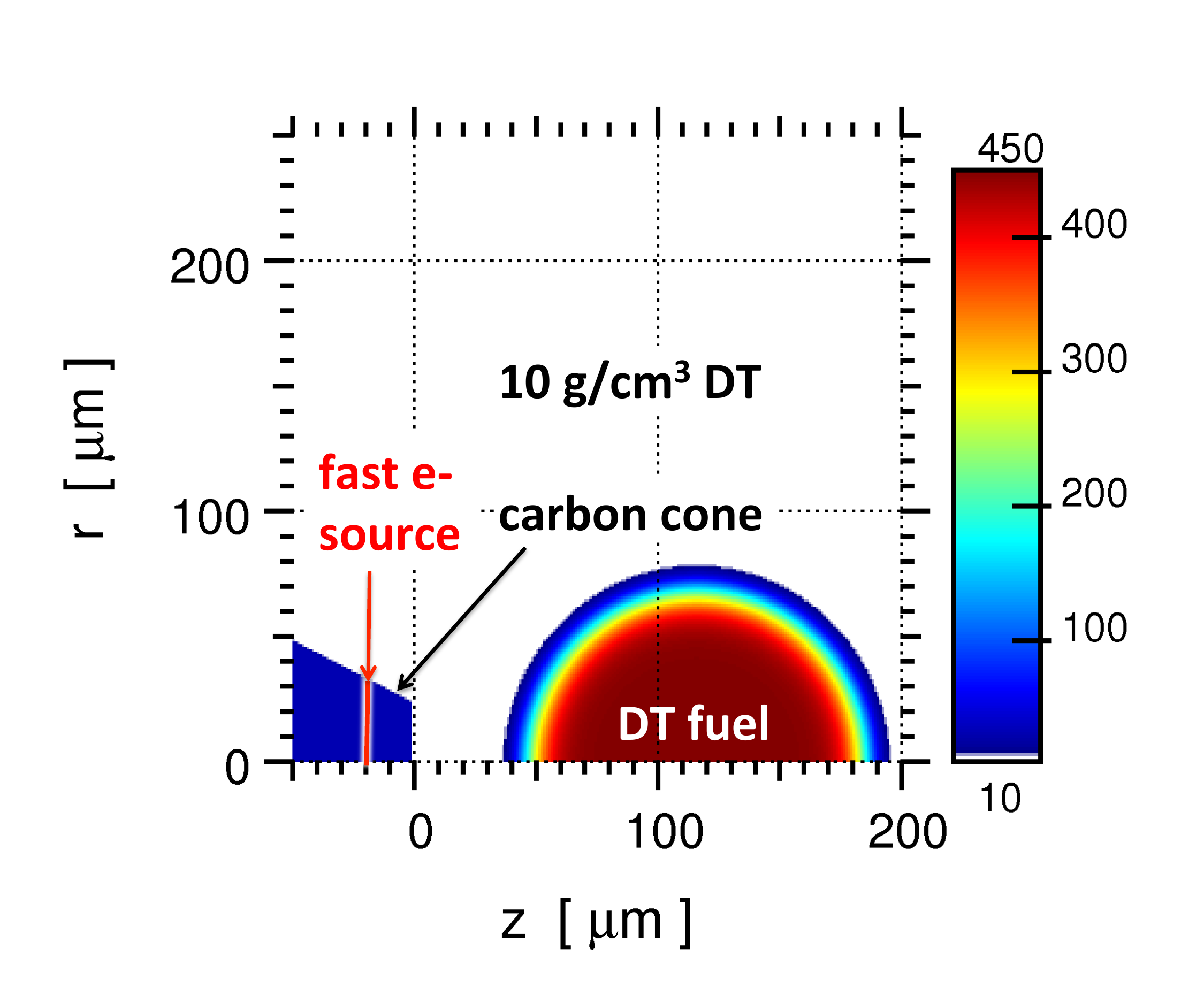}

\caption{(Color online) Initial target density in g/cm$^{3}$. The red line
indicates the source cylinder where fast electrons are injected. The
8 g/cm$^{3}$ carbon cone is colored blue for clarity.}

\label{fig:initdens}
\end{figure}

We consider a spherical assembly of equimolar DT fuel, relevant for
high-gain IFE uses. It is depicted in Fig.~\ref{fig:initdens}. The DT mass density is $\rho[\mathrm{g/cm^{3}]}=10+440\exp[-(R/70\,\mu\mathrm{m})^{12}]$
where $R$ is the distance from $(r,z)=(0,117)\,\micron$. This gives,
for $\rho>100$, an aerial density of $\rho R=3.0$ g/cm$^{2}$ and
mass $m=0.57$ mg. With the simple burn-up estimate $f=\rho R/(\rho R+6)=1/3$,
we obtain a total fusion yield (neutron and $\alpha$) of $Y=fm\cdot338$
MJ/mg = 64 MJ. Igniting such targets at a rate of 16 Hz would provide
1 GW of gross fusion power. A flat-tip carbon cone is located $\approx$50
$\mu$m to the left of the dense DT fuel. The cone density of 8 g/cm$^{3}$
(2.3x solid) was chosen so that, when fully ionized, the total pressure
is the same in the cone and 10 g/cm$^{3}$ DT. All materials are initially
at a temperature of 100 eV. 

Simulation parameters were as follows. Both codes used a uniform mesh
with 1 $\mu$m cell size. We leave the question of beam-plasma micro-instabilities
(e.g., resistive filamentation \citep{gremillet-fil-pop-2002,cottrill-instab-pop-2008}
or electro-thermal \citep{haines-electrotherm-pop-1981}) to future
work. The Hydra domain extended (in $\mu$m) from $(r,z)=(0,-50)$
to $(200,250)$ while Zuma ran on the sub-domain $(0,-50)$ to $(120,250)$.
The Zuma timestep is set by ensuring no fast electrons cross more
than one zone per step ($c\Delta t<\max[\Delta r,\Delta z])$ and
resolving the electron cyclotron frequency. Since Zuma does not support
light-wave propagation, there is no light-wave Courant stability condition. We
used $\Delta t=0.5-1$ fs, which gives $c\Delta t/\Delta z=0.15-0.3$
and $\omega_{ce}\Delta t=1$ for $B=57-114$ MG ($\omega_{ce}\equiv eB/m_{e}$
is the non-relativistic cyclotron frequency). The Hydra timestep was
variable, and the coupling timestep was 0.2 ps for the first 1 ps
and 0.5 ps subsequently.

\begin{table*}
\begin{tabular}{|c|c|c|c|c|c|}
\hline 
Case & Description & $E_{\mathrm{fast}}$ low  &
$E_{\mathrm{fast}}$ high  & Yield low  & Yield high \tabularnewline
 &  & {[}kJ{]} & {[}kJ{]} & {[}MJ{]} & {[}MJ{]}\tabularnewline
\hline 
\hline 
\texttt{DQ0} & $E=$1.5 MeV, $\Delta\theta=0$, no ang. scat. or E/B & 15.8 & 18.5 & 0.217 & 58.1\tabularnewline
\hline 
\texttt{DQ10\_mono} & $E=$1.5 MeV, $\Delta\theta=10^{\circ}$, no E/B & 25.4 & 30.4 & 0.189 & 56.8\tabularnewline
\hline 
\texttt{DQ10\_noEB} & PIC-based $dN/dE$, $\Delta\theta=10^{\circ}$, no E/B & 81.0 & 102 & 0.928 & 54.9\tabularnewline
\hline 
\texttt{DQ10} & $\Delta\theta=10^{\circ},\, B_{z}=0$, full Ohm's law & 121 & 132 & 0.426 & 48.7\tabularnewline
\hline 
\texttt{DQ90} & $\Delta\theta=90^{\circ},\, B_{z}=0$ & 949 & N/A & 6.82E-4 & N/A\tabularnewline
\hline 
\texttt{DQ90\_36} & \texttt{DQ90} but $\rspot=36\,\micron$ & 1270 & N/A & 0.0144 & N/A\tabularnewline
\hline 
\texttt{BZ30} & $B_{z}=30$ uniform & 211 & 237 & 0.538 & 52.3\tabularnewline
\hline 
\texttt{BZ50} & $B_{z}=50$ uniform & 106 & 132 & 2.66 & 54.0\tabularnewline
\hline 
\texttt{BZ30-75} & $B_{z}=30-75$ & 316 & N/A & 0.0523 & N/A\tabularnewline
\hline 
\texttt{BZ50-75} & $B_{z}=50-75$ & 158 & 211 & 0.412 & 53.9\tabularnewline
\hline 
\texttt{BZ0-50} & $B_{z}=0.1-50$ quickly in $z$ & 211 & N/A & 0.0785 & N/A\tabularnewline
\hline 
\texttt{BZ30pipe} & $B_{z}=30$ hollow pipe & 290 & 316 & 0.276 & 48.5\tabularnewline
\hline 
\texttt{BZ50pipe} & $B_{z}=50$ hollow pipe & 132 & 158 & 0.532 & 52.4\tabularnewline
\hline 
\texttt{BZ50pipeA} & \texttt{BZ50pipe} but thinner pipe & 185 & 211 & 0.378 & 49.4\tabularnewline
\hline 
\end{tabular}\caption{Ignition properties for various Zuma-Hydra cases, all for
  the initial plasma conditions shown in Fig.~\ref{fig:initdens}. Cases starting
  with \texttt{DQ} have no initial magnetic field, and the number indicates the
  divergence parameter $\Delta\theta$ in degrees. Those starting with
  \texttt{BZ} use $\Delta\theta=90^\circ$ and have an initial axial magnetic
  field, with the numbers related to the strength in MG. All runs except \texttt{DQ90\_36} have $\rspot=18\,\micron$.
  The first three runs have no E or B fields, while the others do and
  use the full Ohm's law. The (low, high) $E_{\mathrm{fast}}$ are, respectively,
  the (largest, smallest) energy which (did not, did) ignite.}
\label{tab:runs}
\end{table*}

The fast electron source was injected at $z=-20\,\mu$m with an intensity
profile $I_{\mathrm{fast}}(r,t)=I_{0f}\exp\left[-(1/2)(r/\rspot)^{8}\right]f(t)$
with $f(t)$ a flattop from 0.5 to 18.5 ps with 0.5 ps linear ramps.
Unless specified, we use $\rspot=18\,\mu\mathrm{m}$. The total injected
fast electron energy, $\Efast$, is proportional to $I_{0f}$; in
particular, for our runs, $\Efast\mathrm{[kJ]}=I_{0f}/5.77\times10^{18}$
W/cm$^{2}$. As discussed in Sec. \ref{sec:source}, we assume $I_{0f}$
is 0.52 times the laser intensity $I_{0L}$, which we need to find
the ponderomotive temperature $T_{p}$ and energy spectrum. We consider
a 527 nm (2nd harmonic of Nd:glass) laser wavelength, since this lowers
$T_{p}\propto\lambda_{0}$ compared to first harmonic light (but is
technologically more challenging).

\subsection{Results with artificially-collimated source}

\begin{figure}
\includegraphics[width=3in]{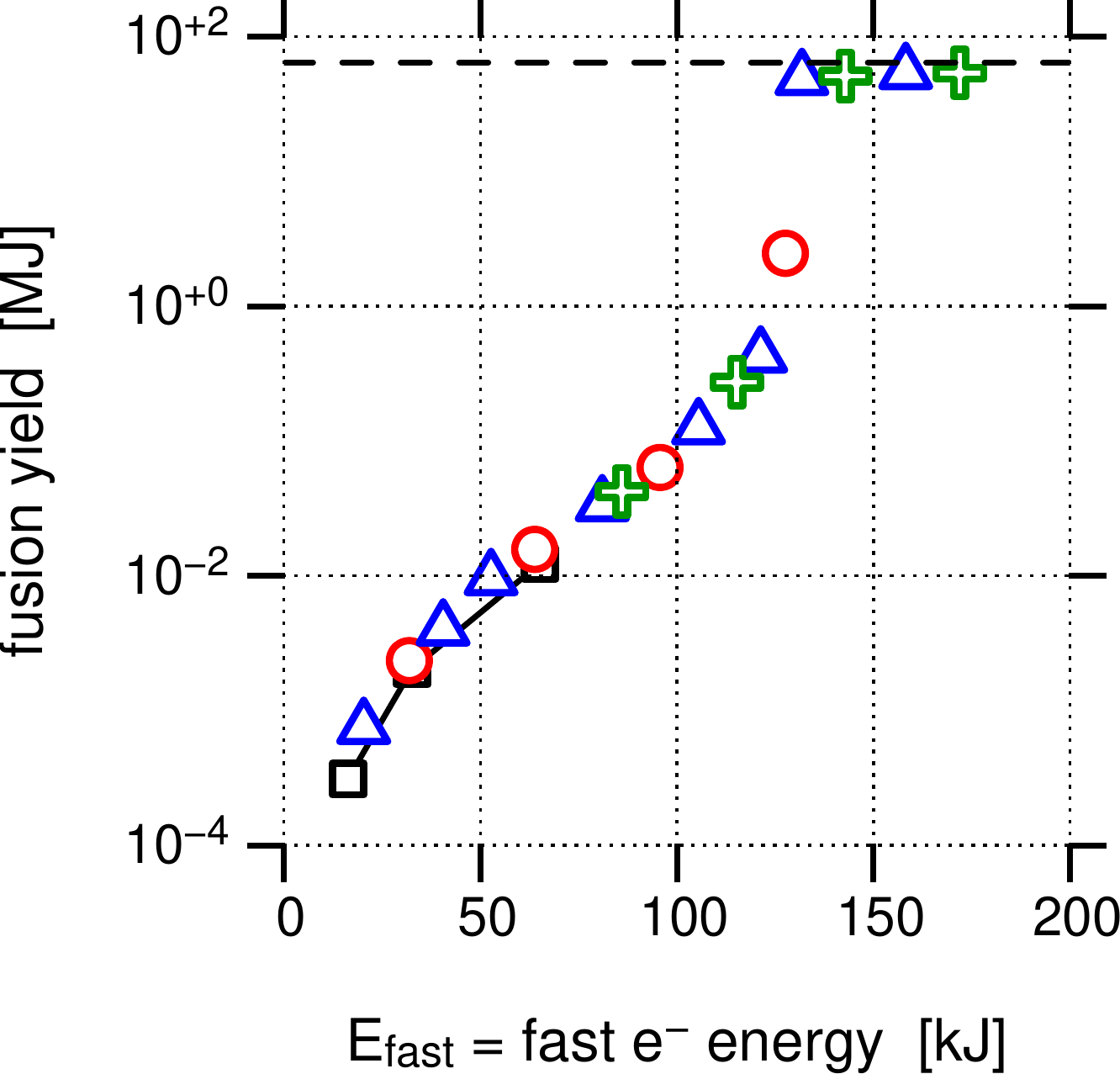}

\caption{(Color online) Fusion yield vs. total injected fast electron energy,
for Zuma-Hydra runs with an artificially collimated source $\Delta\theta=10^{\circ}$.
$\rspot=10\,\micron$ for black squares with solid line, 14 $\micron$
for red circles, 18 $\micron$ for blue triangles (case \texttt{DQ10}),
and 23 $\micron$ for green crosses. In this and subsequent plots,
the dashed black line at 64 MJ is the ideal fusion yield described
in the text. The blue triangle with $\Efast=132$ kJ is the lowest
value that we deem to have ignited.}

\label{fig:noByielddq10}
\end{figure}

The fusion yield for the PIC-based energy spectrum and an artificially
collimated source ($\Delta\theta=10^{\circ}$) is plotted against $\Efast$ for
several values of $\rspot$ in Fig.~\ref{fig:noByielddq10}.  The points lie on a
somewhat universal curve. This is due to two competing effects, both of which
are discussed later in this section. First, the hot spots are in the ``width $>$
depth'' regime \citep{atzeni-fastig-pop-2007}, where increasing the hot spot radius raises
the required deposited heat for ignition. On
the other hand, increasing the source area for fixed power decreases the energy
of individual electrons, and leads to more effective stopping in the hot spot
depth. We do not expect a strong dependence on $\rspot$ for situations where
source divergence has been mitigated. We use $\rspot=18\,\mu$m in subsequent
runs, since this ignites for the lowest $\Efast$ of 132 kJ.

\begin{figure}
\includegraphics[width=3in]{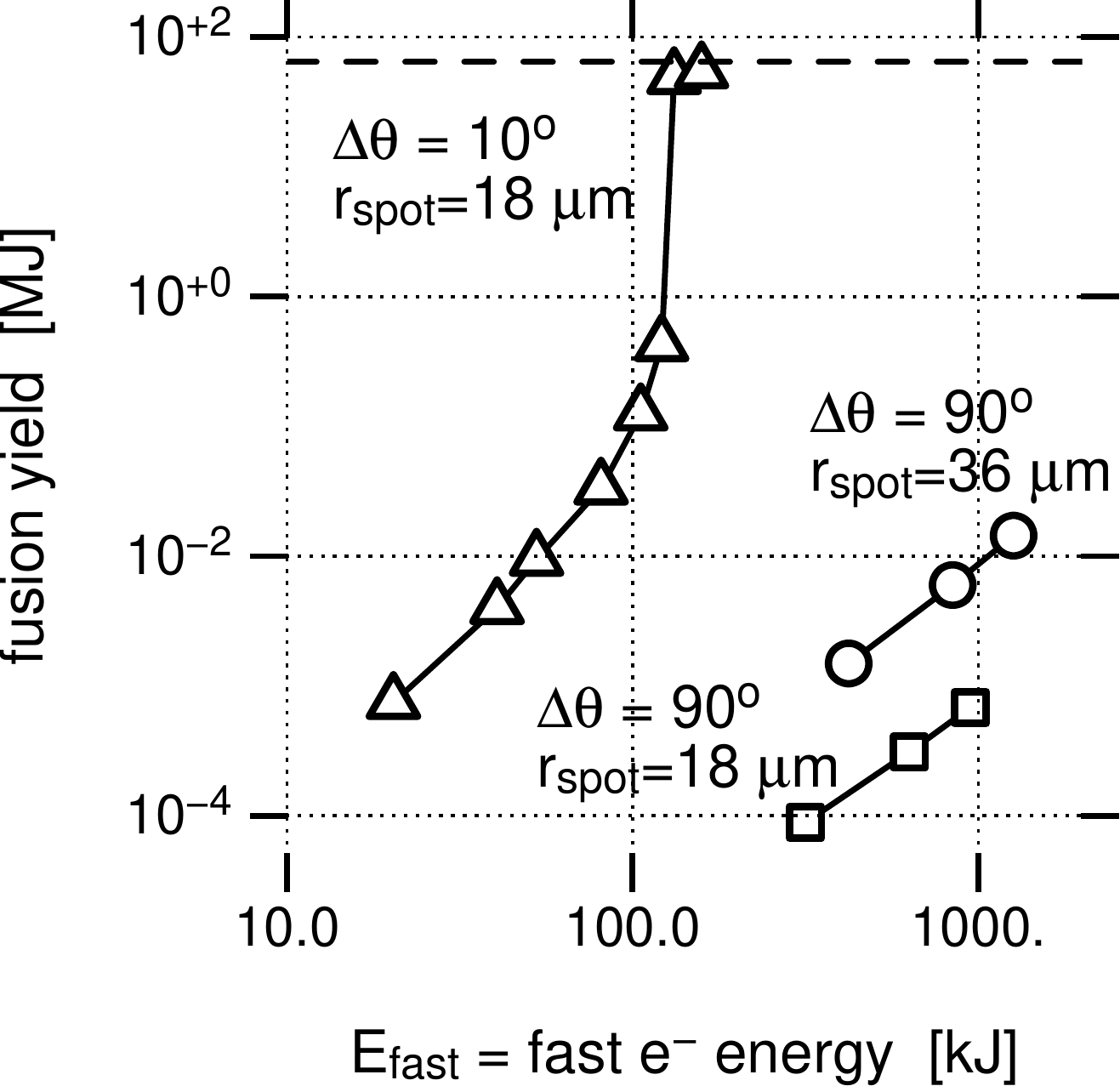}

\caption{Fusion yield vs. total injected fast electron energy, for Zuma-Hydra
runs for artificially collimated source $\Delta\theta=10^{\circ}$
with $\rspot=18\,\micron$ (triangles, case \texttt{DQ10}), PIC-based
source divergence $\Delta\theta=90^{\circ}$ with $\rspot=18\,\micron$
(squares, case \texttt{DQ90}), and $\Delta\theta=90^{\circ}$ with
$\rspot=36\,\micron$ (circles, case \texttt{DQ90\_36}).}

\label{fig:noByield}
\end{figure}

For a collimated source, an estimate of the minimum ignition energy
that must be delivered to the hot spot is given by Atzeni\emph{ et
al.}~in Ref.~\onlinecite{atzeni-fastig-pop-2007}. They performed 2D rad-hydro
simulations of idealized, spherical fuel assemblies heated by a cylindrical
beam of mono-energetic, forward-going particles which fully stop in a prescribed penetration depth.
The result is
\begin{eqnarray}
E_\mathrm{ig} &=& E_{\mathrm{opt}}F(\rho\Delta z, \rho r), \\
E_{\mathrm{opt}} &=& 140\left[\frac{\rho}{100\,\mathrm{g/cm}^{3}}\right]^{-1.85}\mathrm{kJ.}
\end{eqnarray}
$F>1$ if the hot spot transverse radius satisfies $\rho r>0.6$ g/cm$^{2}$ (``width $>$
depth'' regime), or the depth satisfies $\rho\Delta z>\rho\Delta z_{\mathrm{opt}}=1.2$
g/cm$^{2}$. For our peak density of 450 g/cm$^{3}$, $E_{\mathrm{opt}}=8.7$
kJ, and Atzeni finds an optimal hot-spot radius of $r_{\mathrm{opt}}=14\,\micron$
and pulse length of $t_{\mathrm{opt}}=15$ ps. 

The minimum $\Efast$ which ignited in the\texttt{ DQ10} series was
132 kJ, which is 15x $E_{\mathrm{opt}}$. We can understand this with
the simplified runs listed at the top of Table I. First, \texttt{DQ0}
has a perfectly collimated source $(\Delta\theta=0)$, a mono-energetic
1.5 MeV spectrum, no angular scattering, and no E or B fields. This
ignites for 18.4 kJ, or $2.1E_{\mathrm{opt}}$. This reflects our
spot shape and temporal pulse being larger than Atzeni's optimal values,
and deposition in the low-density DT and carbon cone (which Atzeni
did not include); we also did not optimize the 1.5 MeV source energy.
The series \texttt{DQ10\_mono} uses our small but nonzero $\Delta\theta=10^{\circ}$
and includes angular scattering, but no E or B fields; we now obtain
ignition for $3.5E_{\mathrm{opt}}$. We adopt the PIC-based energy
spectrum in \texttt{DQ10\_noEB} but still include no E or B fields.
This raises the ignition energy by another factor of 3.4, or $12E_{\mathrm{opt}}$.
Finally, turning on E and B fields with the full Ohm's law costs another
1.3x, bringing us to $15E_{\mathrm{opt}}$. The role of the energy
spectrum and Ohm's law for an artificially collimated source is discussed
in more detail in \citep{strozzi-ifsa-epjwc-2012}.

\begin{figure}

\includegraphics[width=3in]{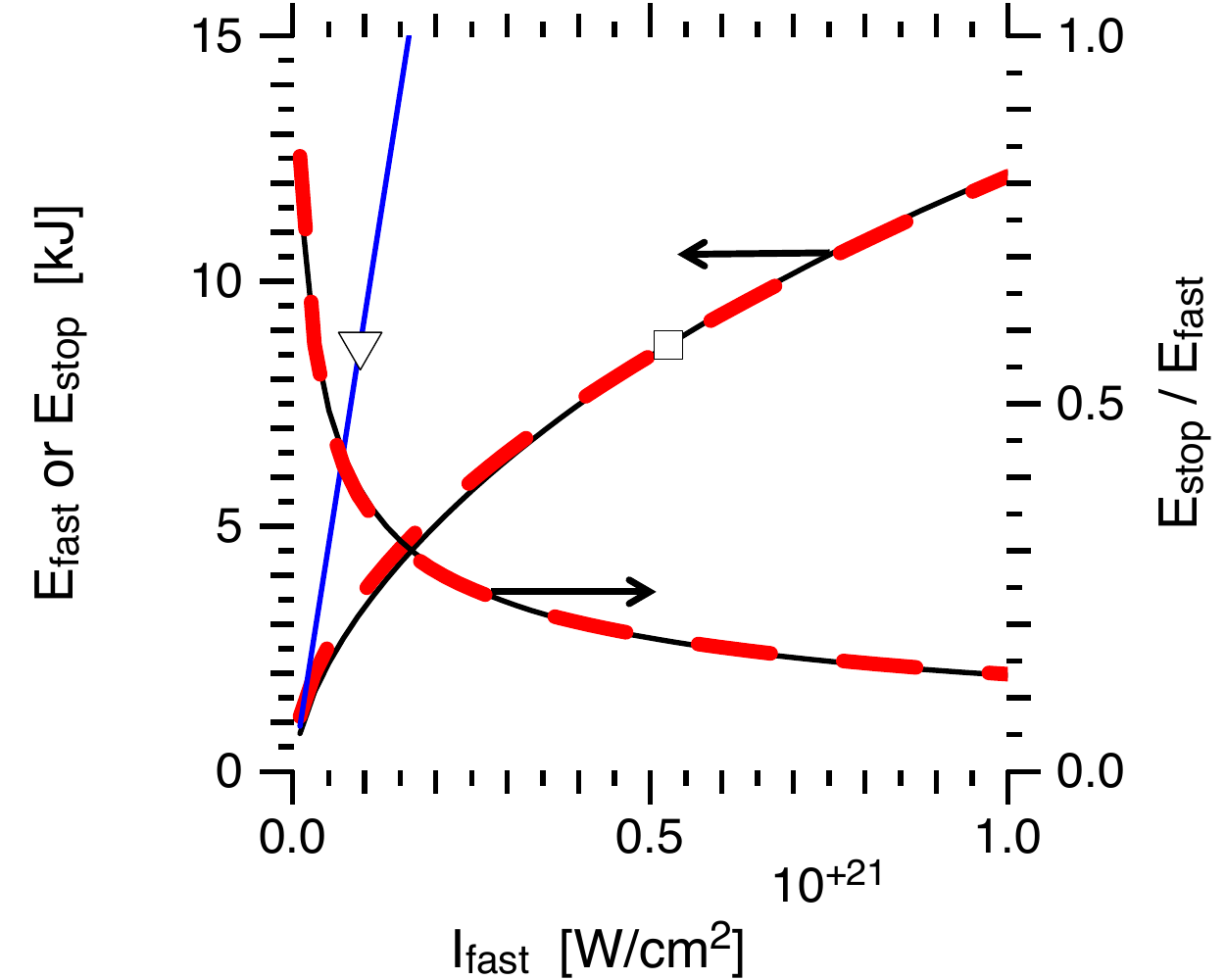}\caption{Fast electron coupling to optimal Atzeni hot spot described in text,
with our PIC-based energy spectrum. Blue: total fast electron energy
$\Efast$. Solid black: exact $E_{\mathrm{stop}}(=\Efast$ stopped
in hot spot), and $\Estop/\Efast$ coupled to hot spot. Thick dashed
red: approximate forms from Eq.~(\ref{eq:fstop}). The triangle and
square indicate points where $E_{\mathrm{stop}}$ equals the optimal
ignition energy of 8.7 kJ.}

\label{fig:Eopt}
\end{figure}

We now estimate the effect of the PIC-based energy spectrum on ignition
energy. Let $E_{\mathrm{stop}}$ be the energy delivered by a perfectly
collimated beam with Atzeni's optimal parameters to a hot spot of
depth $\rho\Delta z_{\mathrm{opt}}$. The total fast electron energy
$E_{\mathrm{fast}}=\alpha I_{0f}$ (with $\alpha=\pi r_{\mathrm{opt}}^{2}t_{\mathrm{opt}}$)
is controlled by the fast electron intensity $I_{0f}$. We consider
$\lambda_{0}=527$ nm and only collisional stopping (no angular scattering)
of the fast electrons. The fraction of kinetic energy lost by a fast
electron of kinetic energy $E$ in the hot spot is well fit by $f=\min(1,E_{DT}/E)$
where $E_{DT}=1.3$ MeV reflects the stopping in the DT hot spot.
Integrated over our PIC-based energy spectrum, the ratio $\Estop/E_{\mathrm{fast}}$
is approximately fit by
\begin{equation}
E_{\mathrm{stop}}/E_{\mathrm{fast}}\approx(1+I_{0f}/I_{0S})^{-0.48}\qquad I_{0S}=1.5\cdot10^{19}\,\mathrm{W/cm^{2}}.\label{eq:fstop}
\end{equation}
Figure \ref{fig:Eopt} shows how these formulas apply to our PIC-based
energy spectrum.

For $I_{0f}\gg I_{0S}$ , which our runs satisfy, we obtain

\begin{eqnarray}
E_{\mathrm{stop}} &\approx& \alpha I_{0f}^{0.52}I_{0S}^{0.48}\label{eq:Estop}, \\
\Efast &\approx& \frac{\Estop^{1.92}}{(\alpha I_{0S})^{0.92}}.
\end{eqnarray}
This is very close to what one finds with a ponderomotively-scaled
energy spectrum by assuming all electrons lose $E_{DT}$: 
\begin{equation}
\Efast\approx\frac{\Estop^{2}}{\alpha I_{0S}^{*}}\,\,\,\,\,\,\,\,\,\, I_{0S}^{*}\lambda_{0}^{2}=13.7\,\mathrm{GW}\frac{I_{0f}}{I_{0L}}\left[\frac{E_{DT}}{m_{e}c^{2}}\frac{T_{p}}{<E>}\right]^{2}.
\end{equation}
Using the values found above for the bracketed quantities and $\lambda_{0}=527$
nm, we find $I_{0S}^{*}=1.6\times10^{19}$ W/cm$^{2}$. This is very
close to the fitted value given above. The upshot is that, due to
partial stopping of fast electrons, the required short-pulse ignitor laser energy $\propto\Efast$
scales roughly as square of the hot-spot energy. In addition, $\Efast$
can be decreased by raising $I_{0S}$, which would happen if the electron
stopping power were higher than our current value (e.g., due to micro-instabilities
\citep{yabuuchi-stopping-njp-2009} or N-particle correlated stopping
\citep{bret-Nstopping-jpp-2008}).

From Eq.~(\ref{eq:fstop}), achieving $\Estop=E_{\textrm{{opt}}}$
of 8.7 kJ requires $\Efast=5.6\Estop=49$ kJ. This factor of 5.6x
is 1.6 times larger than the 3.4x we found in going from case \texttt{DQ10\_mono}
to \texttt{DQ10\_noEB}, which entailed going from the mono-energetic
to PIC-based energy spectrum. We conjecture this is because \texttt{DQ10\_mono}
is already sub-optimal enough (ignites for $3.5E_{\mathrm{opt}}$)
that we do not suffer the largest possible penalty for using the PIC-based
energy spectrum. This implies more idealized targets like \texttt{DQ0}
would pay closer to the full penalty of 5.6x.

\subsection{Results with PIC-based, divergent source}

Figure \ref{fig:noByield} presents the results for the cases \texttt{DQ90
($\rspot=18\,\mu$m)} and \texttt{DQ90\_36 ($\rspot=36\,\mu$m)} with
the PIC-based source divergence $\Delta\theta=90^{\circ}$, as well
as case \texttt{DQ10} with an artificially collimated source of $\Delta\theta=10^{\circ}$.
The PIC-based source is far from igniting, even for electron source
energies $>$ 1 MJ. Figure \ref{fig:gall} shows the greatly increased
current divergence for \texttt{DQ90}. Note also the filaments that
develop at large radius. Their nature and effect on beam propagation
and fuel coupling should be further examined in future work. With
the realistic divergence of $\Delta\theta=90^{\circ}$, the yield
for the same $\Efast$ is higher for the larger spot radius. This
is generally the case for divergent sources, where the benefit of
reduced laser intensity (and lower energy electrons, which stop more
efficiently in the hot spot) outweighs the cost of increased spot
size. Divergent sources thus ignite in the so-called ``width $>$ depth
regime'', where the hot spot has $\rho r$ above the optimal value
of 0.6 g/cm$^{2}$. The source spot size that minimizes $\Efast$
depends on details, like the cone-fuel standoff distance. In any case,
its value will be unacceptably large for reactor purposes, so we turn
our attention below to mitigating source divergence. 

\section{FAST ELECTRON confinement with imposed magnetic fields}

We now attempt to recover the artificial-collimation 132 kJ ignition
energy, with the realistic source divergence, by imposing various
initial magnetic fields. This can be achieved with an axial field
$B_{z}$ with no axial variation and strength $\sim$50 MG. However,
axial variation in $B_{z}$ leads to a radial field $B_{r}$ and a
$v\times B$ force in the $z$ direction (i.e., magnetic mirroring),
as well as finite standoff distance from the source region to the
confining field. We find that mirroring greatly reduces the benefit
of magnetic fields. A magnetic pipe, with a $B_{z}$ that peaks at
finite radius, does not suffer from the mirroring problem.

We wish to specify an arbitrary $B_{z}(r,z)$ in cylindrical coordinates,
with no dependence on azimuth $\phi$. This can be accomplished by
a vector potential $\vec{A}=A_{\phi}(r,z)\hat{\phi}$, which by construction
satisfies the Coulomb gauge condition $\nabla\cdot\vec{A}=0$. The
magnetic field $\vec{B}=\nabla\times\vec{A}$ automatically satisfies
$\nabla\cdot\vec{B}=0$. In particular, $B_{z}=(1/r)\partial_{r}(rA_{\phi})$
and $B_{r}=-\partial_{z}A_{\phi}$. This allows us to solve for $A_{\phi}:$
\begin{equation}
A_{\phi}=\frac{1}{r}\int_{0}^{r}dr'\, r'B_{z}(r',z).\label{eq:Aphi}
\end{equation}
As $r\rightarrow0$, if $B_{z}\rightarrow kr^{p}$ then $A_{\phi}\rightarrow k(p+2)^{-1}r^{p+1}$.
Since $B_{r}$ scales with $r$ like $A_{\phi}$, as long as $p\geq-1$,
Eq. (\ref{eq:Aphi}) guarantees $B_{r}(r=0)=0$. This is physically
necessary, since the radial direction is ill-defined at $r=0$. We
can find the current $\vec{J}=J_{\phi}\hat{\phi}$ needed to maintain
the magnetic field from Amp\`ere's law without displacement current:
\begin{equation}
\mu_{0}J_{\phi}=\partial_{z}B_{r}-\partial_{r}B_{z}=\nabla^{2}(A_{\phi}\hat{\phi)}.
\end{equation}
We estimate the magnitude of $J_{\phi}\sim B/\mu_{0}L$, and taking
$B\sim10$ MG, $L\sim10\,\micron$ gives $J_{\phi}\sim10^{14}$ A/m$^{2}$.
The fast electron current is of order $n_{cr}ec$, which for 527 nm
light is $1.9\times10^{17}$ A/m$^{2}$. Although substantial, the
currents implied by our imposed fields are much less than the fast
electron current. They may compete with the much smaller net (fast
plus background) current.

\subsection{High $B_{z}$ in fast electron source region}

\begin{figure}
\includegraphics[height=3in]{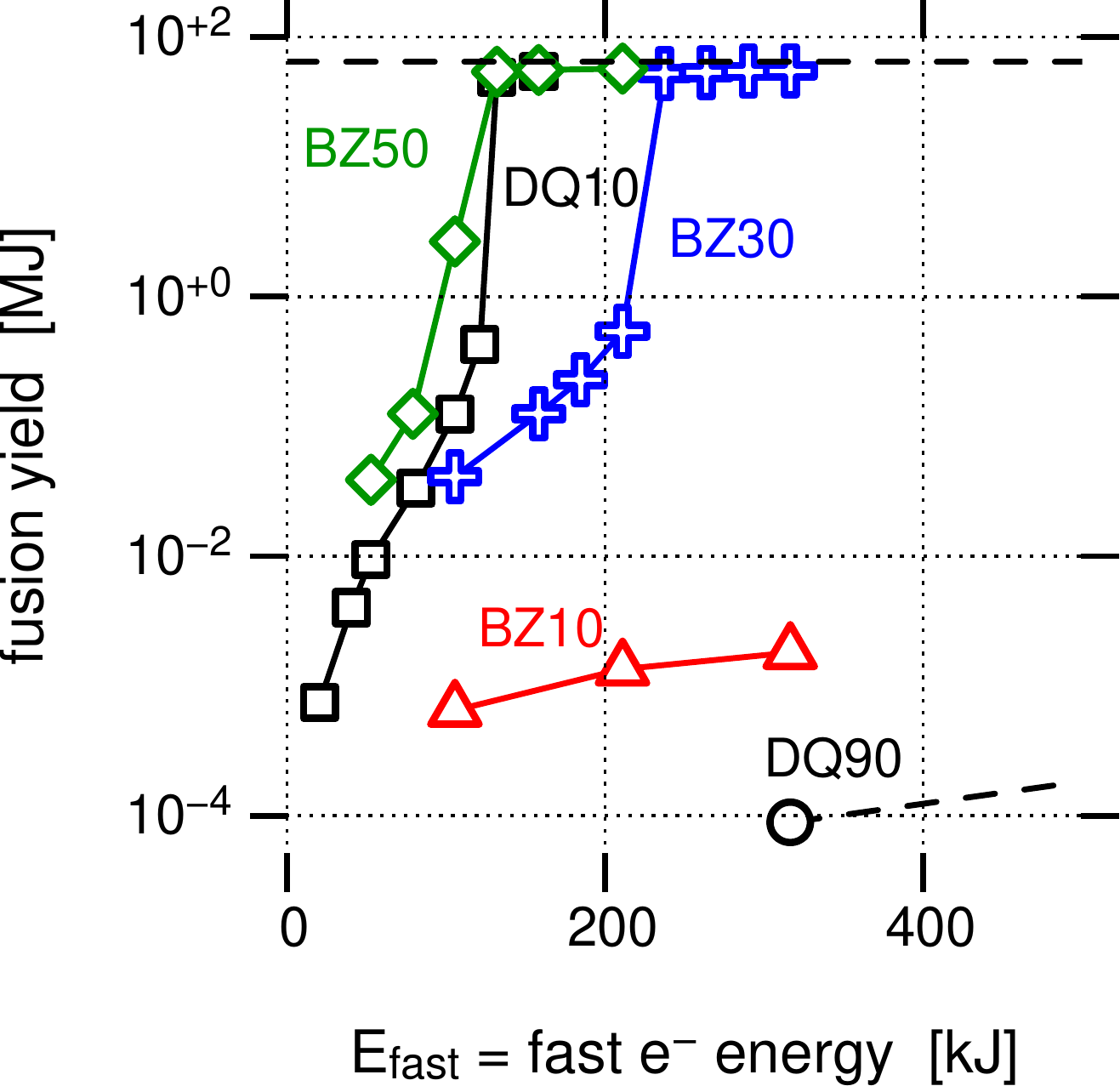}

\caption{(Color online) Fusion yield for runs with $\rspot=18\,\micron$. Black
solid line with squares is for an artificially collimated source $\Delta\theta=10^{\circ}$
(case \texttt{DQ10}). All other cases use the PIC-based source divergence
($\Delta\theta=90^{\circ})$. Black dashed line with circles is for
no imposed B field (case \texttt{DQ10\_18}). The other cases have
a ``uniform'' initial $B_{z}$ given by Eq. (\ref{eq:Bzunif}),
with $B_{z1}=10$ MG (red triangles), 30 MG (blue crosses, case \texttt{BZ30}),
and 50 MG (green diamonds, case \texttt{BZ50}).}
\label{fig:Bzunifyield}
\end{figure}

\begin{figure}
\includegraphics[height=3in]{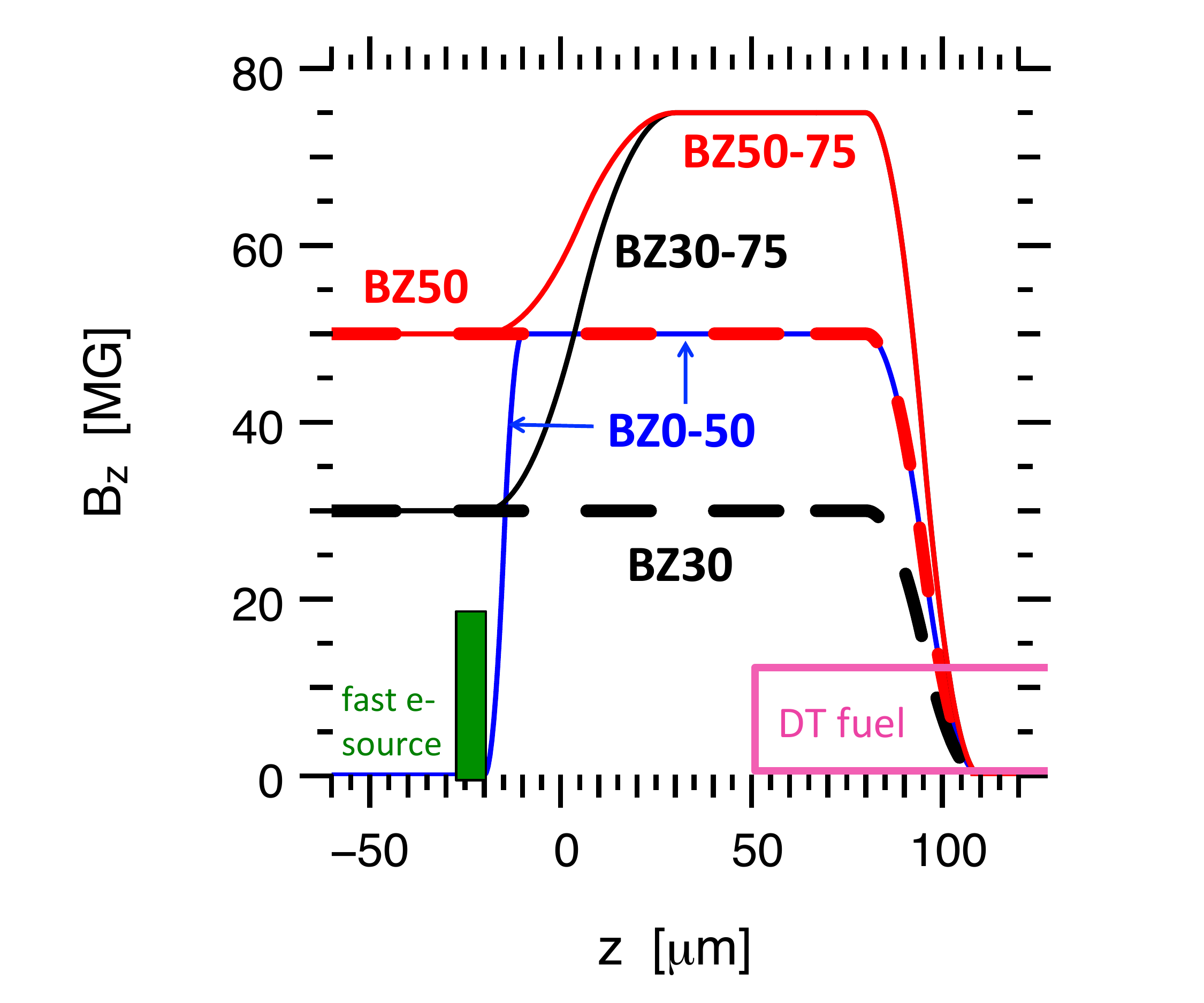}

\caption{(Color online) \label{fig:Bzprof_mirr}$B_{z}$ profiles at $r=0$.
Thick black dash: case \texttt{BZ30}, thick red dash: case \texttt{BZ50}, black
solid: case \texttt{BZ30-75}, red solid: case \texttt{BZ50-75}, blue
solid: case \texttt{BZ0-50}.}

\end{figure}

We utilize initial magnetic fields of the form

\begin{equation}
B_{z}=B_{z0}+(B_{z1}-B_{z0})H(z)\exp\left[-\left((r-r_{0})/\Delta r\right)^{8}\right].\label{eq:Bzunif}
\end{equation}
Except for the magnetic pipe configurations discussed below, $r_{0}=0$ and $\Delta r=50\,\mu$m. We first consider a $B_{z}$ that we call ``uniform,'' since it
does not vary between the source region and dense fuel.
$B_{z}(z)$ for several cases is plotted in Fig.~\ref{fig:Bzprof_mirr}.
For the uniform case, $H(z)=1$ for $z<80\,\micron$, with a piecewise-parabolic
ramp to zero for $z>110\,\micron$. $B_{z0}=0.1$ MG is the uncompressed
seed field, and we vary the peak compressed field $B_{z1}$. Setting
$H=1$ slightly decreases the ignition energy, but may be less realistic
than our $H(z)$. Figure \ref{fig:Bzunifyield} shows the fusion yield
for the PIC-based source divergence ($\Delta\theta=90^{\circ})$ and
various values of $B_{z1}$. An initial field of $B_{z1}=10$ MG gives
better coupling than the unmagnetized cases, but would still lead
to an unacceptable ignition energy. A field of 30 MG (case \texttt{BZ30})
gives about 2x the ignition energy as an artificially collimated source
($\Delta\theta=10^{\circ})$, while 50 MG (case \texttt{BZ50}) gives
essentially the same coupling. 

The pressure and fast electron current profiles in Fig.~\ref{fig:gall}
illustrate the improvement due to the imposed field. The current is
much more confined in the case \texttt{BZ50} than without the field,
although not as much as in the artificial-collimation case \texttt{DQ10}.
The loss of confinement at $z\approx100\,\micron$ is due to the end
of the high-field region (see Fig.~\ref{fig:Bzprof_mirr}). Note
also the appearance of current to the left of the fast electron injection
plane at $z=-20\,\micron$. We call this the reflected current. It
is due to the plasma being slightly diamagnetic, and reducing the
imposed $B_{z}$ somewhat during the course of the run. The reflected
current becomes enhanced in runs with significant mirroring, such
as case \texttt{BZ0-50}.

We now turn to the effect of more realistic initial field geometries.
It is plausible to compress the field to the desired strength, $\gtrsim$50
MG, in a fast-ignition fuel-assembly implosion \citep{tabak-dpp-2010}.
However, it will not be uniform. In particular, to the extent the
MHD frozen-in law is followed, the axial field compression will follow
the radial compression of matter. Standard schemes of fuel assembly
around a cone tip will thus result in the largest field being located
between the cone tip and dense fuel. Moreover, the purpose of the
cone is to provide a plasma-free region so the short-pulse laser converts
to fast electrons near the fuel. The shell motion down the outer cone
surface launches a strong shock in the cone, which must not reach
the inner cone surface before the short-pulse laser fires (to avoid
a rarefaction that would fill the cone interior). In standard schemes,
the short-pulse laser thus converts to fast electrons in a region
with essentially the uncompressed, seed magnetic field. The field
may be enhanced somewhat by resistive diffusion of compressed field
into and through the cone material, but we expect the cone and surrounding
DT to be sufficiently conducting to prevent significant diffusion.

The upshot is the fast electrons must transit from their birth region
of low field to a region of high field in front of the cone. This
poses two separate challenges. First, the fast electrons may be reflected
axially by the magnetic mirror effect. Also, they must travel a finite
standoff distance before the compressed field can impede their radial
motion.

\begin{figure}
\includegraphics[height=3in]{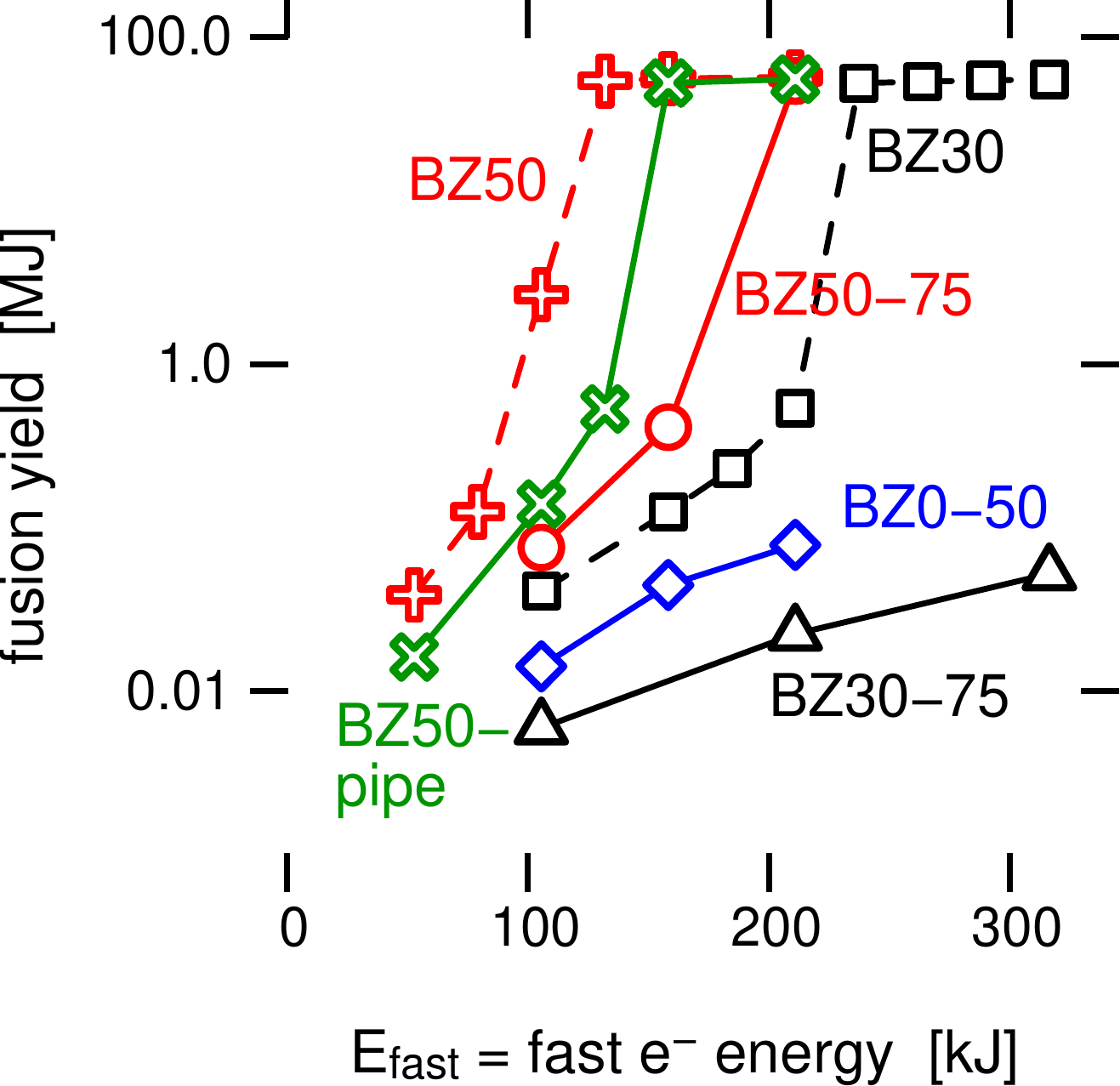}

\caption{\label{fig:Bzyield_mirr}(Color online) Fusion yield for runs from
Fig.~\ref{fig:Bzprof_mirr}. The curves have the same meaning as
in that figure. solid green X's is case \texttt{BZ50pipe}. The other curves have the same meaning as
in Fig.~\ref{fig:Bzprof_mirr}: black dash squares: case \texttt{BZ30}, red dash crosses: case \texttt{BZ50}, black
solid triangles: case \texttt{BZ30-75}, red solid circles: case \texttt{BZ50-75}, blue
solid diamonds: case \texttt{BZ0-50}.}

\end{figure}

We consider the role of mirroring with no standoff, by modifying $H(z)$.
We first increase $B_{z}$ to 75 MG at $z=30\,\micron$ (located between
the cone tip and the dense fuel) while keeping $B_{z}$ fixed at 30
MG (case \texttt{BZ30-75}) or 50 MG (case \texttt{BZ50-75}) in the
source region. The new $B_{z}$ profiles are plotted in Fig. \ref{fig:Bzprof_mirr}.
Table I and Fig.~\ref{fig:Bzyield_mirr} show that the energy needed
to ignite increases slightly for case \texttt{BZ50-75} but substantially
for case \texttt{BZ30-75}. This demonstrates the significant impact
of mirroring for modest (1.5-2.5x) increases in $B_{z}$ between the
source and fuel regions.

\begin{figure}
\includegraphics[height=3in]{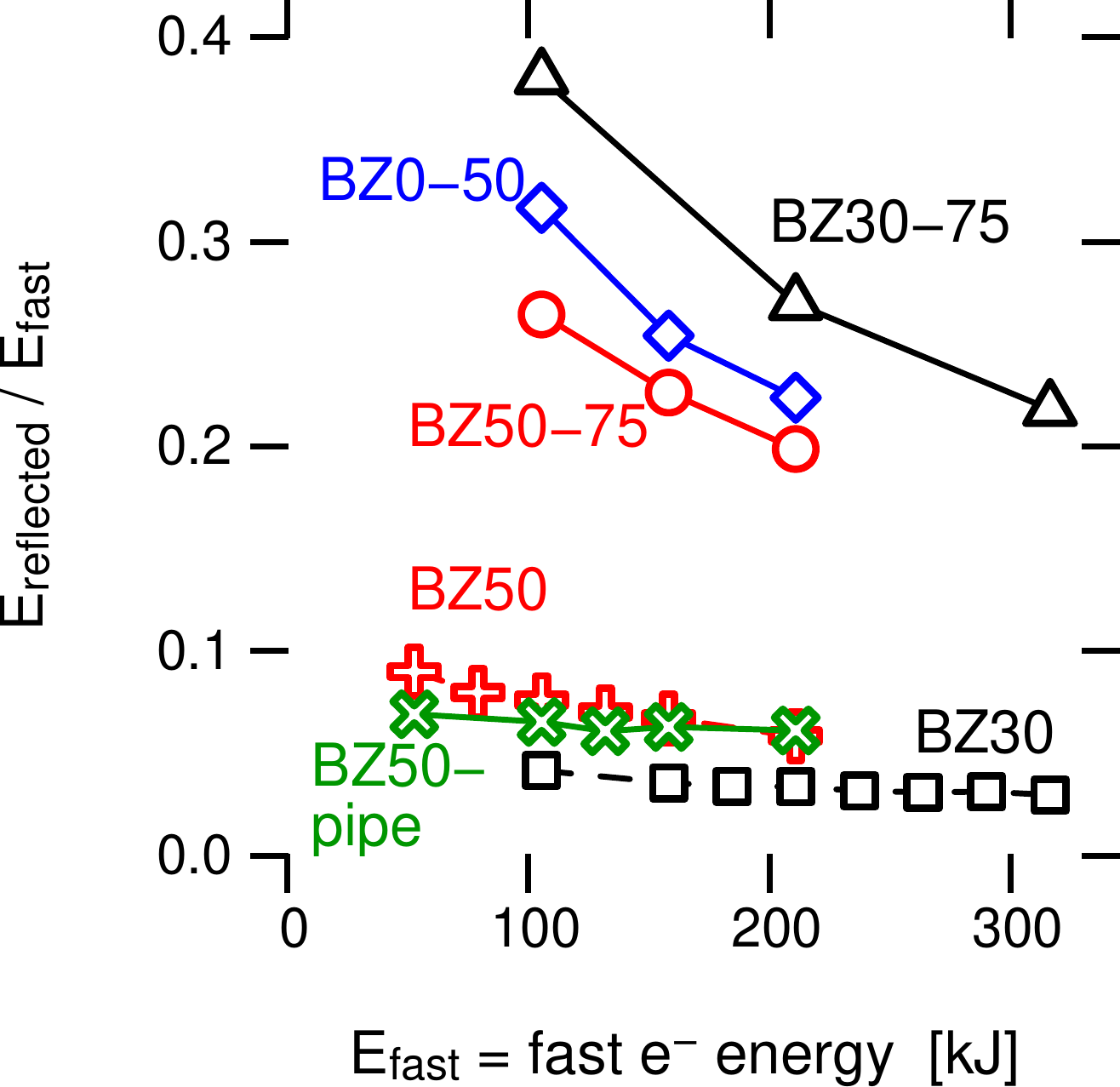}

\caption{\label{fig:Bzrefl_mirr}(Color online) Reflected fraction, i.e.\ the fast electron
energy reaching left $z$ boundary, divided by $\Efast$. The runs and curves are
as in Fig.~\ref{fig:Bzyield_mirr}: solid green X's: case \texttt{BZ50pipe},
black dash squares: case \texttt{BZ30}, red dash crosses: case \texttt{BZ50}, black
solid triangles: case \texttt{BZ30-75}, red solid circles: case \texttt{BZ50-75}, blue
solid diamonds: case \texttt{BZ0-50}.}
\end{figure}

To demonstrate further the impact of mirroring, we plot in Fig. \ref{fig:Bzrefl_mirr}
the reflected fraction, or the ratio of the fast electron energy reaching
the left edge of the domain ($z=-50\,\micron)$ to $\Efast$. For
fixed $\Efast$, the reflected fraction is quite small for the uniform
field profiles, but substantial for the non-uniform ones. The reflected
fraction actually understates the effect of mirroring. The low-energy
electrons deposit more of their energy in the hot spot, and are also
more magnetized (and thus more likely to mirror). So, the reflected
electrons would have more effectively heated the hot spot than typical
electrons.

\subsection{Low $B_{z}$ in fast electron source region: the magnetic pipe}

\begin{figure*}
\includegraphics[width=5.5in]{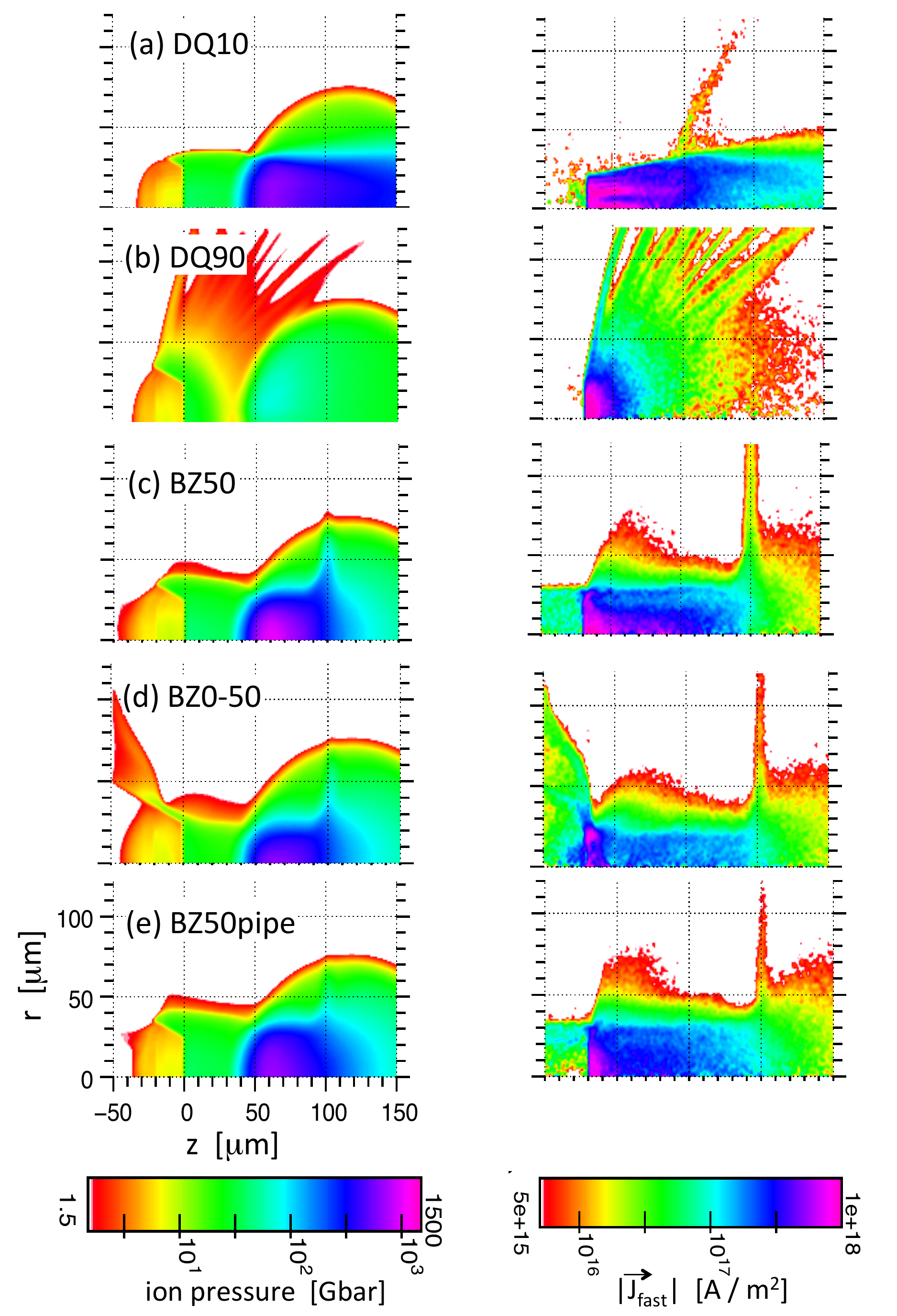}
\caption{(Color online) Ion pressure (left) and fast electron current density
$|\protect\overrightarrow{J}_{\mathrm{fast}}|$ (right) at time 10
ps (middle of fast electron time pulse) for cases, from top to bottom:
\texttt{DQ10}, \texttt{DQ90}, \texttt{BZ50}, \texttt{BZ0-50}, and\texttt{
BZ50pipe}. All cases have $\Efast=158$ kJ, except \texttt{DQ90} has
$\Efast=317$ kJ.}
\label{fig:gall}
\end{figure*}

We now turn to the situation where the fast electrons are born in
the uncompressed seed field. First we consider case \texttt{BZ0-50},
where the field rises quickly in $z$, so that standoff is minimized.
Figure \ref{fig:Bzprof_mirr} shows the $B_{z}$ profile. The electrons
are still subject to the mirror force, which results in an ignition
energy of $\Efast>211$ kJ (blue curve in Fig.~\ref{fig:Bzyield_mirr}). Runs with higher $\Efast$ encountered
numerical difficulties, which we are studying. Figure \ref{fig:Bzrefl_mirr}
depicts the reflected fraction, indicating substantial mirroring in
this case. Figure \ref{fig:gall} shows the increase in the reflected
current in case \texttt{BZ0-50} compared to \texttt{BZ50}.

To remedy mirroring, we propose a hollow magnetic pipe, which is free
of high field at small radius. Fast electrons are reflected by the
pipe as they move outward radially, but do not experience a mirror
force in $z$ ($\partial_{z}B_{z}=B_{r}=0$ inside the pipe). A certain
product of field strength times length is needed to reflect an electron,
and can be estimated for planar (not cylindrical) geometry by \citep{robinson-switchyard-pop-2007}
\begin{equation}
BL>K\gamma\beta(1-\cos\theta),\qquad\qquad K\equiv\frac{m_{e}c}{e}=17.0\,\mathrm{MG}\cdot\mu\mathrm{m}.
\end{equation}
For our PIC-based $\Delta\theta=90^{\circ}$ angle spectrum, $\left\langle 1-\cos\theta\right\rangle =0.43$.
The ignition energy for case \texttt{DQ10} (artificial collimation)
was 132 kJ, which gives an average electron energy of 8.5 MeV. This
requires $BL>129$ MG$\cdot\mu$m to reflect. Although lower-energy
particles are easier to reflect and stop more fully in the DT hot spot,
our spectrum does not contain much energy there. For $B=50$ MG, we
need a pipe of thickness 2.6 $\mu$m. We consider pipes that are thicker
than this, which substantially reduce the ignition energy over the
no-field case. The results shown here establish the feasibility of
the pipe. We are exploring thinner, more optimal pipes, which entails
variation of other parameters like spot size.

\begin{figure}
\includegraphics[width=3in]{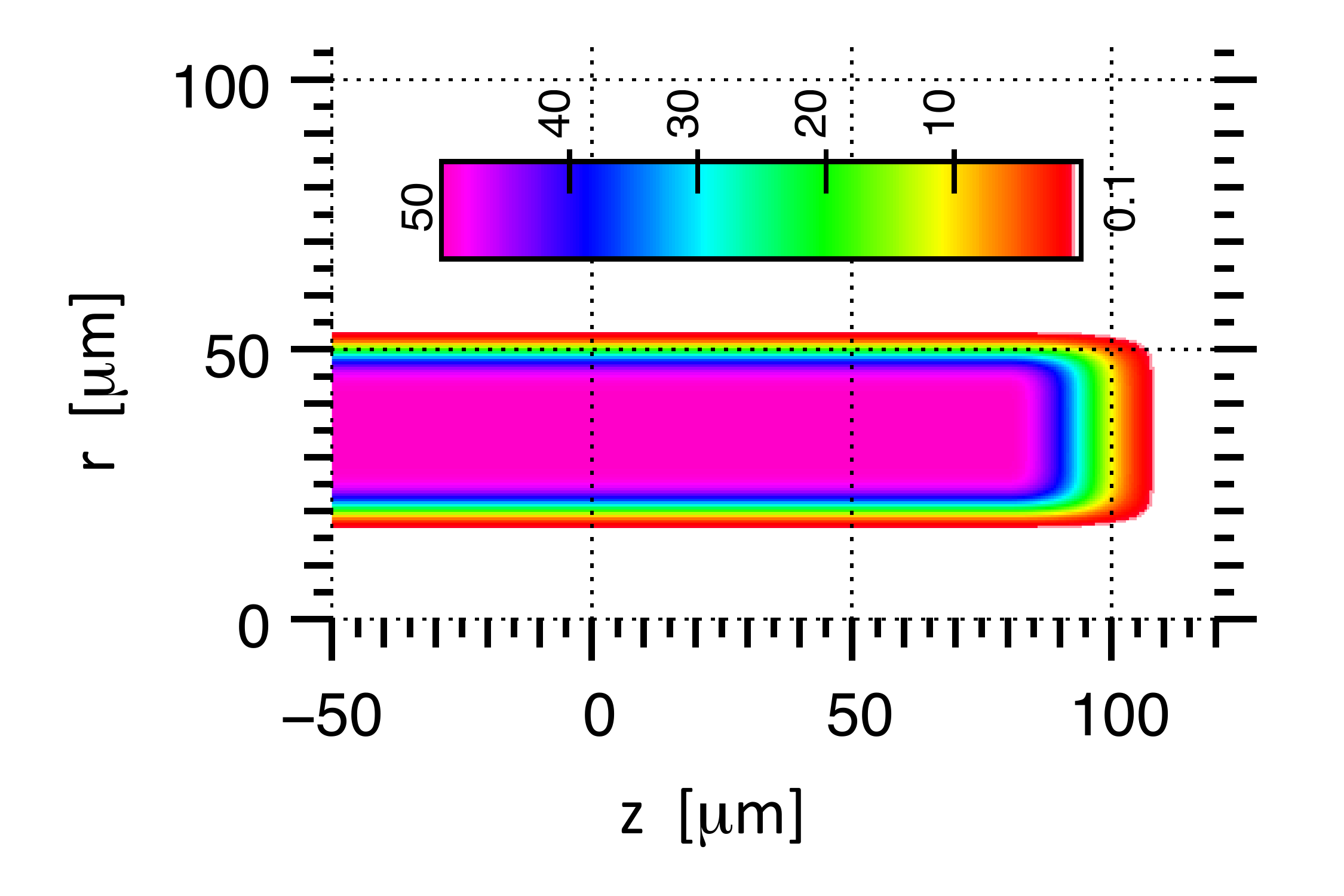}

\caption{(Color online) Initial $B_{z}$ profile, in MG, for case \texttt{BZ50pipe}.}
\label{fig:Bzprof_pipe}
\end{figure}

\begin{figure}
\includegraphics[width=2.3in]{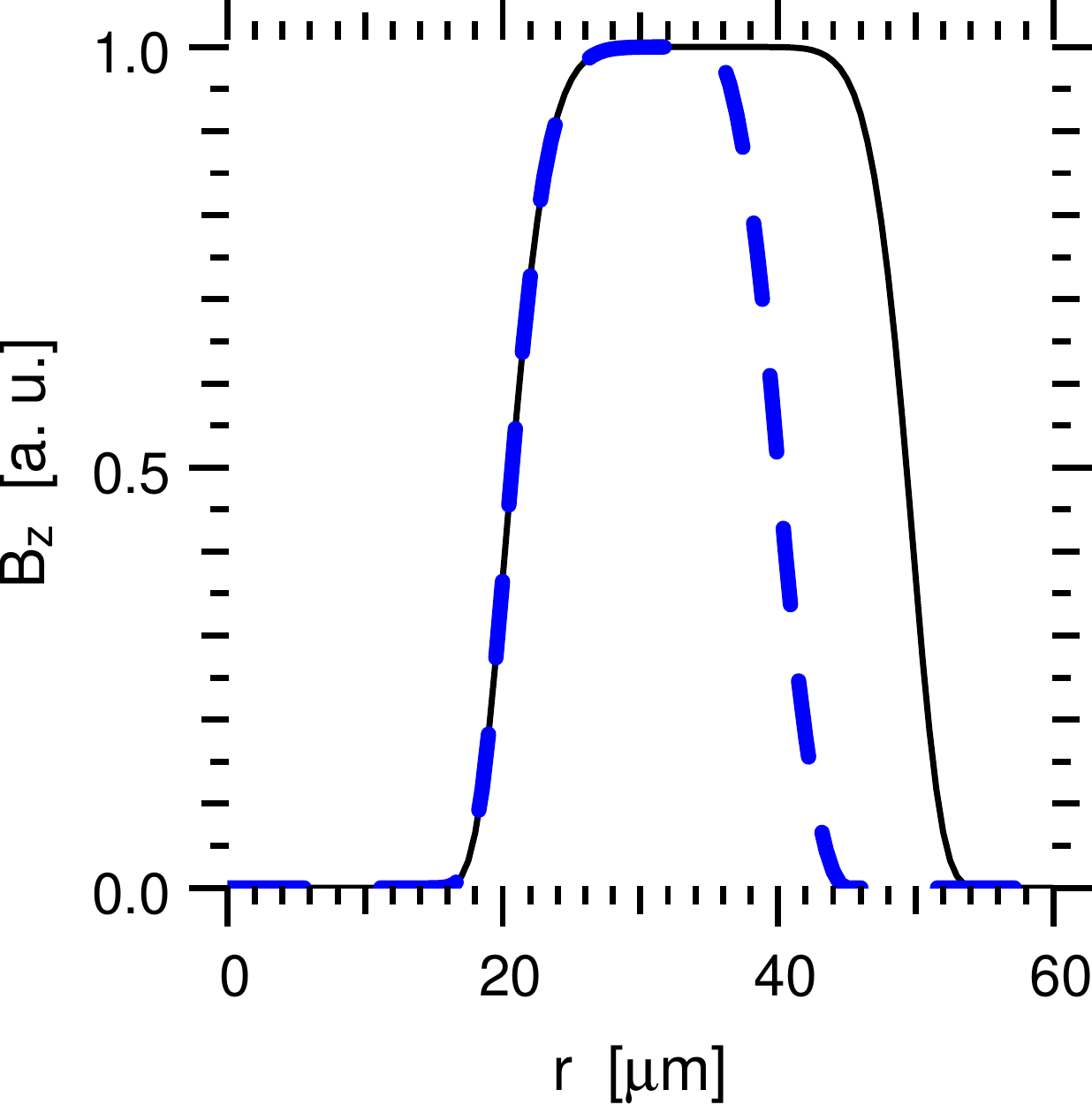}

\caption{(Color online) Radial $B_{z}$ profiles for pipe cases \texttt{BZ30pipe}
and \texttt{BZ50pipe} (solid black), and \texttt{BZ50pipeA} (thick
dashed blue). }

\label{fig:piper}
\end{figure}

The initial field for the case \texttt{BZ50pipe} is given by Eq.~(\ref{eq:Bzunif})
with $r_{0}=35\,\mu$m and $\Delta r=15\,\mu$m, and is displayed
in Fig.~\ref{fig:Bzprof_pipe}. This gives roughly the same ignition
energy as the uniform $B_{z}$ field, shown as the solid green line in Fig.~\ref{fig:Bzyield_mirr}.
The mirror effect, as measured by the reflected fraction, is about
the same as for the uniform case \texttt{BZ50} (see Fig.~\ref{fig:Bzrefl_mirr}).
We studied pipes with a peak field of 30 and 50 MG (cases \texttt{BZ30pipe}
and \texttt{BZ50pipe}), as well as thin and thick 50 MG pipes (cases
\texttt{BZ50pipe} and \texttt{BZ50pipeA}, see Fig.~\ref{fig:piper}).

\begin{figure}
\includegraphics[width=3in]{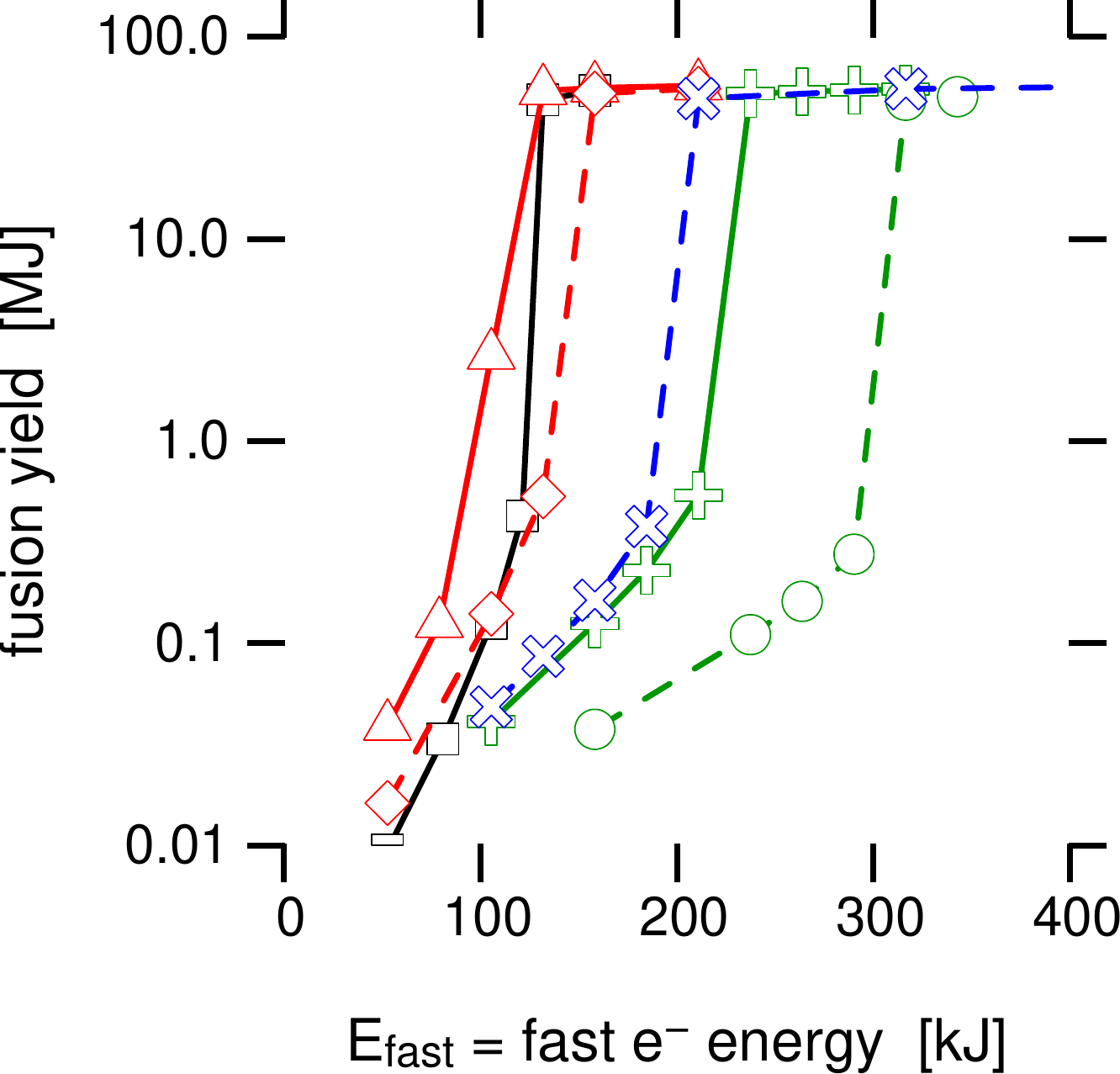}

\caption{(Color online) Fusion yield for run cases \texttt{DQ10} (solid black
squares), \texttt{BZ50} (solid red triangles), \texttt{BZ50pipe} (dashed
red diamonds), \texttt{BZ50pipeA} (dashed blue X's) \texttt{BZ30}
(solid green crosses), and\texttt{ BZ30pipe} (dashed green circles).}
\label{fig:Bzyield_summ}
\end{figure}

We summarize the development of this paper in Fig.~\ref{fig:Bzyield_summ}.
The challenge was to find imposed magnetic field configurations that
recover the performance of an artificially collimated fast electron
source, when using the PIC-based divergent source. A uniform 50 MG
axial field does this, and may even perform slightly better. However,
the more realistic case is for fast electrons to be born in a lower
field and suffer magnetic mirror forces. To circumvent this, we introduced
the hollow magnetic pipe. For a 50 MG peak field, this works essentially
as well as the uniform field. A lower peak field of 30 MG performs
significantly worse than the 50 MG cases, for both the uniform and
pipe configurations.

\section{Conclusion}

In this paper, we have presented recent transport modeling efforts
geared towards a fast ignition ``point design.'' This requires knowledge
of the fast electron source produced by a short-pulse laser. We characterized
the results of a full-PIC 3D simulation with the PSC code in terms
of 1D energy and angle spectra. The energy spectrum is well-matched
by a quasi two-temperature form, which we scale ponderomotively as
we vary $I\lambda^{2}$. The angle spectrum is divergent, and
the PIC data showed only a slight reduction in average angle with
electron energy (justifying our 1D factorization). A more sophisticated
handoff method involving a 4D distribution function, and using more
recent PIC simulations, seems to give qualitatively similar results %
\footnote{C. Bellei, A. J. Kemp, private communication%
}. The major design challenges posed by this source are:
1.\ the electrons are too energetic to fully stop in a DT hot spot, and 2.\ they
are sufficiently divergent that mitigation strategies are required in a point design.

We have developed a transport modeling capability which entails the
hybrid-PIC code Zuma and rad-hydro code Hydra running in tandem. We
detailed the physics contained in Zuma. It is similar to other codes
that use a reduced model to eliminate light and Langmuir waves. Namely,
the displacement current is removed from Amp\`ere's law, and the electric
field is found from Ohm's law (obtained from the background electron
momentum equation). This model is applicable in sufficiently collisional
plasmas, and for time and space scales longer than the plasma frequency
and Debye length.

Zuma-Hydra 2D cylindrical RZ runs on an idealized cone-fuel assembly
were performed. For a perfectly parallel source ($\Delta\theta=0$,
no angular scattering) of mono-energetic 1.5 MeV electrons, ignition
occurred for 18.5 kJ of fast electrons, or 2.1x Atzeni's ideal estimate
of 8.7 kJ. We discussed the impact of (small) divergence, E and B
fields, and the PIC-based energy spectrum, and refer the reader to
Ref.~\onlinecite{strozzi-ifsa-epjwc-2012} for more details, including
the role of different terms in Ohm's law. With the PIC-based energy
spectrum, the full Ohm's law, and an artificially-collimated source
($\Delta\theta=10^{\circ}$), the ignition energy was raised to 132
kJ, or 15x the ideal value.

The realistic angular spectrum was then considered, and raised the
ignition energy to $>$ 1 MJ.
Several mitigation ideas have been proposed, including magnetic fields
produced by resistivity gradients (e.g.~at material interfaces).
While this approach is promising, we chose to examine imposed axial
magnetic fields. An initial, uniform field of 50 MG recovered the
132 kJ ignition energy of the artificially collimated source. Assembling
$\sim10$'s MG field strengths in an ICF implosion, via the frozen-in
law of MHD, is reasonable, and has been demonstrated recently at Omega.

However, a cone-in-shell implosion is not likely to produce a uniform
magnetic field. In particular, the field in the fast electron source
region (inside the cone tip) will not be enhanced much over the seed
value, but would be enhanced in the region between the cone tip and
dense fuel \citep{tabak-dpp-2010}. Fast electrons would therefore
encounter an increasing axial field, and be subject to magnetic mirroring.
Simulations that quantify this effect for a few profiles were shown.
We showed one way to provide confinement but avoid mirroring is a
magnetic pipe, which peaks at a finite radius.

We have started to address the design problem of assembling a pipe
field in an implosion. Inserting an axial structure (a ``wire'')
between the cone tip and fuel, that does not get compressed, is one
way to achieve this. Magnetic confinement schemes based on self-generated
azimuthal fields due to resistivity gradients require a similar structure.
Thus both approaches share some hydro assembly features, and would
mutually benefit from progress in hydro design. One advantage of the
pipe in this regard is that the resistivity (e.g., Z) is irrelevant
for the pipe, as long as it isn't compressed and thereby produce a
large on-axis field. The self-azimuthal fields rely on resistivity
gradients, usually achieved by a high-Z material on-axis. This can
lead to unacceptable fast-electron energy loss or angular scattering
in the wire %
\footnote{H.~D.~Shay, private communication%
}.  Megagauss magnetic fields may lower the ignition threshold by reducing electron
thermal conduction out of the hot spot or, at even higher values, enhancing
alpha deposition.

Integrated hybrid-PIC and rad-hydro simulations offer a powerful new
tool for fast-ignition modeling, and we look forward to them enabling
the emergence of attractive ignition designs.
\begin{acknowledgments}
It is a pleasure to thank A.~A.~Solodov, J.~R.~Davies, B.~I.~Cohen,
H.~D.~Shay, and P.~K.~Patel for fruitful discussions. This work
was performed under the auspices of the U.S. Department of Energy
by Lawrence Livermore National Laboratory under Contract DE-AC52-07NA27344
and partially supported by LDRD 11-SI-002. 
\end{acknowledgments}


\bibliography{fastigb11_arxiv.bbl}

\begin{thebibliography}{61}%
\makeatletter
\providecommand \@ifxundefined [1]{%
 \@ifx{#1\undefined}
}%
\providecommand \@ifnum [1]{%
 \ifnum #1\expandafter \@firstoftwo
 \else \expandafter \@secondoftwo
 \fi
}%
\providecommand \@ifx [1]{%
 \ifx #1\expandafter \@firstoftwo
 \else \expandafter \@secondoftwo
 \fi
}%
\providecommand \natexlab [1]{#1}%
\providecommand \enquote  [1]{``#1''}%
\providecommand \bibnamefont  [1]{#1}%
\providecommand \bibfnamefont [1]{#1}%
\providecommand \citenamefont [1]{#1}%
\providecommand \href@noop [0]{\@secondoftwo}%
\providecommand \href [0]{\begingroup \@sanitize@url \@href}%
\providecommand \@href[1]{\@@startlink{#1}\@@href}%
\providecommand \@@href[1]{\endgroup#1\@@endlink}%
\providecommand \@sanitize@url [0]{\catcode `\\12\catcode `\$12\catcode
  `\&12\catcode `\#12\catcode `\^12\catcode `\_12\catcode `\%12\relax}%
\providecommand \@@startlink[1]{}%
\providecommand \@@endlink[0]{}%
\providecommand \url  [0]{\begingroup\@sanitize@url \@url }%
\providecommand \@url [1]{\endgroup\@href {#1}{\urlprefix }}%
\providecommand \urlprefix  [0]{URL }%
\providecommand \Eprint [0]{\href }%
\providecommand \doibase [0]{http://dx.doi.org/}%
\providecommand \selectlanguage [0]{\@gobble}%
\providecommand \bibinfo  [0]{\@secondoftwo}%
\providecommand \bibfield  [0]{\@secondoftwo}%
\providecommand \translation [1]{[#1]}%
\providecommand \BibitemOpen [0]{}%
\providecommand \bibitemStop [0]{}%
\providecommand \bibitemNoStop [0]{.\EOS\space}%
\providecommand \EOS [0]{\spacefactor3000\relax}%
\providecommand \BibitemShut  [1]{\csname bibitem#1\endcsname}%
\let\auto@bib@innerbib\@empty
\bibitem [{\citenamefont {Tabak}\ \emph {et~al.}(1994)\citenamefont {Tabak},
  \citenamefont {Hammer}, \citenamefont {Glinsky}, \citenamefont {Kruer},
  \citenamefont {Wilks}, \citenamefont {Woodworth}, \citenamefont {Campbell},
  \citenamefont {Perry},\ and\ \citenamefont {Mason}}]{tabak-fi-pop-1994}%
  \BibitemOpen
  \bibfield  {author} {\bibinfo {author} {\bibfnamefont {M.}~\bibnamefont
  {Tabak}}, \bibinfo {author} {\bibfnamefont {J.}~\bibnamefont {Hammer}},
  \bibinfo {author} {\bibfnamefont {M.~E.}\ \bibnamefont {Glinsky}}, \bibinfo
  {author} {\bibfnamefont {W.~L.}\ \bibnamefont {Kruer}}, \bibinfo {author}
  {\bibfnamefont {S.~C.}\ \bibnamefont {Wilks}}, \bibinfo {author}
  {\bibfnamefont {J.}~\bibnamefont {Woodworth}}, \bibinfo {author}
  {\bibfnamefont {E.~M.}\ \bibnamefont {Campbell}}, \bibinfo {author}
  {\bibfnamefont {M.~D.}\ \bibnamefont {Perry}}, \ and\ \bibinfo {author}
  {\bibfnamefont {R.~J.}\ \bibnamefont {Mason}},\ }\href {\doibase
  10.1063/1.870664} {\bibfield  {journal} {\bibinfo  {journal} {Phys. Plasmas}\
  }\textbf {\bibinfo {volume} {1}},\ \bibinfo {pages} {1626} (\bibinfo {year}
  {1994})}\BibitemShut {NoStop}%
\bibitem [{\citenamefont {Basov}, \citenamefont {Gus'kov},\ and\ \citenamefont
  {Feokistov}(1992)}]{basov-fastign-1992}%
  \BibitemOpen
  \bibfield  {author} {\bibinfo {author} {\bibfnamefont {N.~G.}\ \bibnamefont
  {Basov}}, \bibinfo {author} {\bibfnamefont {S.~Y.}\ \bibnamefont {Gus'kov}},
  \ and\ \bibinfo {author} {\bibfnamefont {L.~P.}\ \bibnamefont {Feokistov}},\
  }\href@noop {} {\bibfield  {journal} {\bibinfo  {journal} {J. Sov. Laser
  Res.}\ }\textbf {\bibinfo {volume} {13}},\ \bibinfo {pages} {396} (\bibinfo
  {year} {1992})}\BibitemShut {NoStop}%
\bibitem [{\citenamefont {Kodama}\ \emph {et~al.}(2001)\citenamefont {Kodama},
  \citenamefont {Norreys}, \citenamefont {Mima}, \citenamefont {Dangor},
  \citenamefont {Evans}, \citenamefont {Fujita}, \citenamefont {Kitagawa},
  \citenamefont {Krushelnick}, \citenamefont {Miyakoshi}, \citenamefont
  {Miyanaga}, \citenamefont {Norimatsu}, \citenamefont {Rose}, \citenamefont
  {Shozaki}, \citenamefont {Shigemori}, \citenamefont {Sunahara}, \citenamefont
  {Tampo}, \citenamefont {Tanaka}, \citenamefont {Toyama}, \citenamefont
  {Yamanaka},\ and\ \citenamefont {Zepf}}]{kodama-fastig-nature-2001}%
  \BibitemOpen
  \bibfield  {author} {\bibinfo {author} {\bibfnamefont {R.}~\bibnamefont
  {Kodama}}, \bibinfo {author} {\bibfnamefont {P.~A.}\ \bibnamefont {Norreys}},
  \bibinfo {author} {\bibfnamefont {K.}~\bibnamefont {Mima}}, \bibinfo {author}
  {\bibfnamefont {A.~E.}\ \bibnamefont {Dangor}}, \bibinfo {author}
  {\bibfnamefont {R.~G.}\ \bibnamefont {Evans}}, \bibinfo {author}
  {\bibfnamefont {H.}~\bibnamefont {Fujita}}, \bibinfo {author} {\bibfnamefont
  {Y.}~\bibnamefont {Kitagawa}}, \bibinfo {author} {\bibfnamefont
  {K.}~\bibnamefont {Krushelnick}}, \bibinfo {author} {\bibfnamefont
  {T.}~\bibnamefont {Miyakoshi}}, \bibinfo {author} {\bibfnamefont
  {N.}~\bibnamefont {Miyanaga}}, \bibinfo {author} {\bibfnamefont
  {T.}~\bibnamefont {Norimatsu}}, \bibinfo {author} {\bibfnamefont {S.~J.}\
  \bibnamefont {Rose}}, \bibinfo {author} {\bibfnamefont {T.}~\bibnamefont
  {Shozaki}}, \bibinfo {author} {\bibfnamefont {K.}~\bibnamefont {Shigemori}},
  \bibinfo {author} {\bibfnamefont {A.}~\bibnamefont {Sunahara}}, \bibinfo
  {author} {\bibfnamefont {M.}~\bibnamefont {Tampo}}, \bibinfo {author}
  {\bibfnamefont {K.~A.}\ \bibnamefont {Tanaka}}, \bibinfo {author}
  {\bibfnamefont {Y.}~\bibnamefont {Toyama}}, \bibinfo {author} {\bibfnamefont
  {T.}~\bibnamefont {Yamanaka}}, \ and\ \bibinfo {author} {\bibfnamefont
  {M.}~\bibnamefont {Zepf}},\ }\href@noop {} {\bibfield  {journal} {\bibinfo
  {journal} {Nature}\ }\textbf {\bibinfo {volume} {412}},\ \bibinfo {pages}
  {798} (\bibinfo {year} {2001})}\BibitemShut {NoStop}%
\bibitem [{\citenamefont {Kodama}\ \emph {et~al.}(2002)\citenamefont {Kodama},
  \citenamefont {Shiraga}, \citenamefont {Shigemori}, \citenamefont {Toyama},
  \citenamefont {Fujioka}, \citenamefont {Azechi}, \citenamefont {Fujita},
  \citenamefont {Habara}, \citenamefont {Hall}, \citenamefont {Izawa},
  \citenamefont {Jitsuno}, \citenamefont {Kitagawa}, \citenamefont
  {Krushelnick}, \citenamefont {Lancaster}, \citenamefont {Mima}, \citenamefont
  {Nagai}, \citenamefont {Nakai}, \citenamefont {Nishimura}, \citenamefont
  {Norimatsu}, \citenamefont {Norreys}, \citenamefont {Sakabe}, \citenamefont
  {Tanaka}, \citenamefont {Youssef}, \citenamefont {Zepf},\ and\ \citenamefont
  {Yamanaka}}]{kodama-fastig-nature-2002}%
  \BibitemOpen
  \bibfield  {author} {\bibinfo {author} {\bibfnamefont {R.}~\bibnamefont
  {Kodama}}, \bibinfo {author} {\bibfnamefont {H.}~\bibnamefont {Shiraga}},
  \bibinfo {author} {\bibfnamefont {K.}~\bibnamefont {Shigemori}}, \bibinfo
  {author} {\bibfnamefont {Y.}~\bibnamefont {Toyama}}, \bibinfo {author}
  {\bibfnamefont {S.}~\bibnamefont {Fujioka}}, \bibinfo {author} {\bibfnamefont
  {H.}~\bibnamefont {Azechi}}, \bibinfo {author} {\bibfnamefont
  {H.}~\bibnamefont {Fujita}}, \bibinfo {author} {\bibfnamefont
  {H.}~\bibnamefont {Habara}}, \bibinfo {author} {\bibfnamefont
  {T.}~\bibnamefont {Hall}}, \bibinfo {author} {\bibfnamefont {Y.}~\bibnamefont
  {Izawa}}, \bibinfo {author} {\bibfnamefont {T.}~\bibnamefont {Jitsuno}},
  \bibinfo {author} {\bibfnamefont {Y.}~\bibnamefont {Kitagawa}}, \bibinfo
  {author} {\bibfnamefont {K.~M.}\ \bibnamefont {Krushelnick}}, \bibinfo
  {author} {\bibfnamefont {K.~L.}\ \bibnamefont {Lancaster}}, \bibinfo {author}
  {\bibfnamefont {K.}~\bibnamefont {Mima}}, \bibinfo {author} {\bibfnamefont
  {K.}~\bibnamefont {Nagai}}, \bibinfo {author} {\bibfnamefont
  {M.}~\bibnamefont {Nakai}}, \bibinfo {author} {\bibfnamefont
  {H.}~\bibnamefont {Nishimura}}, \bibinfo {author} {\bibfnamefont
  {T.}~\bibnamefont {Norimatsu}}, \bibinfo {author} {\bibfnamefont {P.~A.}\
  \bibnamefont {Norreys}}, \bibinfo {author} {\bibfnamefont {S.}~\bibnamefont
  {Sakabe}}, \bibinfo {author} {\bibfnamefont {K.~A.}\ \bibnamefont {Tanaka}},
  \bibinfo {author} {\bibfnamefont {A.}~\bibnamefont {Youssef}}, \bibinfo
  {author} {\bibfnamefont {M.}~\bibnamefont {Zepf}}, \ and\ \bibinfo {author}
  {\bibfnamefont {T.}~\bibnamefont {Yamanaka}},\ }\href@noop {} {\bibfield
  {journal} {\bibinfo  {journal} {Nature}\ }\textbf {\bibinfo {volume} {418}},\
  \bibinfo {pages} {933} (\bibinfo {year} {2002})}\BibitemShut {NoStop}%
\bibitem [{\citenamefont {Key}\ \emph {et~al.}(2008)\citenamefont {Key},
  \citenamefont {Adam}, \citenamefont {Akli}, \citenamefont {Borghesi},
  \citenamefont {Chen}, \citenamefont {Evans}, \citenamefont {Freeman},
  \citenamefont {Habara}, \citenamefont {Hatchett}, \citenamefont {Hill},
  \citenamefont {Heron}, \citenamefont {King}, \citenamefont {Kodama},
  \citenamefont {Lancaster}, \citenamefont {MacKinnon}, \citenamefont {Patel},
  \citenamefont {Phillips}, \citenamefont {Romagnani}, \citenamefont {Snavely},
  \citenamefont {Stephens}, \citenamefont {Stoeckl}, \citenamefont {Town},
  \citenamefont {Toyama}, \citenamefont {Zhang}, \citenamefont {Zepf},\ and\
  \citenamefont {Norreys}}]{key-fastign-pop-2008}%
  \BibitemOpen
  \bibfield  {author} {\bibinfo {author} {\bibfnamefont {M.~H.}\ \bibnamefont
  {Key}}, \bibinfo {author} {\bibfnamefont {J.~C.}\ \bibnamefont {Adam}},
  \bibinfo {author} {\bibfnamefont {K.~U.}\ \bibnamefont {Akli}}, \bibinfo
  {author} {\bibfnamefont {M.}~\bibnamefont {Borghesi}}, \bibinfo {author}
  {\bibfnamefont {M.~H.}\ \bibnamefont {Chen}}, \bibinfo {author}
  {\bibfnamefont {R.~G.}\ \bibnamefont {Evans}}, \bibinfo {author}
  {\bibfnamefont {R.~R.}\ \bibnamefont {Freeman}}, \bibinfo {author}
  {\bibfnamefont {H.}~\bibnamefont {Habara}}, \bibinfo {author} {\bibfnamefont
  {S.~P.}\ \bibnamefont {Hatchett}}, \bibinfo {author} {\bibfnamefont {J.~M.}\
  \bibnamefont {Hill}}, \bibinfo {author} {\bibfnamefont {A.}~\bibnamefont
  {Heron}}, \bibinfo {author} {\bibfnamefont {J.~A.}\ \bibnamefont {King}},
  \bibinfo {author} {\bibfnamefont {R.}~\bibnamefont {Kodama}}, \bibinfo
  {author} {\bibfnamefont {K.~L.}\ \bibnamefont {Lancaster}}, \bibinfo {author}
  {\bibfnamefont {A.~J.}\ \bibnamefont {MacKinnon}}, \bibinfo {author}
  {\bibfnamefont {P.}~\bibnamefont {Patel}}, \bibinfo {author} {\bibfnamefont
  {T.}~\bibnamefont {Phillips}}, \bibinfo {author} {\bibfnamefont
  {L.}~\bibnamefont {Romagnani}}, \bibinfo {author} {\bibfnamefont {R.~A.}\
  \bibnamefont {Snavely}}, \bibinfo {author} {\bibfnamefont {R.}~\bibnamefont
  {Stephens}}, \bibinfo {author} {\bibfnamefont {C.}~\bibnamefont {Stoeckl}},
  \bibinfo {author} {\bibfnamefont {R.}~\bibnamefont {Town}}, \bibinfo {author}
  {\bibfnamefont {Y.}~\bibnamefont {Toyama}}, \bibinfo {author} {\bibfnamefont
  {B.}~\bibnamefont {Zhang}}, \bibinfo {author} {\bibfnamefont
  {M.}~\bibnamefont {Zepf}}, \ and\ \bibinfo {author} {\bibfnamefont {P.~A.}\
  \bibnamefont {Norreys}},\ }\href {\doibase DOI:10.1063/1.2834727} {\bibfield
  {journal} {\bibinfo  {journal} {Phys. Plasmas}\ }\textbf {\bibinfo {volume}
  {15}},\ \bibinfo {pages} {022701} (\bibinfo {year} {2008})}\BibitemShut
  {NoStop}%
\bibitem [{\citenamefont {Theobald}\ \emph {et~al.}(2011)\citenamefont
  {Theobald}, \citenamefont {Solodov}, \citenamefont {Stoeckl}, \citenamefont
  {Anderson}, \citenamefont {Betti}, \citenamefont {Boehly}, \citenamefont
  {Craxton}, \citenamefont {Delettrez}, \citenamefont {Dorrer}, \citenamefont
  {Frenje}, \citenamefont {Glebov}, \citenamefont {Habara}, \citenamefont
  {Tanaka}, \citenamefont {Knauer}, \citenamefont {Lauck}, \citenamefont
  {Marshall}, \citenamefont {Marshall}, \citenamefont {Meyerhofer},
  \citenamefont {Nilson}, \citenamefont {Patel}, \citenamefont {Chen},
  \citenamefont {Sangster}, \citenamefont {Seka}, \citenamefont {Sinenian},
  \citenamefont {Ma}, \citenamefont {Beg}, \citenamefont {Giraldez},\ and\
  \citenamefont {Stephens}}]{theobald-fastign-pop-2011}%
  \BibitemOpen
  \bibfield  {author} {\bibinfo {author} {\bibfnamefont {W.}~\bibnamefont
  {Theobald}}, \bibinfo {author} {\bibfnamefont {A.~A.}\ \bibnamefont
  {Solodov}}, \bibinfo {author} {\bibfnamefont {C.}~\bibnamefont {Stoeckl}},
  \bibinfo {author} {\bibfnamefont {K.~S.}\ \bibnamefont {Anderson}}, \bibinfo
  {author} {\bibfnamefont {R.}~\bibnamefont {Betti}}, \bibinfo {author}
  {\bibfnamefont {T.~R.}\ \bibnamefont {Boehly}}, \bibinfo {author}
  {\bibfnamefont {R.~S.}\ \bibnamefont {Craxton}}, \bibinfo {author}
  {\bibfnamefont {J.~A.}\ \bibnamefont {Delettrez}}, \bibinfo {author}
  {\bibfnamefont {C.}~\bibnamefont {Dorrer}}, \bibinfo {author} {\bibfnamefont
  {J.~A.}\ \bibnamefont {Frenje}}, \bibinfo {author} {\bibfnamefont {V.~Y.}\
  \bibnamefont {Glebov}}, \bibinfo {author} {\bibfnamefont {H.}~\bibnamefont
  {Habara}}, \bibinfo {author} {\bibfnamefont {K.~A.}\ \bibnamefont {Tanaka}},
  \bibinfo {author} {\bibfnamefont {J.~P.}\ \bibnamefont {Knauer}}, \bibinfo
  {author} {\bibfnamefont {R.}~\bibnamefont {Lauck}}, \bibinfo {author}
  {\bibfnamefont {F.~J.}\ \bibnamefont {Marshall}}, \bibinfo {author}
  {\bibfnamefont {K.~L.}\ \bibnamefont {Marshall}}, \bibinfo {author}
  {\bibfnamefont {D.~D.}\ \bibnamefont {Meyerhofer}}, \bibinfo {author}
  {\bibfnamefont {P.~M.}\ \bibnamefont {Nilson}}, \bibinfo {author}
  {\bibfnamefont {P.~K.}\ \bibnamefont {Patel}}, \bibinfo {author}
  {\bibfnamefont {H.}~\bibnamefont {Chen}}, \bibinfo {author} {\bibfnamefont
  {T.~C.}\ \bibnamefont {Sangster}}, \bibinfo {author} {\bibfnamefont
  {W.}~\bibnamefont {Seka}}, \bibinfo {author} {\bibfnamefont {N.}~\bibnamefont
  {Sinenian}}, \bibinfo {author} {\bibfnamefont {T.}~\bibnamefont {Ma}},
  \bibinfo {author} {\bibfnamefont {F.~N.}\ \bibnamefont {Beg}}, \bibinfo
  {author} {\bibfnamefont {E.}~\bibnamefont {Giraldez}}, \ and\ \bibinfo
  {author} {\bibfnamefont {R.~B.}\ \bibnamefont {Stephens}},\ }\href {\doibase
  DOI:10.1063/1.3566082} {\bibfield  {journal} {\bibinfo  {journal} {Phys.
  Plasmas}\ }\textbf {\bibinfo {volume} {18}},\ \bibinfo {pages} {056305}
  (\bibinfo {year} {2011})}\BibitemShut {NoStop}%
\bibitem [{\citenamefont {Shiraga}\ \emph {et~al.}(2011)\citenamefont
  {Shiraga}, \citenamefont {Fujioka}, \citenamefont {Nakai}, \citenamefont
  {Watari}, \citenamefont {Nakamura}, \citenamefont {Arikawa}, \citenamefont
  {Hosoda}, \citenamefont {Nagai}, \citenamefont {Koga}, \citenamefont
  {Kikuchi}, \citenamefont {Ishii}, \citenamefont {Sogo}, \citenamefont
  {Shigemori}, \citenamefont {Nishimura}, \citenamefont {Zhang}, \citenamefont
  {Tanabe}, \citenamefont {Ohira}, \citenamefont {Fujii}, \citenamefont
  {Namimoto}, \citenamefont {Sakawa}, \citenamefont {Maegawa}, \citenamefont
  {Ozaki}, \citenamefont {Tanaka}, \citenamefont {Habara}, \citenamefont
  {Iwawaki}, \citenamefont {Shimada}, \citenamefont {Nagatomo}, \citenamefont
  {Johzaki}, \citenamefont {Sunahara}, \citenamefont {Murakami}, \citenamefont
  {Sakagami}, \citenamefont {Taguchi}, \citenamefont {Norimatsu}, \citenamefont
  {Homma}, \citenamefont {Fujimoto}, \citenamefont {Iwamoto}, \citenamefont
  {Miyanaga}, \citenamefont {Kawanaka}, \citenamefont {Jitsuno}, \citenamefont
  {Nakata}, \citenamefont {Tsubakimoto}, \citenamefont {Morio}, \citenamefont
  {Kawasaki}, \citenamefont {Sawai}, \citenamefont {Tsuji}, \citenamefont
  {Murakami}, \citenamefont {Kanabe}, \citenamefont {Kondo}, \citenamefont
  {Sarukura}, \citenamefont {Shimizu}, \citenamefont {Mima},\ and\
  \citenamefont {Azechi}}]{shiraga-gekko-ppcf-2011}%
  \BibitemOpen
  \bibfield  {author} {\bibinfo {author} {\bibfnamefont {H.}~\bibnamefont
  {Shiraga}}, \bibinfo {author} {\bibfnamefont {S.}~\bibnamefont {Fujioka}},
  \bibinfo {author} {\bibfnamefont {M.}~\bibnamefont {Nakai}}, \bibinfo
  {author} {\bibfnamefont {T.}~\bibnamefont {Watari}}, \bibinfo {author}
  {\bibfnamefont {H.}~\bibnamefont {Nakamura}}, \bibinfo {author}
  {\bibfnamefont {Y.}~\bibnamefont {Arikawa}}, \bibinfo {author} {\bibfnamefont
  {H.}~\bibnamefont {Hosoda}}, \bibinfo {author} {\bibfnamefont
  {T.}~\bibnamefont {Nagai}}, \bibinfo {author} {\bibfnamefont
  {M.}~\bibnamefont {Koga}}, \bibinfo {author} {\bibfnamefont {H.}~\bibnamefont
  {Kikuchi}}, \bibinfo {author} {\bibfnamefont {Y.}~\bibnamefont {Ishii}},
  \bibinfo {author} {\bibfnamefont {T.}~\bibnamefont {Sogo}}, \bibinfo {author}
  {\bibfnamefont {K.}~\bibnamefont {Shigemori}}, \bibinfo {author}
  {\bibfnamefont {H.}~\bibnamefont {Nishimura}}, \bibinfo {author}
  {\bibfnamefont {Z.}~\bibnamefont {Zhang}}, \bibinfo {author} {\bibfnamefont
  {M.}~\bibnamefont {Tanabe}}, \bibinfo {author} {\bibfnamefont
  {S.}~\bibnamefont {Ohira}}, \bibinfo {author} {\bibfnamefont
  {Y.}~\bibnamefont {Fujii}}, \bibinfo {author} {\bibfnamefont
  {T.}~\bibnamefont {Namimoto}}, \bibinfo {author} {\bibfnamefont
  {Y.}~\bibnamefont {Sakawa}}, \bibinfo {author} {\bibfnamefont
  {O.}~\bibnamefont {Maegawa}}, \bibinfo {author} {\bibfnamefont
  {T.}~\bibnamefont {Ozaki}}, \bibinfo {author} {\bibfnamefont
  {K.}~\bibnamefont {Tanaka}}, \bibinfo {author} {\bibfnamefont
  {H.}~\bibnamefont {Habara}}, \bibinfo {author} {\bibfnamefont
  {T.}~\bibnamefont {Iwawaki}}, \bibinfo {author} {\bibfnamefont
  {K.}~\bibnamefont {Shimada}}, \bibinfo {author} {\bibfnamefont
  {H.}~\bibnamefont {Nagatomo}}, \bibinfo {author} {\bibfnamefont
  {T.}~\bibnamefont {Johzaki}}, \bibinfo {author} {\bibfnamefont
  {A.}~\bibnamefont {Sunahara}}, \bibinfo {author} {\bibfnamefont
  {M.}~\bibnamefont {Murakami}}, \bibinfo {author} {\bibfnamefont
  {H.}~\bibnamefont {Sakagami}}, \bibinfo {author} {\bibfnamefont
  {T.}~\bibnamefont {Taguchi}}, \bibinfo {author} {\bibfnamefont
  {T.}~\bibnamefont {Norimatsu}}, \bibinfo {author} {\bibfnamefont
  {H.}~\bibnamefont {Homma}}, \bibinfo {author} {\bibfnamefont
  {Y.}~\bibnamefont {Fujimoto}}, \bibinfo {author} {\bibfnamefont
  {A.}~\bibnamefont {Iwamoto}}, \bibinfo {author} {\bibfnamefont
  {N.}~\bibnamefont {Miyanaga}}, \bibinfo {author} {\bibfnamefont
  {J.}~\bibnamefont {Kawanaka}}, \bibinfo {author} {\bibfnamefont
  {T.}~\bibnamefont {Jitsuno}}, \bibinfo {author} {\bibfnamefont
  {Y.}~\bibnamefont {Nakata}}, \bibinfo {author} {\bibfnamefont
  {K.}~\bibnamefont {Tsubakimoto}}, \bibinfo {author} {\bibfnamefont
  {N.}~\bibnamefont {Morio}}, \bibinfo {author} {\bibfnamefont
  {T.}~\bibnamefont {Kawasaki}}, \bibinfo {author} {\bibfnamefont
  {K.}~\bibnamefont {Sawai}}, \bibinfo {author} {\bibfnamefont
  {K.}~\bibnamefont {Tsuji}}, \bibinfo {author} {\bibfnamefont
  {H.}~\bibnamefont {Murakami}}, \bibinfo {author} {\bibfnamefont
  {T.}~\bibnamefont {Kanabe}}, \bibinfo {author} {\bibfnamefont
  {K.}~\bibnamefont {Kondo}}, \bibinfo {author} {\bibfnamefont
  {N.}~\bibnamefont {Sarukura}}, \bibinfo {author} {\bibfnamefont
  {T.}~\bibnamefont {Shimizu}}, \bibinfo {author} {\bibfnamefont
  {K.}~\bibnamefont {Mima}}, \ and\ \bibinfo {author} {\bibfnamefont
  {H.}~\bibnamefont {Azechi}},\ }\href
  {http://stacks.iop.org/0741-3335/53/i=12/a=124029} {\bibfield  {journal}
  {\bibinfo  {journal} {Plasma Phys. Controlled Fusion}\ }\textbf {\bibinfo
  {volume} {53}},\ \bibinfo {pages} {124029} (\bibinfo {year}
  {2011})}\BibitemShut {NoStop}%
\bibitem [{\citenamefont {Fujioka}\ \emph {et~al.}(2011)\citenamefont
  {Fujioka}, \citenamefont {Azechi}, \citenamefont {Shiraga}, \citenamefont
  {Miyanaga}, \citenamefont {Norimatsu}, \citenamefont {Sarukura},
  \citenamefont {Nagatomo}, \citenamefont {Johzaki},\ and\ \citenamefont
  {Sunahara}}]{fujioka-sofe-2011}%
  \BibitemOpen
  \bibfield  {author} {\bibinfo {author} {\bibfnamefont {S.}~\bibnamefont
  {Fujioka}}, \bibinfo {author} {\bibfnamefont {H.}~\bibnamefont {Azechi}},
  \bibinfo {author} {\bibfnamefont {H.}~\bibnamefont {Shiraga}}, \bibinfo
  {author} {\bibfnamefont {N.}~\bibnamefont {Miyanaga}}, \bibinfo {author}
  {\bibfnamefont {T.}~\bibnamefont {Norimatsu}}, \bibinfo {author}
  {\bibfnamefont {N.}~\bibnamefont {Sarukura}}, \bibinfo {author}
  {\bibfnamefont {H.}~\bibnamefont {Nagatomo}}, \bibinfo {author}
  {\bibfnamefont {T.}~\bibnamefont {Johzaki}}, \ and\ \bibinfo {author}
  {\bibfnamefont {A.}~\bibnamefont {Sunahara}},\ }in\ \href {\doibase
  10.1109/SOFE.2011.6052352} {\emph {\bibinfo {booktitle} {Fusion Engineering
  (SOFE), 2011 IEEE/NPSS 24th Symposium on}}}\ (\bibinfo {year} {2011})\ pp.\
  \bibinfo {pages} {1 --4}\BibitemShut {NoStop}%
\bibitem [{\citenamefont {Baton}\ \emph {et~al.}(2008)\citenamefont {Baton},
  \citenamefont {Koenig}, \citenamefont {Fuchs}, \citenamefont
  {Benuzzi-Mounaix}, \citenamefont {Guillou}, \citenamefont {Loupias},
  \citenamefont {Vinci}, \citenamefont {Gremillet}, \citenamefont {Rousseaux},
  \citenamefont {Drouin}, \citenamefont {Lefebvre}, \citenamefont {Dorchies},
  \citenamefont {Fourment}, \citenamefont {Santos}, \citenamefont {Batani},
  \citenamefont {Morace}, \citenamefont {Redaelli}, \citenamefont
  {Nakatsutsumi}, \citenamefont {Kodama}, \citenamefont {Nishida},
  \citenamefont {Ozaki}, \citenamefont {Norimatsu}, \citenamefont {Aglitskiy},
  \citenamefont {Atzeni},\ and\ \citenamefont {Schiavi}}]{baton-cone-pop-2008}%
  \BibitemOpen
  \bibfield  {author} {\bibinfo {author} {\bibfnamefont {S.~D.}\ \bibnamefont
  {Baton}}, \bibinfo {author} {\bibfnamefont {M.}~\bibnamefont {Koenig}},
  \bibinfo {author} {\bibfnamefont {J.}~\bibnamefont {Fuchs}}, \bibinfo
  {author} {\bibfnamefont {A.}~\bibnamefont {Benuzzi-Mounaix}}, \bibinfo
  {author} {\bibfnamefont {P.}~\bibnamefont {Guillou}}, \bibinfo {author}
  {\bibfnamefont {B.}~\bibnamefont {Loupias}}, \bibinfo {author} {\bibfnamefont
  {T.}~\bibnamefont {Vinci}}, \bibinfo {author} {\bibfnamefont
  {L.}~\bibnamefont {Gremillet}}, \bibinfo {author} {\bibfnamefont
  {C.}~\bibnamefont {Rousseaux}}, \bibinfo {author} {\bibfnamefont
  {M.}~\bibnamefont {Drouin}}, \bibinfo {author} {\bibfnamefont
  {E.}~\bibnamefont {Lefebvre}}, \bibinfo {author} {\bibfnamefont
  {F.}~\bibnamefont {Dorchies}}, \bibinfo {author} {\bibfnamefont
  {C.}~\bibnamefont {Fourment}}, \bibinfo {author} {\bibfnamefont {J.~J.}\
  \bibnamefont {Santos}}, \bibinfo {author} {\bibfnamefont {D.}~\bibnamefont
  {Batani}}, \bibinfo {author} {\bibfnamefont {A.}~\bibnamefont {Morace}},
  \bibinfo {author} {\bibfnamefont {R.}~\bibnamefont {Redaelli}}, \bibinfo
  {author} {\bibfnamefont {M.}~\bibnamefont {Nakatsutsumi}}, \bibinfo {author}
  {\bibfnamefont {R.}~\bibnamefont {Kodama}}, \bibinfo {author} {\bibfnamefont
  {A.}~\bibnamefont {Nishida}}, \bibinfo {author} {\bibfnamefont
  {N.}~\bibnamefont {Ozaki}}, \bibinfo {author} {\bibfnamefont
  {T.}~\bibnamefont {Norimatsu}}, \bibinfo {author} {\bibfnamefont
  {Y.}~\bibnamefont {Aglitskiy}}, \bibinfo {author} {\bibfnamefont
  {S.}~\bibnamefont {Atzeni}}, \ and\ \bibinfo {author} {\bibfnamefont
  {A.}~\bibnamefont {Schiavi}},\ }\href {\doibase 10.1063/1.2903054} {\bibfield
   {journal} {\bibinfo  {journal} {Phys. Plasmas}\ }\textbf {\bibinfo {volume}
  {15}},\ \bibinfo {eid} {042706} (\bibinfo {year} {2008})}\BibitemShut
  {NoStop}%
\bibitem [{\citenamefont {MacPhee}\ \emph {et~al.}(2010)\citenamefont
  {MacPhee}, \citenamefont {Divol}, \citenamefont {Kemp}, \citenamefont {Akli},
  \citenamefont {Beg}, \citenamefont {Chen}, \citenamefont {Chen},
  \citenamefont {Hey}, \citenamefont {Fedosejevs}, \citenamefont {Freeman},
  \citenamefont {Henesian}, \citenamefont {Key}, \citenamefont {Le~Pape},
  \citenamefont {Link}, \citenamefont {Ma}, \citenamefont {Mackinnon},
  \citenamefont {Ovchinnikov}, \citenamefont {Patel}, \citenamefont {Phillips},
  \citenamefont {Stephens}, \citenamefont {Tabak}, \citenamefont {Town},
  \citenamefont {Tsui}, \citenamefont {Van~Woerkom}, \citenamefont {Wei},\ and\
  \citenamefont {Wilks}}]{macphee-prepulse-prl-2010}%
  \BibitemOpen
  \bibfield  {author} {\bibinfo {author} {\bibfnamefont {A.~G.}\ \bibnamefont
  {MacPhee}}, \bibinfo {author} {\bibfnamefont {L.}~\bibnamefont {Divol}},
  \bibinfo {author} {\bibfnamefont {A.~J.}\ \bibnamefont {Kemp}}, \bibinfo
  {author} {\bibfnamefont {K.~U.}\ \bibnamefont {Akli}}, \bibinfo {author}
  {\bibfnamefont {F.~N.}\ \bibnamefont {Beg}}, \bibinfo {author} {\bibfnamefont
  {C.~D.}\ \bibnamefont {Chen}}, \bibinfo {author} {\bibfnamefont
  {H.}~\bibnamefont {Chen}}, \bibinfo {author} {\bibfnamefont {D.~S.}\
  \bibnamefont {Hey}}, \bibinfo {author} {\bibfnamefont {R.~J.}\ \bibnamefont
  {Fedosejevs}}, \bibinfo {author} {\bibfnamefont {R.~R.}\ \bibnamefont
  {Freeman}}, \bibinfo {author} {\bibfnamefont {M.}~\bibnamefont {Henesian}},
  \bibinfo {author} {\bibfnamefont {M.~H.}\ \bibnamefont {Key}}, \bibinfo
  {author} {\bibfnamefont {S.}~\bibnamefont {Le~Pape}}, \bibinfo {author}
  {\bibfnamefont {A.}~\bibnamefont {Link}}, \bibinfo {author} {\bibfnamefont
  {T.}~\bibnamefont {Ma}}, \bibinfo {author} {\bibfnamefont {A.~J.}\
  \bibnamefont {Mackinnon}}, \bibinfo {author} {\bibfnamefont {V.~M.}\
  \bibnamefont {Ovchinnikov}}, \bibinfo {author} {\bibfnamefont {P.~K.}\
  \bibnamefont {Patel}}, \bibinfo {author} {\bibfnamefont {T.~W.}\ \bibnamefont
  {Phillips}}, \bibinfo {author} {\bibfnamefont {R.~B.}\ \bibnamefont
  {Stephens}}, \bibinfo {author} {\bibfnamefont {M.}~\bibnamefont {Tabak}},
  \bibinfo {author} {\bibfnamefont {R.}~\bibnamefont {Town}}, \bibinfo {author}
  {\bibfnamefont {Y.~Y.}\ \bibnamefont {Tsui}}, \bibinfo {author}
  {\bibfnamefont {L.~D.}\ \bibnamefont {Van~Woerkom}}, \bibinfo {author}
  {\bibfnamefont {M.~S.}\ \bibnamefont {Wei}}, \ and\ \bibinfo {author}
  {\bibfnamefont {S.~C.}\ \bibnamefont {Wilks}},\ }\href {\doibase
  10.1103/PhysRevLett.104.055002} {\bibfield  {journal} {\bibinfo  {journal}
  {Phys. Rev. Lett.}\ }\textbf {\bibinfo {volume} {104}},\ \bibinfo {pages}
  {055002} (\bibinfo {year} {2010})}\BibitemShut {NoStop}%
\bibitem [{\citenamefont {Ma}\ \emph {et~al.}(2012)\citenamefont {Ma},
  \citenamefont {Sawada}, \citenamefont {Patel}, \citenamefont {Chen},
  \citenamefont {Divol}, \citenamefont {Higginson}, \citenamefont {Kemp},
  \citenamefont {Key}, \citenamefont {Larson}, \citenamefont {Le~Pape},
  \citenamefont {Link}, \citenamefont {MacPhee}, \citenamefont {McLean},
  \citenamefont {Ping}, \citenamefont {Stephens}, \citenamefont {Wilks},\ and\
  \citenamefont {Beg}}]{ma-conewire-prl-2012}%
  \BibitemOpen
  \bibfield  {author} {\bibinfo {author} {\bibfnamefont {T.}~\bibnamefont
  {Ma}}, \bibinfo {author} {\bibfnamefont {H.}~\bibnamefont {Sawada}}, \bibinfo
  {author} {\bibfnamefont {P.~K.}\ \bibnamefont {Patel}}, \bibinfo {author}
  {\bibfnamefont {C.~D.}\ \bibnamefont {Chen}}, \bibinfo {author}
  {\bibfnamefont {L.}~\bibnamefont {Divol}}, \bibinfo {author} {\bibfnamefont
  {D.~P.}\ \bibnamefont {Higginson}}, \bibinfo {author} {\bibfnamefont {A.~J.}\
  \bibnamefont {Kemp}}, \bibinfo {author} {\bibfnamefont {M.~H.}\ \bibnamefont
  {Key}}, \bibinfo {author} {\bibfnamefont {D.~J.}\ \bibnamefont {Larson}},
  \bibinfo {author} {\bibfnamefont {S.}~\bibnamefont {Le~Pape}}, \bibinfo
  {author} {\bibfnamefont {A.}~\bibnamefont {Link}}, \bibinfo {author}
  {\bibfnamefont {A.~G.}\ \bibnamefont {MacPhee}}, \bibinfo {author}
  {\bibfnamefont {H.~S.}\ \bibnamefont {McLean}}, \bibinfo {author}
  {\bibfnamefont {Y.}~\bibnamefont {Ping}}, \bibinfo {author} {\bibfnamefont
  {R.~B.}\ \bibnamefont {Stephens}}, \bibinfo {author} {\bibfnamefont {S.~C.}\
  \bibnamefont {Wilks}}, \ and\ \bibinfo {author} {\bibfnamefont {F.~N.}\
  \bibnamefont {Beg}},\ }\href {\doibase 10.1103/PhysRevLett.108.115004}
  {\bibfield  {journal} {\bibinfo  {journal} {Phys. Rev. Lett.}\ }\textbf
  {\bibinfo {volume} {108}},\ \bibinfo {pages} {115004} (\bibinfo {year}
  {2012})}\BibitemShut {NoStop}%
\bibitem [{\citenamefont {Larson}, \citenamefont {Tabak},\ and\ \citenamefont
  {Ma}(2010)}]{larson-zuma-dpp-2010}%
  \BibitemOpen
  \bibfield  {author} {\bibinfo {author} {\bibfnamefont {D.}~\bibnamefont
  {Larson}}, \bibinfo {author} {\bibfnamefont {M.}~\bibnamefont {Tabak}}, \
  and\ \bibinfo {author} {\bibfnamefont {T.}~\bibnamefont {Ma}},\ }\href@noop
  {} {\bibfield  {journal} {\bibinfo  {journal} {Bull. Am. Phys. Soc.}\
  }\textbf {\bibinfo {volume} {55}} (\bibinfo {year} {2010})},\ \bibinfo {note}
  {poster JP9 119, APS-DPP 2010, Atlanta, USA}\BibitemShut {NoStop}%
\bibitem [{\citenamefont {Marinak}\ \emph {et~al.}(2001)\citenamefont
  {Marinak}, \citenamefont {Kerbel}, \citenamefont {Gentile}, \citenamefont
  {Jones}, \citenamefont {Munro}, \citenamefont {Pollaine}, \citenamefont
  {Dittrich},\ and\ \citenamefont {Haan}}]{marinak-hydra-pop-2001}%
  \BibitemOpen
  \bibfield  {author} {\bibinfo {author} {\bibfnamefont {M.~M.}\ \bibnamefont
  {Marinak}}, \bibinfo {author} {\bibfnamefont {G.~D.}\ \bibnamefont {Kerbel}},
  \bibinfo {author} {\bibfnamefont {N.~A.}\ \bibnamefont {Gentile}}, \bibinfo
  {author} {\bibfnamefont {O.}~\bibnamefont {Jones}}, \bibinfo {author}
  {\bibfnamefont {D.}~\bibnamefont {Munro}}, \bibinfo {author} {\bibfnamefont
  {S.}~\bibnamefont {Pollaine}}, \bibinfo {author} {\bibfnamefont {T.~R.}\
  \bibnamefont {Dittrich}}, \ and\ \bibinfo {author} {\bibfnamefont {S.~W.}\
  \bibnamefont {Haan}},\ }\href@noop {} {\bibfield  {journal} {\bibinfo
  {journal} {Phys. Plasmas}\ }\textbf {\bibinfo {volume} {8}},\ \bibinfo
  {pages} {2275} (\bibinfo {year} {2001})}\BibitemShut {NoStop}%
\bibitem [{\citenamefont {Bonitz}\ \emph {et~al.}(2006)\citenamefont {Bonitz},
  \citenamefont {Bertsch}, \citenamefont {Filinov},\ and\ \citenamefont
  {Ruhl}}]{bonitz-psc-manybody}%
  \BibitemOpen
  \bibfield  {author} {\bibinfo {author} {\bibfnamefont {M.}~\bibnamefont
  {Bonitz}}, \bibinfo {author} {\bibfnamefont {G.}~\bibnamefont {Bertsch}},
  \bibinfo {author} {\bibfnamefont {V.}~\bibnamefont {Filinov}}, \ and\
  \bibinfo {author} {\bibfnamefont {H.}~\bibnamefont {Ruhl}},\ }\enquote
  {\bibinfo {title} {Introduction to computational methods in many body
  physics},}\ \ (\bibinfo  {publisher} {Rinton Press},\ \bibinfo {address}
  {Princeton, NJ},\ \bibinfo {year} {2006})\ Chap.~\bibinfo {chapter}
  {2}\BibitemShut {NoStop}%
\bibitem [{\citenamefont {Kemp}, \citenamefont {Cohen},\ and\ \citenamefont
  {Divol}(2010)}]{kemp-pop-2010}%
  \BibitemOpen
  \bibfield  {author} {\bibinfo {author} {\bibfnamefont {A.~J.}\ \bibnamefont
  {Kemp}}, \bibinfo {author} {\bibfnamefont {B.~I.}\ \bibnamefont {Cohen}}, \
  and\ \bibinfo {author} {\bibfnamefont {L.}~\bibnamefont {Divol}},\ }\href
  {\doibase 10.1063/1.3312825} {\bibfield  {journal} {\bibinfo  {journal}
  {Phys. Plasmas}\ }\textbf {\bibinfo {volume} {17}},\ \bibinfo {eid} {056702}
  (\bibinfo {year} {2010})}\BibitemShut {NoStop}%
\bibitem [{\citenamefont {Wilks}\ \emph {et~al.}(1992)\citenamefont {Wilks},
  \citenamefont {Kruer}, \citenamefont {Tabak},\ and\ \citenamefont
  {Langdon}}]{wilks-pond-prl-1992}%
  \BibitemOpen
  \bibfield  {author} {\bibinfo {author} {\bibfnamefont {S.~C.}\ \bibnamefont
  {Wilks}}, \bibinfo {author} {\bibfnamefont {W.~L.}\ \bibnamefont {Kruer}},
  \bibinfo {author} {\bibfnamefont {M.}~\bibnamefont {Tabak}}, \ and\ \bibinfo
  {author} {\bibfnamefont {A.~B.}\ \bibnamefont {Langdon}},\ }\href {\doibase
  10.1103/PhysRevLett.69.1383} {\bibfield  {journal} {\bibinfo  {journal}
  {Phys. Rev. Lett.}\ }\textbf {\bibinfo {volume} {69}},\ \bibinfo {pages}
  {1383} (\bibinfo {year} {1992})}\BibitemShut {NoStop}%
\bibitem [{\citenamefont {Ren}\ \emph {et~al.}(2006)\citenamefont {Ren},
  \citenamefont {Tzoufras}, \citenamefont {Tonge}, \citenamefont {Mori},
  \citenamefont {Tsung}, \citenamefont {Fiore}, \citenamefont {Fonseca},
  \citenamefont {Silva}, \citenamefont {Adam},\ and\ \citenamefont
  {Heron}}]{ren-aps05-pop-2006}%
  \BibitemOpen
  \bibfield  {author} {\bibinfo {author} {\bibfnamefont {C.}~\bibnamefont
  {Ren}}, \bibinfo {author} {\bibfnamefont {M.}~\bibnamefont {Tzoufras}},
  \bibinfo {author} {\bibfnamefont {J.}~\bibnamefont {Tonge}}, \bibinfo
  {author} {\bibfnamefont {W.~B.}\ \bibnamefont {Mori}}, \bibinfo {author}
  {\bibfnamefont {F.~S.}\ \bibnamefont {Tsung}}, \bibinfo {author}
  {\bibfnamefont {M.}~\bibnamefont {Fiore}}, \bibinfo {author} {\bibfnamefont
  {R.~A.}\ \bibnamefont {Fonseca}}, \bibinfo {author} {\bibfnamefont {L.~O.}\
  \bibnamefont {Silva}}, \bibinfo {author} {\bibfnamefont {J.-C.}\ \bibnamefont
  {Adam}}, \ and\ \bibinfo {author} {\bibfnamefont {A.}~\bibnamefont {Heron}},\
  }\href {\doibase 10.1063/1.2173617} {\bibfield  {journal} {\bibinfo
  {journal} {Phys. Plasmas}\ }\textbf {\bibinfo {volume} {13}},\ \bibinfo {eid}
  {056308} (\bibinfo {year} {2006})}\BibitemShut {NoStop}%
\bibitem [{\citenamefont {Adam}, \citenamefont {H\'eron},\ and\ \citenamefont
  {Laval}(2006)}]{adam-shortpulse-prl-2006}%
  \BibitemOpen
  \bibfield  {author} {\bibinfo {author} {\bibfnamefont {J.~C.}\ \bibnamefont
  {Adam}}, \bibinfo {author} {\bibfnamefont {A.}~\bibnamefont {H\'eron}}, \
  and\ \bibinfo {author} {\bibfnamefont {G.}~\bibnamefont {Laval}},\ }\href
  {\doibase 10.1103/PhysRevLett.97.205006} {\bibfield  {journal} {\bibinfo
  {journal} {Phys. Rev. Lett.}\ }\textbf {\bibinfo {volume} {97}},\ \bibinfo
  {pages} {205006} (\bibinfo {year} {2006})}\BibitemShut {NoStop}%
\bibitem [{\citenamefont {Honrubia}\ \emph {et~al.}(2006)\citenamefont
  {Honrubia}, \citenamefont {Alfons\'in}, \citenamefont {Alonso}, \citenamefont
  {P\'erez},\ and\ \citenamefont {Cerrada}}]{honrubia-heating-lpb-2006}%
  \BibitemOpen
  \bibfield  {author} {\bibinfo {author} {\bibfnamefont {J.~J.}\ \bibnamefont
  {Honrubia}}, \bibinfo {author} {\bibfnamefont {C.}~\bibnamefont
  {Alfons\'in}}, \bibinfo {author} {\bibfnamefont {L.}~\bibnamefont {Alonso}},
  \bibinfo {author} {\bibfnamefont {B.}~\bibnamefont {P\'erez}}, \ and\
  \bibinfo {author} {\bibfnamefont {J.~A.}\ \bibnamefont {Cerrada}},\
  }\href@noop {} {\bibfield  {journal} {\bibinfo  {journal} {Laser Part.
  Beams}\ }\textbf {\bibinfo {volume} {24}},\ \bibinfo {pages} {217} (\bibinfo
  {year} {2006})}\BibitemShut {NoStop}%
\bibitem [{\citenamefont {Stephens}\ \emph {et~al.}(2004)\citenamefont
  {Stephens}, \citenamefont {Snavely}, \citenamefont {Aglitskiy}, \citenamefont
  {Amiranoff}, \citenamefont {Andersen}, \citenamefont {Batani}, \citenamefont
  {Baton}, \citenamefont {Cowan}, \citenamefont {Freeman}, \citenamefont
  {Hall}, \citenamefont {Hatchett}, \citenamefont {Hill}, \citenamefont {Key},
  \citenamefont {King}, \citenamefont {Koch}, \citenamefont {Koenig},
  \citenamefont {MacKinnon}, \citenamefont {Lancaster}, \citenamefont
  {Martinolli}, \citenamefont {Norreys}, \citenamefont {Perelli-Cippo},
  \citenamefont {Rabec Le~Gloahec}, \citenamefont {Rousseaux}, \citenamefont
  {Santos},\ and\ \citenamefont {Scianitti}}]{stephens-fastign-pre-2004}%
  \BibitemOpen
  \bibfield  {author} {\bibinfo {author} {\bibfnamefont {R.~B.}\ \bibnamefont
  {Stephens}}, \bibinfo {author} {\bibfnamefont {R.~A.}\ \bibnamefont
  {Snavely}}, \bibinfo {author} {\bibfnamefont {Y.}~\bibnamefont {Aglitskiy}},
  \bibinfo {author} {\bibfnamefont {F.}~\bibnamefont {Amiranoff}}, \bibinfo
  {author} {\bibfnamefont {C.}~\bibnamefont {Andersen}}, \bibinfo {author}
  {\bibfnamefont {D.}~\bibnamefont {Batani}}, \bibinfo {author} {\bibfnamefont
  {S.~D.}\ \bibnamefont {Baton}}, \bibinfo {author} {\bibfnamefont
  {T.}~\bibnamefont {Cowan}}, \bibinfo {author} {\bibfnamefont {R.~R.}\
  \bibnamefont {Freeman}}, \bibinfo {author} {\bibfnamefont {T.}~\bibnamefont
  {Hall}}, \bibinfo {author} {\bibfnamefont {S.~P.}\ \bibnamefont {Hatchett}},
  \bibinfo {author} {\bibfnamefont {J.~M.}\ \bibnamefont {Hill}}, \bibinfo
  {author} {\bibfnamefont {M.~H.}\ \bibnamefont {Key}}, \bibinfo {author}
  {\bibfnamefont {J.~A.}\ \bibnamefont {King}}, \bibinfo {author}
  {\bibfnamefont {J.~A.}\ \bibnamefont {Koch}}, \bibinfo {author}
  {\bibfnamefont {M.}~\bibnamefont {Koenig}}, \bibinfo {author} {\bibfnamefont
  {A.~J.}\ \bibnamefont {MacKinnon}}, \bibinfo {author} {\bibfnamefont {K.~L.}\
  \bibnamefont {Lancaster}}, \bibinfo {author} {\bibfnamefont {E.}~\bibnamefont
  {Martinolli}}, \bibinfo {author} {\bibfnamefont {P.}~\bibnamefont {Norreys}},
  \bibinfo {author} {\bibfnamefont {E.}~\bibnamefont {Perelli-Cippo}}, \bibinfo
  {author} {\bibfnamefont {M.}~\bibnamefont {Rabec Le~Gloahec}}, \bibinfo
  {author} {\bibfnamefont {C.}~\bibnamefont {Rousseaux}}, \bibinfo {author}
  {\bibfnamefont {J.~J.}\ \bibnamefont {Santos}}, \ and\ \bibinfo {author}
  {\bibfnamefont {F.}~\bibnamefont {Scianitti}},\ }\href {\doibase
  10.1103/PhysRevE.69.066414} {\bibfield  {journal} {\bibinfo  {journal} {Phys.
  Rev. E}\ }\textbf {\bibinfo {volume} {69}},\ \bibinfo {pages} {066414}
  (\bibinfo {year} {2004})}\BibitemShut {NoStop}%
\bibitem [{\citenamefont {Westover}\ \emph {et~al.}(2011)\citenamefont
  {Westover}, \citenamefont {Chen}, \citenamefont {Patel}, \citenamefont {Key},
  \citenamefont {McLean},\ and\ \citenamefont {Beg}}]{westover-dpp-2011}%
  \BibitemOpen
  \bibfield  {author} {\bibinfo {author} {\bibfnamefont {B.}~\bibnamefont
  {Westover}}, \bibinfo {author} {\bibfnamefont {C.}~\bibnamefont {Chen}},
  \bibinfo {author} {\bibfnamefont {P.}~\bibnamefont {Patel}}, \bibinfo
  {author} {\bibfnamefont {M.}~\bibnamefont {Key}}, \bibinfo {author}
  {\bibfnamefont {H.}~\bibnamefont {McLean}}, \ and\ \bibinfo {author}
  {\bibfnamefont {F.}~\bibnamefont {Beg}},\ }\href
  {http://meetings.aps.org/link/BAPS.2011.DPP.JO6.8} {\bibfield  {journal}
  {\bibinfo  {journal} {Bull. Am. Phys. Soc.}\ }\textbf {\bibinfo {volume}
  {56}} (\bibinfo {year} {2011})},\ \bibinfo {note} {oral JO6.8, APS-DPP 2011,
  Salt Lake, USA}\BibitemShut {NoStop}%
\bibitem [{\citenamefont {Strozzi}\ \emph {et~al.}(2011)\citenamefont
  {Strozzi}, \citenamefont {Tabak}, \citenamefont {Larson}, \citenamefont
  {Marinak}, \citenamefont {Key}, \citenamefont {Divol}, \citenamefont {Kemp},
  \citenamefont {Bellei},\ and\ \citenamefont
  {Shay}}]{strozzi-ifsa-epjwc-2012}%
  \BibitemOpen
  \bibfield  {author} {\bibinfo {author} {\bibfnamefont {D.~J.}\ \bibnamefont
  {Strozzi}}, \bibinfo {author} {\bibfnamefont {M.}~\bibnamefont {Tabak}},
  \bibinfo {author} {\bibfnamefont {D.~J.}\ \bibnamefont {Larson}}, \bibinfo
  {author} {\bibfnamefont {M.~M.}\ \bibnamefont {Marinak}}, \bibinfo {author}
  {\bibfnamefont {M.~H.}\ \bibnamefont {Key}}, \bibinfo {author} {\bibfnamefont
  {L.}~\bibnamefont {Divol}}, \bibinfo {author} {\bibfnamefont {A.~J.}\
  \bibnamefont {Kemp}}, \bibinfo {author} {\bibfnamefont {C.}~\bibnamefont
  {Bellei}}, \ and\ \bibinfo {author} {\bibfnamefont {H.~D.}\ \bibnamefont
  {Shay}},\ }\href@noop {} {\bibfield  {journal} {\bibinfo  {journal}
  {Submitted to Eur. Phys. J.: Web Conf.}\ } (\bibinfo {year}
  {2011})}\BibitemShut {NoStop}%
\bibitem [{\citenamefont {Nicola\"\i}\ \emph {et~al.}(2011)\citenamefont
  {Nicola\"\i}, \citenamefont {Feugeas}, \citenamefont {Regan}, \citenamefont
  {Olazabal-Loum\'e}, \citenamefont {Breil}, \citenamefont {Dubroca},
  \citenamefont {Morreeuw},\ and\ \citenamefont
  {Tikhonchuk}}]{nicolai-xport-pre-2011}%
  \BibitemOpen
  \bibfield  {author} {\bibinfo {author} {\bibfnamefont {P.}~\bibnamefont
  {Nicola\"\i}}, \bibinfo {author} {\bibfnamefont {J.-L.}\ \bibnamefont
  {Feugeas}}, \bibinfo {author} {\bibfnamefont {C.}~\bibnamefont {Regan}},
  \bibinfo {author} {\bibfnamefont {M.}~\bibnamefont {Olazabal-Loum\'e}},
  \bibinfo {author} {\bibfnamefont {J.}~\bibnamefont {Breil}}, \bibinfo
  {author} {\bibfnamefont {B.}~\bibnamefont {Dubroca}}, \bibinfo {author}
  {\bibfnamefont {J.-P.}\ \bibnamefont {Morreeuw}}, \ and\ \bibinfo {author}
  {\bibfnamefont {V.}~\bibnamefont {Tikhonchuk}},\ }\href {\doibase
  10.1103/PhysRevE.84.016402} {\bibfield  {journal} {\bibinfo  {journal} {Phys.
  Rev. E}\ }\textbf {\bibinfo {volume} {84}},\ \bibinfo {pages} {016402}
  (\bibinfo {year} {2011})}\BibitemShut {NoStop}%
\bibitem [{\citenamefont {Robinson}\ and\ \citenamefont
  {Sherlock}(2007)}]{robinson-switchyard-pop-2007}%
  \BibitemOpen
  \bibfield  {author} {\bibinfo {author} {\bibfnamefont {A.~P.~L.}\
  \bibnamefont {Robinson}}\ and\ \bibinfo {author} {\bibfnamefont
  {M.}~\bibnamefont {Sherlock}},\ }\href {\doibase DOI:10.1063/1.2768317}
  {\bibfield  {journal} {\bibinfo  {journal} {Phys. Plasmas}\ }\textbf
  {\bibinfo {volume} {14}},\ \bibinfo {pages} {083105} (\bibinfo {year}
  {2007})}\BibitemShut {NoStop}%
\bibitem [{\citenamefont {Knauer}\ \emph {et~al.}(2010)\citenamefont {Knauer},
  \citenamefont {Gotchev}, \citenamefont {Chang}, \citenamefont {Meyerhofer},
  \citenamefont {Polomarov}, \citenamefont {Betti}, \citenamefont {Frenje},
  \citenamefont {Li}, \citenamefont {Manuel}, \citenamefont {Petrasso},
  \citenamefont {Rygg},\ and\ \citenamefont
  {S\'{e}guin}}]{knauer-bfield-pop-2010}%
  \BibitemOpen
  \bibfield  {author} {\bibinfo {author} {\bibfnamefont {J.~P.}\ \bibnamefont
  {Knauer}}, \bibinfo {author} {\bibfnamefont {O.~V.}\ \bibnamefont {Gotchev}},
  \bibinfo {author} {\bibfnamefont {P.~Y.}\ \bibnamefont {Chang}}, \bibinfo
  {author} {\bibfnamefont {D.~D.}\ \bibnamefont {Meyerhofer}}, \bibinfo
  {author} {\bibfnamefont {O.}~\bibnamefont {Polomarov}}, \bibinfo {author}
  {\bibfnamefont {R.}~\bibnamefont {Betti}}, \bibinfo {author} {\bibfnamefont
  {J.~A.}\ \bibnamefont {Frenje}}, \bibinfo {author} {\bibfnamefont {C.~K.}\
  \bibnamefont {Li}}, \bibinfo {author} {\bibfnamefont {M.~J.-E.}\ \bibnamefont
  {Manuel}}, \bibinfo {author} {\bibfnamefont {R.~D.}\ \bibnamefont
  {Petrasso}}, \bibinfo {author} {\bibfnamefont {J.~R.}\ \bibnamefont {Rygg}},
  \ and\ \bibinfo {author} {\bibfnamefont {F.~H.}\ \bibnamefont {S\'{e}guin}},\
  }\href {\doibase 10.1063/1.3416557} {\bibfield  {journal} {\bibinfo
  {journal} {Phys. Plasmas}\ }\textbf {\bibinfo {volume} {17}},\ \bibinfo {eid}
  {056318} (\bibinfo {year} {2010})}\BibitemShut {NoStop}%
\bibitem [{\citenamefont {Chang}\ \emph {et~al.}(2011)\citenamefont {Chang},
  \citenamefont {Fiksel}, \citenamefont {Hohenberger}, \citenamefont {Knauer},
  \citenamefont {Betti}, \citenamefont {Marshall}, \citenamefont {Meyerhofer},
  \citenamefont {S\'eguin},\ and\ \citenamefont
  {Petrasso}}]{chang-sphere-prl-2011}%
  \BibitemOpen
  \bibfield  {author} {\bibinfo {author} {\bibfnamefont {P.~Y.}\ \bibnamefont
  {Chang}}, \bibinfo {author} {\bibfnamefont {G.}~\bibnamefont {Fiksel}},
  \bibinfo {author} {\bibfnamefont {M.}~\bibnamefont {Hohenberger}}, \bibinfo
  {author} {\bibfnamefont {J.~P.}\ \bibnamefont {Knauer}}, \bibinfo {author}
  {\bibfnamefont {R.}~\bibnamefont {Betti}}, \bibinfo {author} {\bibfnamefont
  {F.~J.}\ \bibnamefont {Marshall}}, \bibinfo {author} {\bibfnamefont {D.~D.}\
  \bibnamefont {Meyerhofer}}, \bibinfo {author} {\bibfnamefont {F.~H.}\
  \bibnamefont {S\'eguin}}, \ and\ \bibinfo {author} {\bibfnamefont {R.~D.}\
  \bibnamefont {Petrasso}},\ }\href {\doibase 10.1103/PhysRevLett.107.035006}
  {\bibfield  {journal} {\bibinfo  {journal} {Phys. Rev. Lett.}\ }\textbf
  {\bibinfo {volume} {107}},\ \bibinfo {pages} {035006} (\bibinfo {year}
  {2011})}\BibitemShut {NoStop}%
\bibitem [{\citenamefont {Hohenberger}\ \emph {et~al.}(2011)\citenamefont
  {Hohenberger} \emph {et~al.}}]{hohenberger-bfield-dpp-2011}%
  \BibitemOpen
  \bibfield  {author} {\bibinfo {author} {\bibfnamefont {M.}~\bibnamefont
  {Hohenberger}} \emph {et~al.},\ }\href@noop {} {\bibfield  {journal}
  {\bibinfo  {journal} {Bull. Am. Phys. Soc.}\ }\textbf {\bibinfo {volume}
  {56}} (\bibinfo {year} {2011})}\BibitemShut {NoStop}%
\bibitem [{\citenamefont {Cohen}, \citenamefont {Kemp},\ and\ \citenamefont
  {Divol}(2010)}]{cohen-psc-jcp-2010}%
  \BibitemOpen
  \bibfield  {author} {\bibinfo {author} {\bibfnamefont {B.~I.}\ \bibnamefont
  {Cohen}}, \bibinfo {author} {\bibfnamefont {A.~J.}\ \bibnamefont {Kemp}}, \
  and\ \bibinfo {author} {\bibfnamefont {L.}~\bibnamefont {Divol}},\ }\href
  {\doibase 10.1016/j.jcp.2010.03.001} {\bibfield  {journal} {\bibinfo
  {journal} {J. Comput. Phys.}\ }\textbf {\bibinfo {volume} {229}},\ \bibinfo
  {pages} {4591} (\bibinfo {year} {2010})}\BibitemShut {NoStop}%
\bibitem [{\citenamefont {Kemp}\ \emph {et~al.}(2006)\citenamefont {Kemp},
  \citenamefont {Sentoku}, \citenamefont {Sotnikov},\ and\ \citenamefont
  {Wilks}}]{kemp-collisions-prl-2006}%
  \BibitemOpen
  \bibfield  {author} {\bibinfo {author} {\bibfnamefont {A.~J.}\ \bibnamefont
  {Kemp}}, \bibinfo {author} {\bibfnamefont {Y.}~\bibnamefont {Sentoku}},
  \bibinfo {author} {\bibfnamefont {V.}~\bibnamefont {Sotnikov}}, \ and\
  \bibinfo {author} {\bibfnamefont {S.~C.}\ \bibnamefont {Wilks}},\ }\href
  {\doibase 10.1103/PhysRevLett.97.235001} {\bibfield  {journal} {\bibinfo
  {journal} {Phys. Rev. Lett.}\ }\textbf {\bibinfo {volume} {97}},\ \bibinfo
  {pages} {235001} (\bibinfo {year} {2006})}\BibitemShut {NoStop}%
\bibitem [{Note1()}]{Note1}%
  \BibitemOpen
  \bibinfo {note} {By source intensity, we mean the injected kinetic energy per
  time, per transverse area in the injection plane. This differs from the $z$
  flux of kinetic energy.}\BibitemShut {Stop}%
\bibitem [{\citenamefont {Welch}\ \emph {et~al.}(2006)\citenamefont {Welch},
  \citenamefont {Rose}, \citenamefont {Cuneo}, \citenamefont {Campbell},\ and\
  \citenamefont {Mehlhorn}}]{welch-lsp-pop-2006}%
  \BibitemOpen
  \bibfield  {author} {\bibinfo {author} {\bibfnamefont {D.~R.}\ \bibnamefont
  {Welch}}, \bibinfo {author} {\bibfnamefont {D.~V.}\ \bibnamefont {Rose}},
  \bibinfo {author} {\bibfnamefont {M.~E.}\ \bibnamefont {Cuneo}}, \bibinfo
  {author} {\bibfnamefont {R.~B.}\ \bibnamefont {Campbell}}, \ and\ \bibinfo
  {author} {\bibfnamefont {T.~A.}\ \bibnamefont {Mehlhorn}},\ }\href {\doibase
  10.1063/1.2207587} {\bibfield  {journal} {\bibinfo  {journal} {Phys.
  Plasmas}\ }\textbf {\bibinfo {volume} {13}},\ \bibinfo {eid} {063105}
  (\bibinfo {year} {2006})}\BibitemShut {NoStop}%
\bibitem [{\citenamefont {Quesnel}\ and\ \citenamefont
  {Mora}(1998)}]{quesnel-elec-pre-1998}%
  \BibitemOpen
  \bibfield  {author} {\bibinfo {author} {\bibfnamefont {B.}~\bibnamefont
  {Quesnel}}\ and\ \bibinfo {author} {\bibfnamefont {P.}~\bibnamefont {Mora}},\
  }\href {\doibase 10.1103/PhysRevE.58.3719} {\bibfield  {journal} {\bibinfo
  {journal} {Phys. Rev. E}\ }\textbf {\bibinfo {volume} {58}},\ \bibinfo
  {pages} {3719} (\bibinfo {year} {1998})}\BibitemShut {NoStop}%
\bibitem [{\citenamefont {Debayle}\ \emph {et~al.}(2010)\citenamefont
  {Debayle}, \citenamefont {Honrubia}, \citenamefont {d'Humi\`eres},\ and\
  \citenamefont {Tikhonchuk}}]{debayle-diverge-pre-2010}%
  \BibitemOpen
  \bibfield  {author} {\bibinfo {author} {\bibfnamefont {A.}~\bibnamefont
  {Debayle}}, \bibinfo {author} {\bibfnamefont {J.~J.}\ \bibnamefont
  {Honrubia}}, \bibinfo {author} {\bibfnamefont {E.}~\bibnamefont
  {d'Humi\`eres}}, \ and\ \bibinfo {author} {\bibfnamefont {V.~T.}\
  \bibnamefont {Tikhonchuk}},\ }\href@noop {} {\bibfield  {journal} {\bibinfo
  {journal} {Phys. Rev. E}\ }\textbf {\bibinfo {volume} {82}},\ \bibinfo
  {pages} {036405} (\bibinfo {year} {2010})}\BibitemShut {NoStop}%
\bibitem [{\citenamefont {Gremillet}, \citenamefont {Bonnaud},\ and\
  \citenamefont {Amiranoff}(2002)}]{gremillet-fil-pop-2002}%
  \BibitemOpen
  \bibfield  {author} {\bibinfo {author} {\bibfnamefont {L.}~\bibnamefont
  {Gremillet}}, \bibinfo {author} {\bibfnamefont {G.}~\bibnamefont {Bonnaud}},
  \ and\ \bibinfo {author} {\bibfnamefont {F.}~\bibnamefont {Amiranoff}},\
  }\href {\doibase 10.1063/1.1432994} {\bibfield  {journal} {\bibinfo
  {journal} {Phys. Plasmas}\ }\textbf {\bibinfo {volume} {9}},\ \bibinfo
  {pages} {941} (\bibinfo {year} {2002})}\BibitemShut {NoStop}%
\bibitem [{\citenamefont {Honrubia}\ and\ \citenamefont {ter
  Vehn}(2009)}]{honrubia-ppcf-2009}%
  \BibitemOpen
  \bibfield  {author} {\bibinfo {author} {\bibfnamefont {J.~J.}\ \bibnamefont
  {Honrubia}}\ and\ \bibinfo {author} {\bibfnamefont {J.~M.}\ \bibnamefont {ter
  Vehn}},\ }\href@noop {} {\bibfield  {journal} {\bibinfo  {journal} {Plasma
  Phys. Controlled Fusion}\ ,\ \bibinfo {pages} {014008}} (\bibinfo {year}
  {2009})}\BibitemShut {NoStop}%
\bibitem [{\citenamefont {Davies}(2002)}]{davies-mc-pre-2002}%
  \BibitemOpen
  \bibfield  {author} {\bibinfo {author} {\bibfnamefont {J.~R.}\ \bibnamefont
  {Davies}},\ }\href {\doibase 10.1103/PhysRevE.65.026407} {\bibfield
  {journal} {\bibinfo  {journal} {Phys. Rev. E}\ }\textbf {\bibinfo {volume}
  {65}},\ \bibinfo {pages} {026407} (\bibinfo {year} {2002})}\BibitemShut
  {NoStop}%
\bibitem [{\citenamefont {Langdon}\ and\ \citenamefont
  {Barnes}(1985)}]{langdon-imppic-1985}%
  \BibitemOpen
  \bibfield  {author} {\bibinfo {author} {\bibfnamefont {A.~B.}\ \bibnamefont
  {Langdon}}\ and\ \bibinfo {author} {\bibfnamefont {D.~C.}\ \bibnamefont
  {Barnes}},\ }in\ \href@noop {} {\emph {\bibinfo {booktitle} {Multiple Time
  Scales}}},\ \bibinfo {editor} {edited by\ \bibinfo {editor} {\bibfnamefont
  {J.~U.}\ \bibnamefont {Brackbill}}\ and\ \bibinfo {editor} {\bibfnamefont
  {B.~I.}\ \bibnamefont {Cohen}}}\ (\bibinfo  {publisher} {Academic Press,
  Inc},\ \bibinfo {address} {Orlando, FL},\ \bibinfo {year} {1985})\ pp.\
  \bibinfo {pages} {335--375}\BibitemShut {NoStop}%
\bibitem [{\citenamefont {Hewett}\ and\ \citenamefont
  {Langdon}(1987)}]{hewett-imppic-jcp-1987}%
  \BibitemOpen
  \bibfield  {author} {\bibinfo {author} {\bibfnamefont {D.~W.}\ \bibnamefont
  {Hewett}}\ and\ \bibinfo {author} {\bibfnamefont {A.~B.}\ \bibnamefont
  {Langdon}},\ }\href@noop {} {\bibfield  {journal} {\bibinfo  {journal} {J.
  Comput. Phys.}\ }\textbf {\bibinfo {volume} {72}},\ \bibinfo {pages} {121}
  (\bibinfo {year} {1987})}\BibitemShut {NoStop}%
\bibitem [{\citenamefont {Drouin}\ \emph {et~al.}(2010)\citenamefont {Drouin},
  \citenamefont {Gremillet}, \citenamefont {Adam},\ and\ \citenamefont
  {H\'eron}}]{drouin-pic-jcp-2010}%
  \BibitemOpen
  \bibfield  {author} {\bibinfo {author} {\bibfnamefont {M.}~\bibnamefont
  {Drouin}}, \bibinfo {author} {\bibfnamefont {L.}~\bibnamefont {Gremillet}},
  \bibinfo {author} {\bibfnamefont {J.-C.}\ \bibnamefont {Adam}}, \ and\
  \bibinfo {author} {\bibfnamefont {A.}~\bibnamefont {H\'eron}},\ }\href
  {\doibase 10.1016/j.jcp.2010.03.015} {\bibfield  {journal} {\bibinfo
  {journal} {J. Comput. Phys.}\ }\textbf {\bibinfo {volume} {229}},\ \bibinfo
  {pages} {4781 } (\bibinfo {year} {2010})}\BibitemShut {NoStop}%
\bibitem [{\citenamefont {Solodov}\ and\ \citenamefont
  {Betti}(2008)}]{solodov-stop-pop-2008}%
  \BibitemOpen
  \bibfield  {author} {\bibinfo {author} {\bibfnamefont {A.~A.}\ \bibnamefont
  {Solodov}}\ and\ \bibinfo {author} {\bibfnamefont {R.}~\bibnamefont
  {Betti}},\ }\href {\doibase 10.1063/1.2903890} {\bibfield  {journal}
  {\bibinfo  {journal} {Phys. Plasmas}\ }\textbf {\bibinfo {volume} {15}},\
  \bibinfo {eid} {042707} (\bibinfo {year} {2008})}\BibitemShut {NoStop}%
\bibitem [{\citenamefont {Atzeni}, \citenamefont {Schiavi},\ and\ \citenamefont
  {Davies}(2009)}]{atzeni-ppcf-2009}%
  \BibitemOpen
  \bibfield  {author} {\bibinfo {author} {\bibfnamefont {S.}~\bibnamefont
  {Atzeni}}, \bibinfo {author} {\bibfnamefont {A.}~\bibnamefont {Schiavi}}, \
  and\ \bibinfo {author} {\bibfnamefont {J.~R.}\ \bibnamefont {Davies}},\
  }\href {\doibase {10.1088/0741-3335/51/1/015016}} {\bibfield  {journal}
  {\bibinfo  {journal} {{Plasma Phys. Controlled Fusion}}\ }\textbf {\bibinfo
  {volume} {{51}}} (\bibinfo {year} {{2009}}),\
  {10.1088/0741-3335/51/1/015016}}\BibitemShut {NoStop}%
\bibitem [{\citenamefont {Davies}(2008)}]{davies-stopping-dpp-2008}%
  \BibitemOpen
  \bibfield  {author} {\bibinfo {author} {\bibfnamefont {J.}~\bibnamefont
  {Davies}},\ }\href@noop {} {\bibfield  {journal} {\bibinfo  {journal} {Bull.
  Am. Phys. Soc.}\ }\textbf {\bibinfo {volume} {53}} (\bibinfo {year}
  {2008})}\BibitemShut {NoStop}%
\bibitem [{\citenamefont {Lemons}\ \emph {et~al.}(2009)\citenamefont {Lemons},
  \citenamefont {Winske}, \citenamefont {Daughton},\ and\ \citenamefont
  {Albright}}]{lemons-collision-jcp-2009}%
  \BibitemOpen
  \bibfield  {author} {\bibinfo {author} {\bibfnamefont {D.~S.}\ \bibnamefont
  {Lemons}}, \bibinfo {author} {\bibfnamefont {D.}~\bibnamefont {Winske}},
  \bibinfo {author} {\bibfnamefont {W.}~\bibnamefont {Daughton}}, \ and\
  \bibinfo {author} {\bibfnamefont {B.}~\bibnamefont {Albright}},\ }\href@noop
  {} {\bibfield  {journal} {\bibinfo  {journal} {J. Comput. Phys.}\ }\textbf
  {\bibinfo {volume} {228}},\ \bibinfo {pages} {1391} (\bibinfo {year}
  {2009})}\BibitemShut {NoStop}%
\bibitem [{\citenamefont {Manheimer}, \citenamefont {Lampe},\ and\
  \citenamefont {Joyce}(1997)}]{manheimer-pic-jcp-1997}%
  \BibitemOpen
  \bibfield  {author} {\bibinfo {author} {\bibfnamefont {W.}~\bibnamefont
  {Manheimer}}, \bibinfo {author} {\bibfnamefont {M.}~\bibnamefont {Lampe}}, \
  and\ \bibinfo {author} {\bibfnamefont {G.}~\bibnamefont {Joyce}},\
  }\href@noop {} {\bibfield  {journal} {\bibinfo  {journal} {J. Comput. Phys.}\
  }\textbf {\bibinfo {volume} {138}},\ \bibinfo {pages} {563} (\bibinfo {year}
  {1997})}\BibitemShut {NoStop}%
\bibitem [{\citenamefont {Takizuka}\ and\ \citenamefont
  {Abe}(1977)}]{takizuka-pic-jcp-1977}%
  \BibitemOpen
  \bibfield  {author} {\bibinfo {author} {\bibfnamefont {T.}~\bibnamefont
  {Takizuka}}\ and\ \bibinfo {author} {\bibfnamefont {H.}~\bibnamefont {Abe}},\
  }\href@noop {} {\bibfield  {journal} {\bibinfo  {journal} {J. Comput. Phys.}\
  }\textbf {\bibinfo {volume} {25}},\ \bibinfo {pages} {205} (\bibinfo {year}
  {1977})}\BibitemShut {NoStop}%
\bibitem [{\citenamefont {Cohen}, \citenamefont {Dimits},\ and\ \citenamefont
  {Strozzi}(2012)}]{cohen-collisions-jcp-2012}%
  \BibitemOpen
  \bibfield  {author} {\bibinfo {author} {\bibfnamefont {B.~I.}\ \bibnamefont
  {Cohen}}, \bibinfo {author} {\bibfnamefont {A.~M.}\ \bibnamefont {Dimits}}, \
  and\ \bibinfo {author} {\bibfnamefont {D.~J.}\ \bibnamefont {Strozzi}},\
  }\href@noop {} {\bibfield  {journal} {\bibinfo  {journal} {J. Comput. Phys.
  (submitted)}\ } (\bibinfo {year} {2012})}\BibitemShut {NoStop}%
\bibitem [{\citenamefont {{International Commission on Radiation Units and
  Measurements}}(1984)}]{icru37}%
  \BibitemOpen
  \bibfield  {author} {\bibinfo {author} {\bibnamefont {{International
  Commission on Radiation Units and Measurements}}},\ }\href@noop {} {\enquote
  {\bibinfo {title} {Stopping powers for electrons and positrons},}\ }\bibinfo
  {type} {Tech. Rep.}\ \bibinfo {number} {37}\ (\bibinfo {address} {Bethesda,
  MD, USA},\ \bibinfo {year} {1984})\BibitemShut {NoStop}%
\bibitem [{Note2()}]{Note2}%
  \BibitemOpen
  \bibinfo {note} {The analogous formula, Eq.~(24) in Ref.~\protect
  \rev@citealpnum {atzeni-ppcf-2009}, contains a typo in the powers of $\beta $
  and $\gamma $}\BibitemShut {NoStop}%
\bibitem [{\citenamefont {Epperlein}\ and\ \citenamefont
  {Haines}(1986)}]{epperlein-xport-pof-1986}%
  \BibitemOpen
  \bibfield  {author} {\bibinfo {author} {\bibfnamefont {E.~M.}\ \bibnamefont
  {Epperlein}}\ and\ \bibinfo {author} {\bibfnamefont {M.~G.}\ \bibnamefont
  {Haines}},\ }\href@noop {} {\bibfield  {journal} {\bibinfo  {journal} {Phys.
  Fluids}\ }\textbf {\bibinfo {volume} {29}},\ \bibinfo {pages} {1029}
  (\bibinfo {year} {1986})}\BibitemShut {NoStop}%
\bibitem [{\citenamefont {Lee}\ and\ \citenamefont
  {More}(1984)}]{lee-more-pof-1984}%
  \BibitemOpen
  \bibfield  {author} {\bibinfo {author} {\bibfnamefont {Y.~T.}\ \bibnamefont
  {Lee}}\ and\ \bibinfo {author} {\bibfnamefont {R.~M.}\ \bibnamefont {More}},\
  }\href {\doibase 10.1063/1.864744} {\bibfield  {journal} {\bibinfo  {journal}
  {Phys. Fluids}\ }\textbf {\bibinfo {volume} {27}},\ \bibinfo {pages} {1273}
  (\bibinfo {year} {1984})}\BibitemShut {NoStop}%
\bibitem [{\citenamefont {Desjarlais}(2001)}]{desjarlais-metal-cpp-2001}%
  \BibitemOpen
  \bibfield  {author} {\bibinfo {author} {\bibfnamefont {M.~P.}\ \bibnamefont
  {Desjarlais}},\ }\href@noop {} {\bibfield  {journal} {\bibinfo  {journal}
  {Contrib. Plasma Phys.}\ }\textbf {\bibinfo {volume} {41}},\ \bibinfo {pages}
  {267} (\bibinfo {year} {2001})}\BibitemShut {NoStop}%
\bibitem [{\citenamefont {Marinak}\ \emph {et~al.}(2010)\citenamefont
  {Marinak}, \citenamefont {Larson}, \citenamefont {Shay},\ and\ \citenamefont
  {Ho}}]{marinak-zuma-dpp-2010}%
  \BibitemOpen
  \bibfield  {author} {\bibinfo {author} {\bibfnamefont {M.~M.}\ \bibnamefont
  {Marinak}}, \bibinfo {author} {\bibfnamefont {D.}~\bibnamefont {Larson}},
  \bibinfo {author} {\bibfnamefont {H.~D.}\ \bibnamefont {Shay}}, \ and\
  \bibinfo {author} {\bibfnamefont {D.}~\bibnamefont {Ho}},\ }\href@noop {}
  {\bibfield  {journal} {\bibinfo  {journal} {Bull. Am. Phys. Soc.}\ }\textbf
  {\bibinfo {volume} {55}} (\bibinfo {year} {2010})},\ \bibinfo {note} {poster
  JP9 106, APS-DPP 2010, Atlanta, USA}\BibitemShut {NoStop}%
\bibitem [{\citenamefont {Grandy}(1999)}]{grandy-overlink-jcp-1999}%
  \BibitemOpen
  \bibfield  {author} {\bibinfo {author} {\bibfnamefont {J.}~\bibnamefont
  {Grandy}},\ }\href {\doibase DOI: 10.1006/jcph.1998.6125} {\bibfield
  {journal} {\bibinfo  {journal} {J. Comput. Phys.}\ }\textbf {\bibinfo
  {volume} {148}},\ \bibinfo {pages} {433 } (\bibinfo {year}
  {1999})}\BibitemShut {NoStop}%
\bibitem [{\citenamefont {Cottrill}\ \emph {et~al.}(2008)\citenamefont
  {Cottrill}, \citenamefont {Langdon}, \citenamefont {Lasinski}, \citenamefont
  {Lund}, \citenamefont {Molvig}, \citenamefont {Tabak}, \citenamefont {Town},\
  and\ \citenamefont {Williams}}]{cottrill-instab-pop-2008}%
  \BibitemOpen
  \bibfield  {author} {\bibinfo {author} {\bibfnamefont {L.~A.}\ \bibnamefont
  {Cottrill}}, \bibinfo {author} {\bibfnamefont {A.~B.}\ \bibnamefont
  {Langdon}}, \bibinfo {author} {\bibfnamefont {B.~F.}\ \bibnamefont
  {Lasinski}}, \bibinfo {author} {\bibfnamefont {S.~M.}\ \bibnamefont {Lund}},
  \bibinfo {author} {\bibfnamefont {K.}~\bibnamefont {Molvig}}, \bibinfo
  {author} {\bibfnamefont {M.}~\bibnamefont {Tabak}}, \bibinfo {author}
  {\bibfnamefont {R.~P.~J.}\ \bibnamefont {Town}}, \ and\ \bibinfo {author}
  {\bibfnamefont {E.~A.}\ \bibnamefont {Williams}},\ }\href {\doibase
  10.1063/1.2953816} {\bibfield  {journal} {\bibinfo  {journal} {Phys.
  Plasmas}\ }\textbf {\bibinfo {volume} {15}},\ \bibinfo {eid} {082108}
  (\bibinfo {year} {2008})}\BibitemShut {NoStop}%
\bibitem [{\citenamefont {Haines}(1981)}]{haines-electrotherm-pop-1981}%
  \BibitemOpen
  \bibfield  {author} {\bibinfo {author} {\bibfnamefont {M.~G.}\ \bibnamefont
  {Haines}},\ }\href {\doibase 10.1103/PhysRevLett.47.917} {\bibfield
  {journal} {\bibinfo  {journal} {Phys. Rev. Lett.}\ }\textbf {\bibinfo
  {volume} {47}},\ \bibinfo {pages} {917} (\bibinfo {year} {1981})}\BibitemShut
  {NoStop}%
\bibitem [{\citenamefont {Atzeni}, \citenamefont {Schiavi},\ and\ \citenamefont
  {Bellei}(2007)}]{atzeni-fastig-pop-2007}%
  \BibitemOpen
  \bibfield  {author} {\bibinfo {author} {\bibfnamefont {S.}~\bibnamefont
  {Atzeni}}, \bibinfo {author} {\bibfnamefont {A.}~\bibnamefont {Schiavi}}, \
  and\ \bibinfo {author} {\bibfnamefont {C.}~\bibnamefont {Bellei}},\ }\href
  {\doibase 10.1063/1.2716682} {\bibfield  {journal} {\bibinfo  {journal}
  {Phys. Plasmas}\ }\textbf {\bibinfo {volume} {14}},\ \bibinfo {eid} {052702}
  (\bibinfo {year} {2007})}\BibitemShut {NoStop}%
\bibitem [{\citenamefont {Yabuuchi}\ \emph {et~al.}(2009)\citenamefont
  {Yabuuchi}, \citenamefont {Das}, \citenamefont {Kumar}, \citenamefont
  {Habara}, \citenamefont {Kaw}, \citenamefont {Kodama}, \citenamefont {Mima},
  \citenamefont {Norreys}, \citenamefont {Sengupta},\ and\ \citenamefont
  {Tanaka}}]{yabuuchi-stopping-njp-2009}%
  \BibitemOpen
  \bibfield  {author} {\bibinfo {author} {\bibfnamefont {T.}~\bibnamefont
  {Yabuuchi}}, \bibinfo {author} {\bibfnamefont {A.}~\bibnamefont {Das}},
  \bibinfo {author} {\bibfnamefont {G.~R.}\ \bibnamefont {Kumar}}, \bibinfo
  {author} {\bibfnamefont {H.}~\bibnamefont {Habara}}, \bibinfo {author}
  {\bibfnamefont {P.~K.}\ \bibnamefont {Kaw}}, \bibinfo {author} {\bibfnamefont
  {R.}~\bibnamefont {Kodama}}, \bibinfo {author} {\bibfnamefont
  {K.}~\bibnamefont {Mima}}, \bibinfo {author} {\bibfnamefont {P.~A.}\
  \bibnamefont {Norreys}}, \bibinfo {author} {\bibfnamefont {S.}~\bibnamefont
  {Sengupta}}, \ and\ \bibinfo {author} {\bibfnamefont {K.~A.}\ \bibnamefont
  {Tanaka}},\ }\href@noop {} {\bibfield  {journal} {\bibinfo  {journal} {New J.
  Phys.}\ }\textbf {\bibinfo {volume} {11}} (\bibinfo {year}
  {2009})}\BibitemShut {NoStop}%
\bibitem [{\citenamefont {Bret}\ and\ \citenamefont
  {Deutsch}(2008)}]{bret-Nstopping-jpp-2008}%
  \BibitemOpen
  \bibfield  {author} {\bibinfo {author} {\bibfnamefont {A.}~\bibnamefont
  {Bret}}\ and\ \bibinfo {author} {\bibfnamefont {C.}~\bibnamefont {Deutsch}},\
  }\href@noop {} {\bibfield  {journal} {\bibinfo  {journal} {J. Plasma Phys.}\
  }\textbf {\bibinfo {volume} {74}},\ \bibinfo {pages} {595} (\bibinfo {year}
  {2008})}\BibitemShut {NoStop}%
\bibitem [{\citenamefont {Tabak}\ \emph {et~al.}(2010)\citenamefont {Tabak},
  \citenamefont {Shay}, \citenamefont {Strozzi}, \citenamefont {Divol},
  \citenamefont {Grote}, \citenamefont {Larson}, \citenamefont {Nuckolls},\
  and\ \citenamefont {Zimmerman}}]{tabak-dpp-2010}%
  \BibitemOpen
  \bibfield  {author} {\bibinfo {author} {\bibfnamefont {M.}~\bibnamefont
  {Tabak}}, \bibinfo {author} {\bibfnamefont {H.}~\bibnamefont {Shay}},
  \bibinfo {author} {\bibfnamefont {D.}~\bibnamefont {Strozzi}}, \bibinfo
  {author} {\bibfnamefont {L.}~\bibnamefont {Divol}}, \bibinfo {author}
  {\bibfnamefont {D.}~\bibnamefont {Grote}}, \bibinfo {author} {\bibfnamefont
  {D.}~\bibnamefont {Larson}}, \bibinfo {author} {\bibfnamefont
  {J.}~\bibnamefont {Nuckolls}}, \ and\ \bibinfo {author} {\bibfnamefont
  {G.}~\bibnamefont {Zimmerman}},\ }\href@noop {} {\bibfield  {journal}
  {\bibinfo  {journal} {Bull. Am. Phys. Soc.}\ }\textbf {\bibinfo {volume}
  {55}} (\bibinfo {year} {2010})},\ \bibinfo {note} {poster JP9.105, APS-DPP
  2010, Chicago, USA}\BibitemShut {NoStop}%
\bibitem [{Note3()}]{Note3}%
  \BibitemOpen
  \bibinfo {note} {C. Bellei, A. J. Kemp, private communication}\BibitemShut
  {NoStop}%
\bibitem [{Note4()}]{Note4}%
  \BibitemOpen
  \bibinfo {note} {H.~D.~Shay, private communication}\BibitemShut {NoStop}%
\end{thebibliography}%


\end{document}